\documentclass[twocolumn,twocolappendix,trackchanges,chicago,appendixfloats,tighten]{aastex631} 

\usepackage{newtxtext,newtxmath}
\usepackage[T1]{fontenc}   
\usepackage{enumerate}
\usepackage{enumitem}
\DeclareRobustCommand{\VAN}[3]{#2}
\let\VANthebibliography\thebibliography
\def\thebibliography{\DeclareRobustCommand{\VAN}[3]{##3}\VANthebibliography}
\usepackage{graphicx}
\usepackage{multirow} 

\usepackage[utf8]{inputenc}
\usepackage{graphicx,color}
\usepackage{mathtools}	
\usepackage{amsmath}
\usepackage{booktabs}
\usepackage{upgreek}
\usepackage{bm}
\usepackage{pstricks} 
\usepackage[normalem]{ulem}
\usepackage{cancel}
    \usepackage[british]{babel}             
\usepackage{commands}

\definecolor{mediumorchid}{rgb}{0.73, 0.33, 0.83}

\newcommand{\G}[1]{\textcolor{mediumorchid}{#1}}

\defcitealias{Chamandy+24}{Paper~I}

\begin{document}    

\title{Galactic magnetic fields II. Applying the model to nearby galaxies}
\author{Rion Glenn Nazareth}
\affiliation{National Institute of Science Education and Research, An OCC of Homi Bhabha National Institute, Bhubaneswar 752050, Odisha, India}
\author{Gayathri Santhosh}
\affiliation{Department of Physics and Astronomy, University of Bologna, via Gobetti 93/2, 40129 Bologna, Italy}
\affiliation{INAF—Istituto di Radioastronomia, via Gobetti 101, 40129 Bologna, Italy}
\affiliation{National Institute of Science Education and Research, An OCC of Homi Bhabha National Institute, Bhubaneswar 752050, Odisha, India}

\author{Luke Chamandy}
\affiliation{National Institute of Science Education and Research, An OCC of Homi Bhabha National Institute, Bhubaneswar 752050, Odisha, India}
\correspondingauthor{Luke Chamandy}
\email{lchamandy@niser.ac.in}

%
%
%
\begin{abstract}
Many spiral galaxies host magnetic fields with energy densities comparable to those of the turbulent and thermal motions of their interstellar gas. However, quantitative comparison between magnetic field properties inferred from observation and those obtained from theoretical modeling has been lacking. In Paper~I we developed a simple, axisymmetric galactic dynamo model that uses various observational data as input. Here we apply our model to calculate radial profiles of azimuthally and vertically averaged magnetic field strength and pitch angle, gas velocity dispersion and scale height, turbulent correlation time and length, and the sizes of supernova remnants for the galaxies M31, M33, M51, and NGC~6946, using input data collected from the literature. Scaling factors are introduced to account for a lack of precision in both theory and observation. Despite the simplicity of our model, its outputs agree fairly well with galaxy properties inferred from observation. Additionally, we find that most of the parameter values are similar between galaxies. We extend the model to predict the magnetic field pitch angles arising from a combination of mean-field dynamo action and the winding up of the random small-scale field owing to the large-scale radial shear. We find their magnitudes to be much smaller than those of the pitch angles measured in polarized radio and far infrared emission. This suggests that effects not included in our model, such as effects associated with spiral arms, are needed to explain the pitch angle values.
\end{abstract}
\keywords
{dynamo -- galaxies: magnetic fields -- galaxies: ISM -- radio continuum: galaxies -- turbulence -- ISM: magnetic fields -- galaxies: individual (NGC~224, NGC~598, NGC~5194, NGC~6946)}
%
%
\section{Introduction} \label{sec:intro}

Interstellar magnetic fields of strength $\sim10\mkG$ inhabit the interstellar gas 
of spiral galaxies, 
implying energy densities comparable to those of turbulent and thermal motions \citep{Beck+19}.
These magnetic fields confine cosmic rays and affect the structure and dynamics 
of the interstellar medium (ISM; e.g.~\citealt{Stil+09,Ntormousi+17,Evirgen+19}).
However, they are challenging to study due to the complexity of the physical processes 
that govern them and the indirectness of observations. 
Galactic dynamo theory 
can explain qualitatively the overall properties of galactic magnetic fields,
but quantitative comparison between theoretical models 
and observations is still rudimentary \citep{Beck+19, Shukurov+Subramanian21,Brandenburg+Ntormousi23}.

One class of theoretical approaches is concerned with modeling 
the magnetic field properties of galaxy populations from a statistical point of view 
\citep[e.g.][hereafter \citetalias{Chamandy+24}]{Rodrigues+15,Sur+18,Basu+18,Rodrigues+19,Jose+24,Chamandy+24}.
Another class involves fairly detailed modeling of specific galaxies like the Milky Way or M31
\citep{Poezd+93,Moss+98c,Moss+01,Moss+07}.
A promising hybrid approach is to model specific galaxies
focusing on a few key averaged quantities, 
like the strength and pitch angle of the mean magnetic field,
that can be compared quantitatively with inferences from observations.
If enough galaxies can be modeled,
it becomes possible to make a statistical comparison.
This approach was employed by \citet{Vaneck+15} 
using an analytic model that assumes a constant gas scale height $h$,
and then by \citet{Chamandy+16} using a somewhat more detailed model 
that includes an exponentially flared disk.
In that paper, the scale height increases with galactocentric radius $r$,
and is scaled from Milky Way data using the ratio of $r_{25}$ values.
They found that this flaring produces much better agreement with the data
than a disk of constant scale height.
In their model, they assumed a constant root-mean-square (rms) turbulent speed of $u=10\kms$
and treated the turbulent correlation time $\tau$ as an adjustable universal constant,
obtaining a best fit value of $\tau\approx14\Myr$, 
which is a few times larger than estimates from simulations
\citep[e.g.][]{Gressel+08b,Hollins+17}.
While the \citet{Chamandy+16} model has the advantage of using only a single adjustable parameter,
it may not be realistic to expect $\tau$, $u$, 
and the correlation length $l=\tau u$ to be universal constants,
and for $h(r)$ to always scale with the Milky Way profile.

Alternatively, one could model the inputs $\tau(r)$, $u(r)$, $l(r)$ and $h(r)$
in terms of observable quantities. 
This is the approach taken in the present work,
using the model presented in \citetalias{Chamandy+24},
which in turn makes use of the results in \citet{Chamandy+Shukurov20} to model 
the parameters of supernova (SN)-driven turbulence.
The above input parameters are modeled in terms of the gas, 
stellar, and star formation rate surface densities, 
$\Sg$, $\Ss$ and $\Sf$, respectively, as well as the gas temperature $T$,
which are sourced from the observational literature.
While this approach circumvents 
the need to make certain assumptions (such as those in \citealt{Chamandy+16}),
one is now forced to introduce extra parameters.
Even so, we take the view that as observations improve, 
the parameters of the model will become better constrained,
and thus this more data-based approach is a step in the right direction.

The paper is organized as follows. 
Section~\ref{sec:model} presents an overview of the model described in \citetalias{Chamandy+24}. 
In addition, we introduce magnetic field observables in Section~\ref{sec:magn_obs} 
to enable direct comparison between our model predictions and observational inferences.
The methods are discussed in Section~\ref{sec:methods},
and the results are presented in Section~\ref{sec:results}.
There we compare the model predictions with inferences made from observation
and discuss the values of the model parameters.
We summarize and provide conclusions in Section~\ref{sec:conclusions}.
\section{Model} \label{sec:model}
In \citetalias{Chamandy+24}, 
three sub-models were presented, 
each with its own prescription for turbulence driving (Models~S, Alt1 and Alt2).
In the present work, we use the fiducial model, Model~S, 
which considers explicitly the expansion of SN remnants (SNRs) and 
the transfer of SNR energy to the ISM \citep[for details, see][]{Chamandy+Shukurov20}.
In addition to including the most physics of the three sub-models, 
we showed in \citetalias{Chamandy+24} that
Model~S leads to scaling relations between 
magnetic field properties and observables 
that better agree with scaling relations inferred from observations than do the other sub-models.
We now summarize Model~S; further details can be found in \citetalias{Chamandy+24}.

The total magnetic field is written as the sum of mean and random parts $\bfB=\meanv{B}+\bfb$,
where overbar represents ensemble or volume averaging.
The random magnetic field energy can be broken up into isotropic and anisotropic contributions.
The key quantities modeled are $l(r)$, $\tau(r)$, $u(r)$, $h(r)$, the mean magnetic field strength $\mean{B}(r)$,
and the strengths of the isotropic $b\iso(r)$ and anisotropic $b\ani(r)$ parts of the random magnetic field. 

\subsection{Turbulence parameters}\label{sec:turb}
The correlation scale is given by
\begin{equation}
  \label{l}
    l=\frac{3}{10}l\SN,
\end{equation} 
which assumes that the turbulent driving scale 
is equal to the maximum radius attained by an SNR before it merges with the ISM,
\begin{equation}
\label{l_SN}
    l\SN=0.14\kpc\;\psi E_{51}^{16/51}n_{0.1}^{-19/51}c_{10}^{-1/3}.
\end{equation}
Here, 
$E_{51}=E\SN/10^{51}\erg$ is the SN energy, 
$n_{0.1}=n/0.1\cmcube$ is the gas number density, 
and $c_{10}=c\sound/10\kms$ is the sound speed. 
A dimensionless free parameter, $\psi$, of order unity is introduced to account for the uncertainty in the model. In our analysis, we assume that $E_{51}$ is fixed at $10^{51}\erg$.
The expression for the rms turbulent speed, 
obtained by equating the energy injection rate of SNe to the dissipation rate,
is
\begin{equation}
  \label{u_noSBs}
  u= \left( \frac{4\uppi}{3}l\SN^3l c\sound^2\nu 
     \right)^{1/3},
\end{equation}
where $\nu$ is the SN rate per unit volume.

In \citetalias{Chamandy+24}, the correlation time $\tau$ is equal 
to the minimum of the eddy turnover time $\tau\eddy$ and the average time 
taken for the flow to renovate after the passage of an SN blast wave $\tau\renov$.
However, 
we find that in cases where $\tau\renov<\tau\eddy$,
the renovation time 
is too small to obtain a good fit to all of the data
at small radii in two of the four galaxies studied.
Therefore, 
we choose to simplify the model of \citetalias{Chamandy+24}
by setting the correlation time equal to the eddy turnover time,
\begin{equation}
  \label{tau_eddy}
  \tau=\frac{l}{u}.    
\end{equation}

The SN rate density is related to the star formation rate surface density by
\begin{equation}
  \label{nu}
  \nu= \frac{\delta\Sigma\sfr}{2hm_\star},
\end{equation}
where $\delta$ is the fraction of stars that evolve to SNe and $m_\star$ is the average stellar mass in the galaxy. 
The mass density of gas is given by
\begin{equation}
  \label{rho}
  \rho= \frac{\Sigma}{2h},
\end{equation}
where $\Sigma$ is the gas surface density.
To convert from mass density to number density we use 
\begin{equation}\label{n}
  n= \frac{\rho}{\mu\mH},
\end{equation}
where $\mH$ is the mass of the hydrogen atom and we adopt $\mu=14/11$. 
\luke{
This value of $\mu$ corresponds to roughly $30\%$ helium fraction by mass
(not counting metals) and an ionization fraction $\ll1$, 
which is appropriate for the HI-emitting gas.
Our results are not sensitive to the choice of $\mu$.
}
Approximating the ISM to be an ideal fluid, the sound speed is given by
\begin{equation}
  \label{cs}
  c\sound= \left(\frac{\gamma_\mathrm{ad} \kB T}{\mu\mH}\right)^{1/2},
\end{equation}
where $\gamma_\mathrm{ad}=1.5$ is adopted for the adiabatic index \citep{Vandenbroucke+13}, $\kB$ is the Boltzmann constant, and $T$ is the gas temperature.

If (cold) molecular gas
is included in addition to (warm) atomic gas, 
then we use a slightly more general form of the equations where $\Sigma$ is the sum of these contributions, 
the mean molecular mass has distinct values in each phase, 
and $c\sound$ is asssumed to be lower in the cold gas (Appendix~\ref{sec:include_mol_gas}).
However, we ultimately chose \textit{not} to include molecular gas in our fiducial model (Section~\ref{sec:molecular}).

Assuming that the 1D velocity dispersion is produced by random motions that are isotropic,
we multiply the 1D gas velocity dispersion 
$\sigma$ by $\sqrt{3}$ to obtain a 3D velocity dispersion.
This can then be compared to $(u^2+c\sound^2)^{1/2}$,
with $u$ given by equation~\eqref{u_noSBs} and $c\sound$ by equation~\eqref{cs}. 
The fit can be improved by adjusting the value of the parameter $\psi$ (equation~\ref{l_SN}), 
which affects $u$.

\subsection{Gas scale height}\label{sec:h}
The scale height $h$ can be estimated from vertical hydrostatic balance, 
as
\begin{equation}
  h\approx \frac{w^2}{3\uppi G(\Sg+\Ss/\zeta)} 
  \label{h}
\end{equation}
where 
\begin{equation}
  \label{w}
  w\equiv(u^2+A^2 c\sound^2)^{1/2},
\end{equation} 
$A$ is a constant of order unity,
$\Ss$ is the surface density of stars, 
and $\zeta$ is a parameter that allows for uncertainty in the model.
We set $A=\sqrt{2}$ since the thermal pressure with $\gamma_\mathrm{ad}=3/2$
is twice the turbulent pressure $\rho u^2/3$.
If stars dominate the total surface density of the disk $\St$,
then $\Ss$ can be replaced by $\St$;
however, dark matter may contribute significantly to $\St$ in some cases. 
In practice, we simply use the best data we can find, 
whether it is for $\Ss$ or $\St$, 
and the adjustable parameter $\zeta$ gives us flexibility to account for 
the associated uncertainty.

\subsection{Mean magnetic field}\label{sec:mean}
For details of the mean-field dynamo model, 
we refer the reader to \citet{Chamandy+14b}, \citet{Chamandy16}, and \citealt{Beck+19}.
We adopt the thin disk, $\alpha$-$\Omega$, and no-$z$ approximations,
and assume circular motion about the galaxy center
($\meanv{U}= r\Omega\bfphihat$) and axisymmetry ($\del/\del\phi=0$).
Here, $\Omega$ denotes the angular rotation rate about the galaxy centre, which varies with radius.
To obtain $\meanv{B}(r)$ in the saturated (steady) state,
we solve a set of coupled non-linear mean-field dynamo equations.
An analytic solution is obtained by setting $\del/\del t=0$ and $\del/\del r=0$,
which is a fairly good approximation for a thin disk \citep{Chamandy16}.
The mean magnetic field strength is given by \citep[c.f.][]{Beck+19}
\begin{equation}
  \label{Bbar}
  \Bbar\equiv |\bm{\Bbar}|= K\frac{\pi l}{h}\left[\left(\frac{D}{D\crit} -1\right)R_\kappa\right]^{1/2}B\eq,
\end{equation}
where  $D$ is the dynamo number, 
$D\crit$ is the critical value of $D$ needed for dynamo action,
and
\begin{equation}\label{Beq}
  B\eq=\beta\sqrt{4\uppi\rho}\,u 
\end{equation}
is the field strength corresponding to energy equipartition with turbulence.
The parameter $\beta$ has been inserted to account for uncertainty in both theory
and observational inference that affects the overall magnetic field strength.
Further, $R_\kappa=\kappa/\eta$, 
where
\begin{equation}\label{eta}
  \eta = \frac{1}{3}\tau u^2
\end{equation} 
is the turbulent magnetic diffusivity and
$\kappa$ is the turbulent diffusivity responsible for the diffusive flux of
the magnetic part of the $\alpha$ effect, 
$\alpha\magn = \frac{1}{3}\tau\,\mean{\bfu\Alf\cdot\bfDel\cro\bfu\Alf}$,
with
$\bfu\Alf = \bfb/\sqrt{4\uppi\rho}$ the Alfv\'{e}n velocity.
\luke{
The quantity $\alpha\magn$ is proportional to the mean small-scale current helicity density,
which is closely related to the mean small-scale magnetic helicity density \citep{Shukurov+06}.
As $\meanv{B}$ grows, $\alpha\magn$ builds up to conserve magnetic helicity,
resulting in quenching and saturation 
\citep[e.g.][]{Pouquet+76,Brandenburg+Subramanian05a,Shukurov+Subramanian21}.
}

Motivated by simulations by \citet{Mitra+10}, we choose $R_\kappa=0.3$, 
though \citet{Gopalakrishnan+Subramanian23}
derive $R_\kappa=(7/9)(1+b^2/B\eq^2)$,
which can, in principle, exceed unity (see their equation~15).
In any case, we also
include the parameter $K$ in order to account for additional uncertainty in the theory, 
as explained in \citetalias{Chamandy+24}.
The parameters $K$ and $R_\kappa$ are thus degenerate
but kept separate because they parameterize different uncertainties; 
for the purpose of this study we fix $R_\kappa$ and then perform a fit to the data to obtain $K$.
\luke{
Given that we could have combined $KR_\kappa^{1/2}$ into a single parameter,
our model does not depend explicitly on $R_\kappa$, nor on $\alpha\magn$.
}

Further, $D= R_\alpha R_\Omega$,
where $R_\alpha\equiv \alpha\kin h/\eta$ and $R_\Omega\equiv -q\Omega h^2/\eta$,
with 
\begin{equation}\label{q}
  q\equiv - \frac{\dd\ln\Omega}{\dd \ln r}, 
\end{equation}
$\alpha\kin = -\tfrac{1}{3}\tau\,\mean{\bfu\cdot(\bfDel\cro\bfu}$),
and $\alpha=\alpha\kin+\alpha\magn$.
The quantity $\alpha\kin$ can be estimated from mean-field theory \citep{Krause+Radler80,Ruzmaikin+88,Shukurov+Subramanian21},
but which expression to use depends on the values of certain parameters 
\citepalias{Chamandy+24}. In this work,
it turns out that, for the models presented, we are always in the regime where
\begin{equation}\label{alpha_k}
  \alpha\kin = \frac{C_\alpha \tau^2 u^2\Omega}{h},
\end{equation}
with $C_\alpha$ a parameter of order unity that accounts for theoretical uncertainty in this estimate.

In the analytic solution, the critical dynamo number is given by
$D\crit= -\left(\uppi/2\right)^5$
and the pitch angle of the mean magnetic field by 
\begin{equation}
  \label{p_B}
  \tan p_B= -\frac{\uppi^2\,\tau\,u^2}{12\,q\,\Omega\, h^2},
\end{equation}
where a negative value means a spiral that is trailing with respect to the galactic rotation.
We also calculate the local (in radius) exponential growth rate,
\begin{equation}
\label{eqn:local_growth_rate_formula}
  \gamma = \frac{\pi^2\tau u^2}{12h^2}\left(\sqrt{\frac{D}{D\crit}} - 1 \right),
\end{equation}
and constrain our solutions to have $\gamma>0$.
However, it should be noted that in the kinematic regime, when $\mean{B}\ll B\eq$, 
the actual growth rate is the \textit{global} growth rate $\Gamma$,
which is comparable to (somewhat less than) the \textit{maximum} value of the local growth rate along $r$
\citep{Ruzmaikin+88,Chamandy+13a,Shukurov+Subramanian21},
so the values of $\gamma$ in the outer galaxy where the dynamo is weaker
underestimate the true growth rate there.

It is possible to obtain a higher-accuracy solution for $\meanv{B}(r)$ 
by keeping $r$-derivatives and solving the dynamo equations numerically
as an initial value problem (simulation).
As this is much more involved and would require running a large suite of simulations 
to explore the parameter space, we leave it for future work.
However, we do not expect the results to differ greatly \citep{Chamandy16}.

\subsection{Random magnetic field}\label{sec:random}
The model for the random (turbulent) magnetic field
is partly motivated by direct numerical fluctuation dynamo simulations from the literature.
Based on the results of \cite{Federrath+11} and other works, 
we choose the rms strength of the isotropic component of the random field 
to be given by
\begin{equation}
  \label{b_iso}
  b\iso= \frac{\xi\f^{1/2} B\eq}{\max(1,\Mach/A)},
\end{equation}
where $\Mach=u/c\sound$ is the turbulent sonic Mach number, 
$A=\sqrt{2}$ is chosen for convenience to be the same constant used in equation~\eqref{w}, 
$c\sound$ is the speed of sound, and $\xi\f=0.4$. 
Anisotropy of the random field is assumed to be produced by large-scale radial shear,
which leads to the following expression 
for the strength of the anisotropic component \citepalias{Chamandy+24}:
\begin{equation}
  \label{b_ani}
  b\ani\equiv \sqrt{b^2-b\iso^2}
    =\frac{b\iso}{\sqrt{3}}
      \left[2q\Omega\tau\left(1+\frac{q\Omega\tau}{2}\right)\right]^{1/2}.
\end{equation}

\begin{table*}
\caption{Distance $d$, inclination angle $i$ and logarithm of the isophotal diameter $\log{d_{25}}$ used in the study. $\log{d_{25}}$ is sourced from HyperLeda (\cite{Makarov+14}).
\label{tab:dist_inc}}
\begin{tabular}{@{}ccccccc@{}}
\hline 
Galaxy &$d$ &	Source  &$i$	   &Source  &  $\log{d_{25}}$\\
&[Mpc]&	 &	[$^\circ$]   && [0.1 arcsec]\\
\hline   
M31	&0.78 $\pm$ 0.04&  \cite{Stanek+98}  &75 $\pm$ 2        &\cite{Chemin+09}    & 3.25 $\pm$ 0.01\\ 
M33	&0.84 $\pm$ 0.01&  \cite{Breuval+23} &56 $\pm$ 1        &\cite{Zaritsky+89}  &2.79 $\pm$ 0.01\\ 
M51 &8.5  $\pm$ 0.7 &  \cite{Vinko+12}   &20 $\pm$ 5        &\cite{Tully+74a}    &2.14 $\pm$ 0.02\\ 
NGC 6946 &7.72 $\pm$ 0.32 &\cite{Anand+18}&38 $\pm$ 2&\cite{Lelli+16}            & 2.06 $\pm$ 0.01\\ 
\hline
  \end{tabular}
\end{table*}

\begin{figure*}
  \centering
  \includegraphics[width = 0.8\textwidth, clip = true,trim={0 0 0 10cm}]{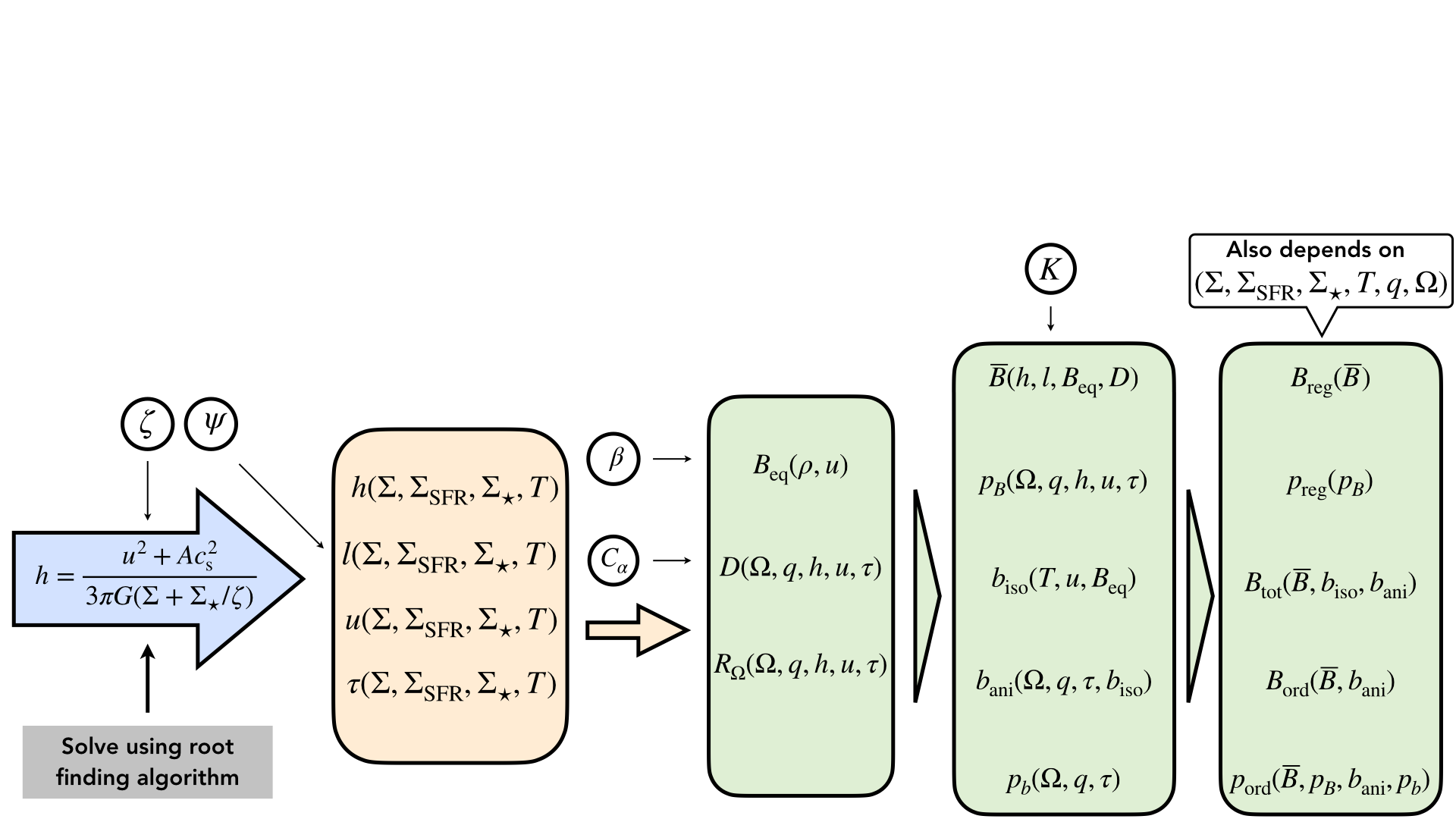}
  \caption{Flowchart demonstrating different elements in the code. The semi-analytical solution obtained for the gas scale height at each radius is used to find the parameters of turbulence and random and mean components of the magnetic field. This is further used to calculate the magnetic observables described in Section~\ref{sec:magn_obs}. Note that the adjustable parameters are depicted in circles. 
  }
  \label{fig:code-flow}
\end{figure*}

\begin{table*}
  \def\arraystretch{1.1}
  \caption{Sources of data used in this study. Wherever multiple sources are available, 
  the source listed first for the corresponding quantity is used.
  For M31, M51, and NGC~6496 we use 
  data for $\St$, and for M33 we use data for $\Ss$.
  To model the star formation rate surface density $\Sf$ of M51 and NGC 6946, 
  \cite{Kumari+20} considers the effect of the diffuse background. 
  In this study, we use the data that includes the diffuse background correction, given in appendix~D of that paper. 
  For the temperature profiles, 
  we use linear fits to data for electron temperature $T$ of the warm ionized gas vs $r$.
  For M51, the temperatures of the three types of ionization zones in \cite{Bresolin+04} are first averaged and then fitted. 
  } 
  \label{tab:datasource}
  \begin{tabular}{@{}ccccc@{}}
\hline 
Quantity		            &M31		 &M33	   &M51		  &NGC~6946		\\
\hline   
\multicolumn{5}{c}{\textit{\textbf{Model inputs}}} \\
\hline

$\Vc(r)$&C.~Carignan, priv.~comm.&\cite{Kam+17}&\cite{Sofue+18}&\cite{Sofue+18}\\           
        &\cite{Chemin+09}&\cite{Koch+18a}&\cite{Sofue+99}         &\cite{Sofue+99}	\\

$\SHI(r)$&C.~Carignan, priv.~comm. &\cite{Kam+17}&\cite{Bigiel+08}&\cite{Bigiel+08}	\\
&\cite{Chemin+09}&&\cite{Kumari+20}&\cite{Kumari+20}\\

$\SHmol(r)$		    &\cite{Nieten+06}&	\cite{Gratier+10}	   &\cite{Bigiel+08}&	\cite{Bigiel+08}	\\

$\Sf(r)$   &\cite{Tabatabaei+Berkhuijsen10}& \cite{Verley+09}		   &\cite{Bigiel+08}&	\cite{Bigiel+08}	\\

$\St(r)$ or $\Ss(r)$ &C.~Carignan, priv.~comm.&\cite{Kam+17}&\cite{Sofue+18}&\cite{Sofue+18}\\ 

$T(r)$&\cite{Tabatabaei+13b}&	\cite{Lin+17} &\cite{Bresolin+04}&\cite{Gusev+13}\\
\hline
\multicolumn{5}{c}{\textit{\textbf{Quantities used for comparison with model outputs}}} \\
\hline
$h(r)$ & \cite{Chamandy+16}& \cite{Chamandy+16}& \cite{Chamandy+16}& \cite{Chamandy+16}\\
 & &\cite{Braun91}& &  \cite{Patra+20}\\
 & && &  \cite{Bacchini+19}\\
 
$\sigma_\mathrm{HI}(r)$&C.~Carignan, priv.~comm.&\cite{Kam+17}&\cite{Hitschfeld+09}&\cite{Boomsma+08}\\
 & && &  \cite{Bacchini+19}\\
 
$B\tot(r)$& \cite{Fletcher+04}& \cite{Tabatabaei+08}& \cite{Fletcher+11}& \cite{Ehle+Beck93}\\
& & & & \cite{Basu+Roy13}\\
& & & & \cite{Beck07}\\

$B\ord(r)$& \cite{Fletcher+04}& \cite{Tabatabaei+08}& \cite{Fletcher+11}& \cite{Ehle+Beck93}\\
& & & & \cite{Beck07}\\

$B\reg(r)$ & \cite{Beck+19}& \cite{Beck+19}& \cite{Beck+19}& \cite{Beck+19}\\

$p\ord(r)$ & \cite{Beck+19}& \cite{Beck+19}& \cite{Beck+19}& \cite{Beck+19}\\ 
&&&\cite{Borlaff+21}&\cite{Borlaff+23} \\ 
&&&\cite{Borlaff+23}&\cite{Surgent+23} \\ 
&&&\cite{Surgent+23}& \\ 
$p\reg(r)$ & \cite{Fletcher+04}, \cite{Beck+19}& \cite{Tabatabaei+08}& \cite{Beck+19}& no data\\

\hline

  \end{tabular}
\end{table*}

\subsection{Theoretical modeling of magnetic field observables}\label{sec:magn_obs}
Observations of magnetic fields are indirect, 
and magnetic field parameters like the strength and pitch angle can be inferred 
from direct observables like Stokes $I$, $Q$ and $U$ only after extensive analysis and modeling.
Below, we refer to observationally derived magnetic field properties as magnetic field \textit{observables}.
A common approach, outlined in \citet{Beck+19}, 
is to model three different types of magnetic field from the observational data,
with strengths $B\reg$, $B\ord$ and $B\tot$:

\begin{itemize}
  \item The regular field $\bfB\reg$ is coherent over large scales.
  Global patterns in the Faraday rotation measure owe their existence to this field component,
  which also contributes to the linearly polarized and total synchrotron emission.
  To obtain $\bfB\reg$, 
  global azimuthal Fourier modes may be fitted to the polarization angles, 
  accounting for Faraday rotation \citep[e.g.][]{Fletcher+04}.
  
  \item However, the regular field is apparently too weak to alone explain 
  the polarized radio emission in galaxies \citep{Beck+19}.
  The strength of the net magnetic field associated with the polarized emission 
  is sometimes referred to as the ``ordered'' field strength $B\ord$. 
  It may contain contributions from 
  both a large-scale (mean) component and an anisotropic small-scale component.
  
  \item The total magnetic field inferred from the total synchrotron emission is $\bfB\tot$,
  which is sensitive to all magnetic field components 
  (random isotropic, random anisotropic and mean).
  Hence $B\tot\ge B\ord\ge B\reg$.
\end{itemize}

Mean-field dynamo models tend to assume averaging 
that satisfies the Reynolds averaging rules \citep[][\S7.2]{Shukurov+Subramanian21}.
By contrast, in observations the magnetic field is averaged implicitly over the telescope beam,
and if Fourier analysis is used to estimate $\bfB\reg$ from observational data,
this involves additional averaging over several telescope beams.
This mismatch between theory and observation complicates comparison \citep{Zhou+18,Beck+19}.
Despite this important caveat, 
it seems reasonable to compare the strength $B\reg$ and pitch angle $p\reg$ 
of the regular magnetic field (derived from observation)
with the strength $\mean{B}$ and pitch angle $p_B$ of the mean field (derived from theory).
The strength of the total field $B\tot$ can be compared 
with the strength of the total field $(\mean{B}^2 + b^2)^{1/2}$ in our theoretical model.
Finally, the strength of the ordered field can be 
compared with that of the mean field plus that part of the random field that exists due to shear,
\begin{equation}\label{Bord}
  B\ord = \left(\mean{B}^2 + b\ani^2\right)^{1/2}.
\end{equation}

\subsection{Predicting the pitch angle associated with polarized emission}
\label{sec:new_pitch}
The degree of linear polarization of radio or far-infrared emission depends on the orientation,
rather than the direction, of field lines.  
Hence, 
a magnetic field whose projection in the plane of the sky is to some degree aligned along a given axis
emits polarized emission even if it has many reversals within a telescope beam \citep[e.g.][]{Jaffe+10,Beck+19},
as expected for a small-scale random field.
By contrast, 
the mean component is generally large-scale and hence fairly uniform within a telescope beam for nearby galaxies,
so it also produces polarized emission.
The polarization angle can be converted to a pitch angle by considering 
the orientation with respect to the radial and azimuthal directions.
For radio observations, the pitch angle should be corrected for Faraday rotation.
Here and in Appendix~\ref{sec:pord} we present new theory for this pitch angle, $p\ord$.

To begin with, we first define $p_0$ to be the pitch angle of a background \textit{isotropic} 
random magnetic field,
\begin{equation}\label{p0}
   p_0 = \arctan\left(\frac{b_{r,0}}{b_{\phi,0}} \right),
\end{equation}
with $b_{r,0}$ and $b_{\phi,0}$ the radial and azimuthal components of this field component,
and $-\pi/2<p_0\le\pi/2$.
Since this field component is isotropic, the probability distribution function of $p_0$ is uniform.

Next, we derive an expression for the pitch angle $p_b$ of the random component of the magnetic field,
which is in general anisotropic. 
Below, we assume that this anisotropy is due to global radial shear.
In Appendix~\ref{sec:shear},
we show that this results in the expressions
\begin{equation}
  b_r = b_{r,0}, \qquad b_\phi \approx b_{\phi,0} - q\Omega \tau b_{r,0},
\end{equation}
where $\tau$ is the turbulent correlation time and $q$ is given by equation~\eqref{q},
and thus,
\begin{equation}\label{pb}
  p_b = \arctan\left[\frac{\tan p_0}{ 1 - q\Omega\tau \tan p_0 }\right].
\end{equation}
Note that if $p_0<0$, then $|p_b|<|p_0|$, 
but if $p_0>0$, then $p_b$ is increased from $p_0$
up to its maximum value of $\pi/2$, before becoming negative for larger $p_0$.
Hence, shear causes the field to rotate toward the $-\bfphihat$ direction in both cases,
and the net effect is to make the mean value of the pitch angle negative
(see the discussion and associated figure in Appendix~\ref{sec:shear}).
In other words, like the spiral arms of the galaxy,
the magnetic field lines tend to form spirals that are \textit{trailing} 
with respect to the sense of the galactic rotation,
which is consistent with observation.

The pitch angle of the ordered field $p\ord$ 
is given by the following expression which is motivated in Appendix~\ref{sec:pord},
\begin{equation}\label{pord}
\begin{split}
  p\ord = &\frac{1}{(2\pi)^{1/2}b\ani}\displaystyle\int_{-\infty}^\infty
     e^{-\frac{\btilde^2}{2b\ani^2}}\left[1 + \frac{2\mean{B}\btilde}{\mean{B}^2+\btilde^2}\cos\left(p_b-p_B\right)\right]\\
     &\times\arctan\left(\frac{\mean{B}\sin p_B + \btilde\sin p_b}{\mean{B}\cos p_B + \btilde\cos p_b}\right)\,d\btilde. 
\end{split}
\end{equation}

\section{Methods} \label{sec:methods}
\subsection{Input data} \label{sec:data}
The distances and inclinations used for each galaxy are listed in Table~\ref{tab:dist_inc}.
Literature references for radially-dependent data used for model inputs and for comparison with model outputs 
are provided in Table~\ref{tab:datasource}.
Data are calibrated to the chosen distance $d$ and inclination $i$,
as explained in Appendix~\ref{sec:di}.%
\footnote{The data can be obtained in tabulated form along with the code for this paper 
at \dataset[10.5281/zenodo.15118869]{https://zenodo.org/records/15118869}, and in the plotted form in the supplementary material.}
The model input data are the gas circular speed $\Vc(r)$, 
the surface mass densities of neutral hydrogen $\SHI(r)$ and molecular hydrogen $\SHmol(r)$
(though we ultimately chose not to include the latter as a contribution 
to the gas surface density $\Sg(r)$ in our fiducial model, 
as explained in Section~\ref{sec:molecular}),
the star formation rate surface density $\Sf(r)$, 
the stellar surface density $\Ss(r)$ 
or total surface density $\St(r)$ (depending on which is available),
and the temperature $T(r)$.
Here, $r=\rtilde d$ is the galactocentric radius and $\rtilde$ is the angular radius.

From the rotation curve, we obtain the angular speed $\Omega(r)=\Vc(r)/r$
and the shear parameter $q\equiv-\dd\ln\Omega/\dd\ln r$. For M31, circular velocity data is fitted using a mass model to obtain the rotation curve.
\cite{Kam+17} fits a tilted-ring model to HI velocity field to obtain the rotation curve of M33. 
\cite{Sofue+99} derives rotation curves for M51 and NGC~6946 using CO and H$\alpha$ emission lines.

We note that \luke{the gas surface density} $\Sg$ 
is typically much smaller than \luke{the stellar surface density} $\Ss$,
but the two are comparable at large radius in M33 (see Section 2 in supplementary material).
To model the gas surface density $\Sg(r)$, 
we multiply the hydrogen surface density by a factor to account for helium.
Let $\ntilde$ be the number of baryons per unit area and $\ntilde\hyd$ the number of hydrogen atoms per unit area.
The mean particle mass is approximately given by $\mu=[\ntilde\hyd +4(\ntilde-\ntilde\hyd)]/\ntilde$
if metals and molecules are neglected.
Solving, we obtain $\ntilde=3\ntilde_\mathrm{HI}/(4-\mu)$.
Since $\Sg=\ntilde\mu\mH$ and the surface density of hydrogen is 
$\Sigma_\mathrm{HI}=\ntilde\hyd\mH$, 
we obtain
\begin{equation}
  \Sg = \frac{3\mu}{4-\mu}\Sigma_\mathrm{HI}, 
  \label{eqn:sigmagas}
\end{equation}
with $\mu=14/11$. The calculation of $\Sg$ when both diffuse and molecular gas 
are included is detailed in Appendix~\ref{sec:include_mol_gas}. 

For each galaxy, 
the code finds the data set with the coarsest radial resolution and rescales the remaining data to that resolution. 
The model for each galaxy uses the common radial range 
for which data for all observables are available for that galaxy.

\subsection{Data that is compared with model outputs}\label{sec:comparison_data}
Model results are compared with observational data, and the adjustable parameters are varied to fit the data.
For this purpose, we make use of data for HI velocity dispersion $\sigma_\mathrm{HI}$,
total magnetic field strength $B\tot$, ordered magnetic field strength $B\ord$, 
regular magnetic field strength $B\reg$, ordered magnetic field pitch angle $p\ord$,
and regular magnetic field pitch angle $p\reg$ (though the latter is not available for NGC~6946).
When possible, we also compare our model results for the gas scale height $h(r)$
with models from the literature.

\subsection{Fitting procedure}\label{sec:fit}
Our primary goal is to ascertain whether our model can explain various data.
To this end, fitting the relatively small number of adjustable parameters 
is done by eye and we do not attempt to fine-tune the model to match the data.
From the model equations and experience of performing the fits,
we are confident that the adjustable parameters are not strongly degenerate.
We did not attempt to fit the model to the data using a statistical technique 
because the model is very approximate and the data have large uncertainties,
but a more rigorous approach would be useful in future work involving a larger number of galaxies.
Note that $h$, $l$, $\tau$, and $u$ are only affected by two of the parameters ($\psi$ and $\zeta$),
which makes the fitting procedure relatively straightforward. 
Moreover, for our fiducial model, we choose to set $\psi=1$, 
which effectively removes the parameter $\psi$ from the model.
In Section~\ref{sec:M31_alt} we consider an alternative model for M31 for which $\psi=2$.

\subsection{Code structure}
\label{sec:code}

The \textsc{python} code developed for this work processes observational data sourced from the references 
listed in Table~\ref{tab:datasource}, generates model predictions,
propagates errors in the observable inputs to obtain uncertainty estimates on the model outputs,
and generates plots comparing solutions from our model and empirical observations. 
The computational workflow is presented in Figure~\ref{fig:code-flow}.
    
The turbulent velocity $u$ depends on the scale height $h$ via the SN rate density $\nu$ (equation~\ref{nu}) and through the turbulent correlation length $l$ (equations~\ref{l}, \ref{l_SN}, \ref{rho}, and \ref{n}). On the other hand, $h$ depends on $u$ from equation~\eqref{h}.
This interdependence leads to a polynomial equation in $h$, 
which is solved for each radius, independently using the \textsc{fsolve} function from \textsc{SciPy} \citep{SciPy-NMeth}.
All other equations are solved algebraically using \textsc{SymPy} \citep{Meurer+17} 
to obtain solutions to the radial variation of the interstellar turbulence and magnetic field parameters.
To ensure that the equations are being solved correctly,
we verified that the scaling relations of \citetalias{Chamandy+24} 
are recovered in the asymptotic limits considered in that work.

Our code includes various options, detailed in the code documentation, 
that facilitate exploration beyond the fiducial model presented here.
 
\begin{figure*} 
    \includegraphics[width=8.3cm,keepaspectratio]{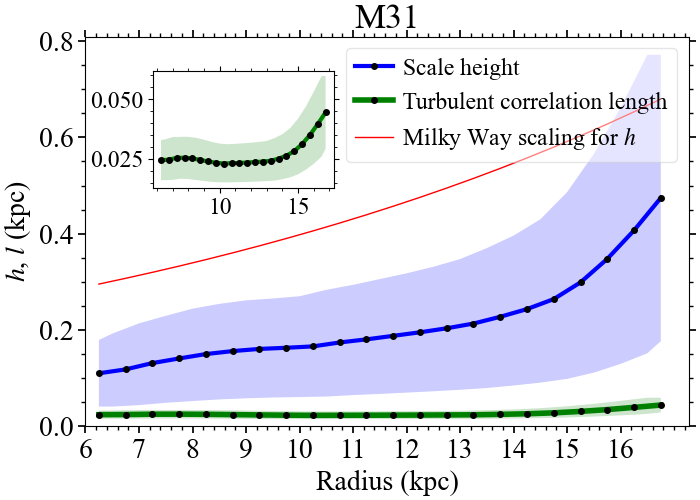}
    \includegraphics[width=8.3cm,keepaspectratio]{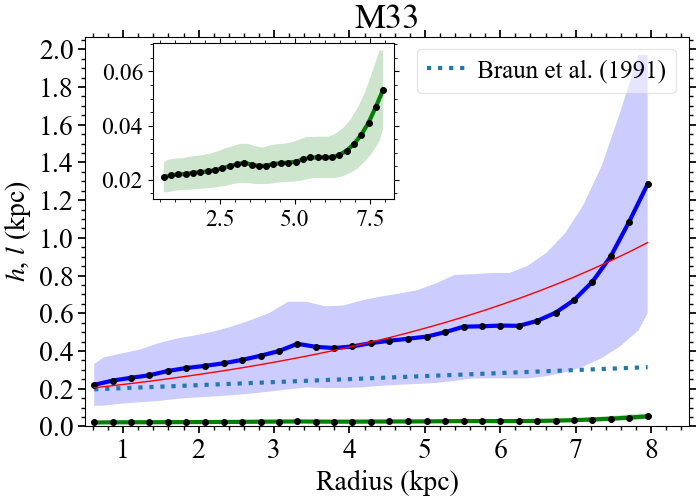}\\
    \includegraphics[width=8.3cm,keepaspectratio]{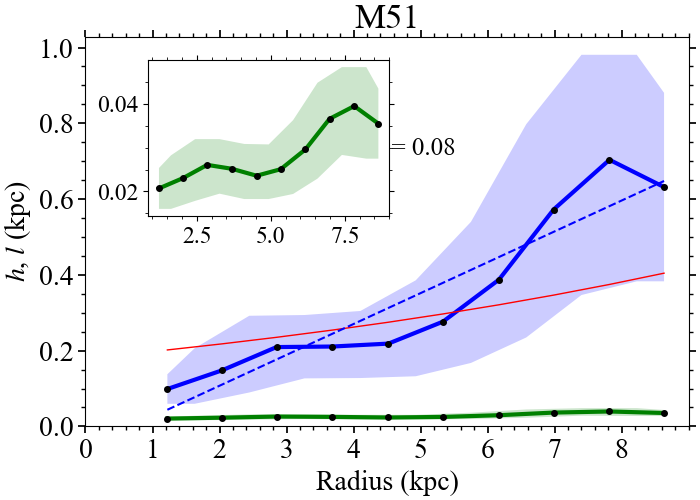}
    \includegraphics[width=8.3cm,keepaspectratio]{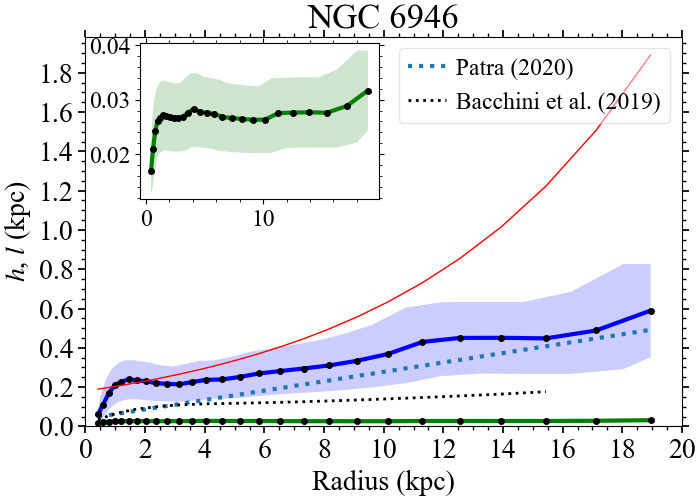}
    \caption{Scale height (blue) and turbulent correlation length (green, magnified in inset) predicted by the model. The shaded region represents the error found using the scaling relations presented in \citetalias{Chamandy+24}. The exponential scale height profile (red) uses distance and $r_{25}=d_{25}/2$ values from Table~\ref{tab:dist_inc}.
    }
    
    \label{fig:h}
\end{figure*}

\begin{figure*}
    \includegraphics[width=8.3cm,keepaspectratio]{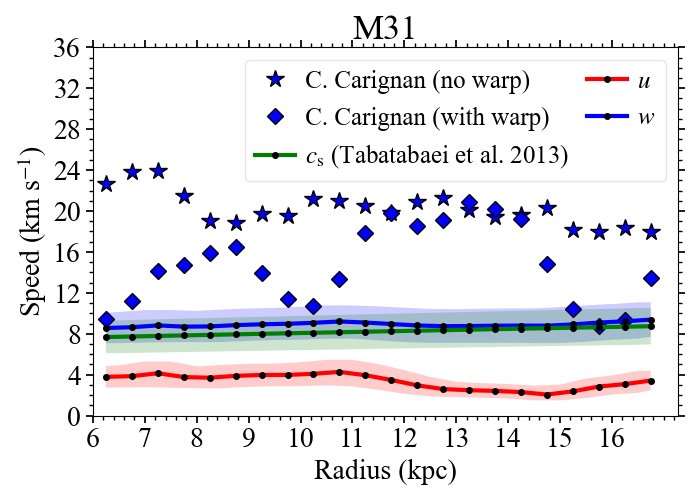}
    \includegraphics[width=8.3cm,keepaspectratio]{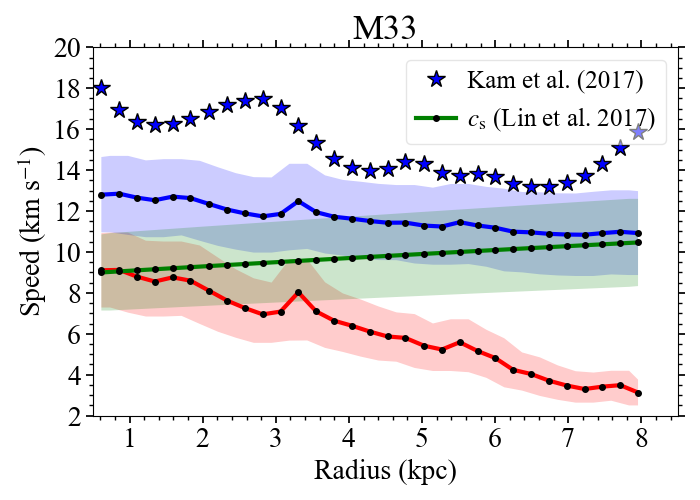}\\
    \includegraphics[width=8.3cm,keepaspectratio]{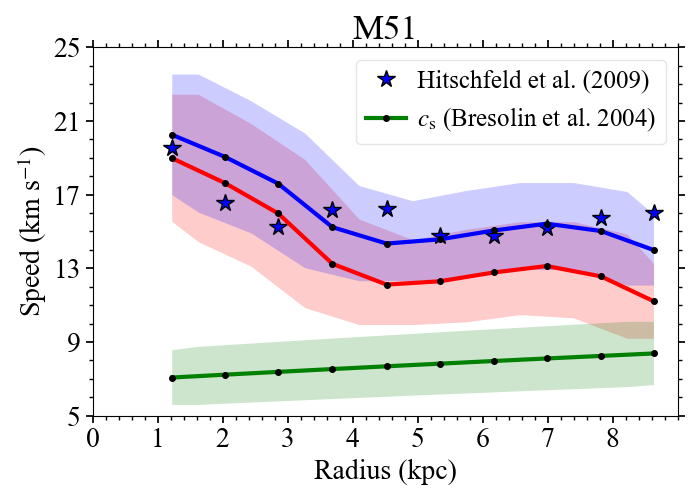}
    \includegraphics[width=8.3cm,keepaspectratio]{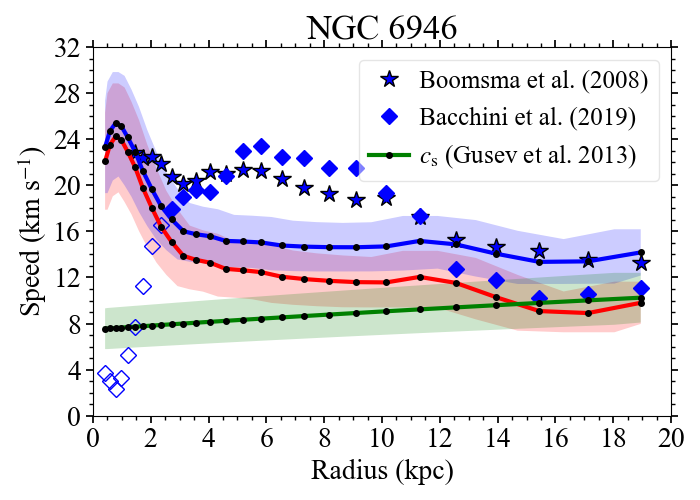}
    \caption{Turbulent velocity $u$ (red), sound speed $c_\mathrm{s}$ (green), and $w = (u^2 + c_\mathrm{s}^2)^{1/2}$ (blue) from the model. Shaded regions show associated errors. The blue points show the velocity dispersion $\sqrt{3}\sigma_{\mathrm{HI}}$, which can be compared to $w$.
    Two sets of points for M31 reflect different assumptions in modeling the observational data (C.~Carignan, priv.~comm.). The first model (stars) assumes disk warping while the second model (diamonds) assumes no warp. Note that data presented for all galaxies use the inclinations assumed in the source papers, not those listed in Table~\ref{tab:dist_inc}. In the last panel, the unfilled diamonds correspond to the data rejected by \cite{Bacchini+19} as described in Section \ref{sec:u}.
    }
    \label{fig:u}
\end{figure*}

\begin{figure*}
    \includegraphics[width=8.3cm,keepaspectratio]{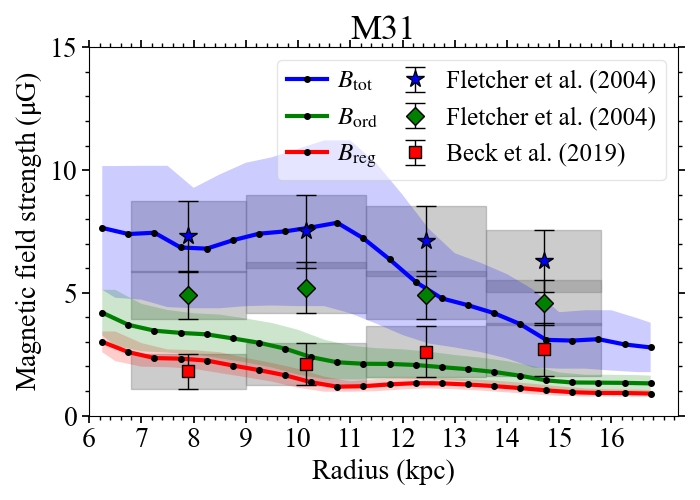}
    \includegraphics[width=8.3cm,keepaspectratio]{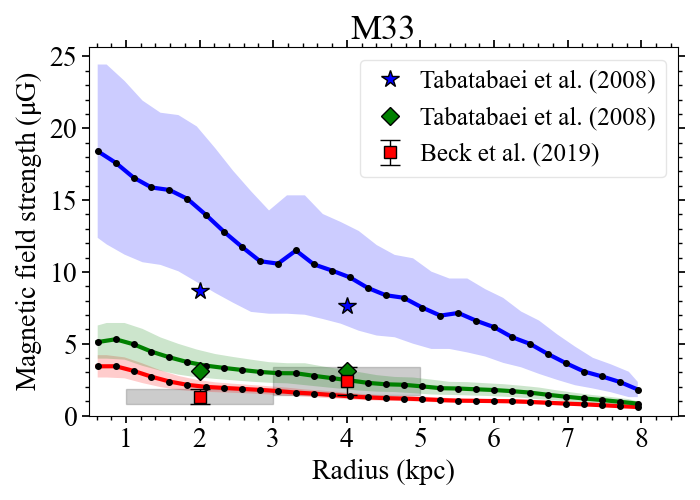}    \\
    \includegraphics[width=8.3cm,keepaspectratio]{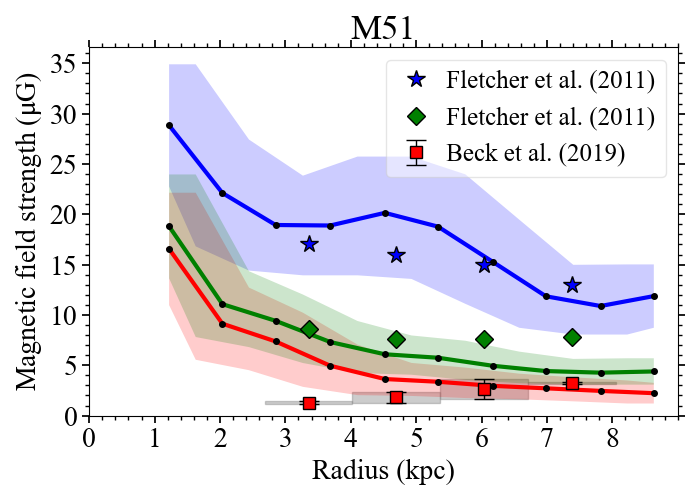}
    \includegraphics[width=8.3cm,keepaspectratio]{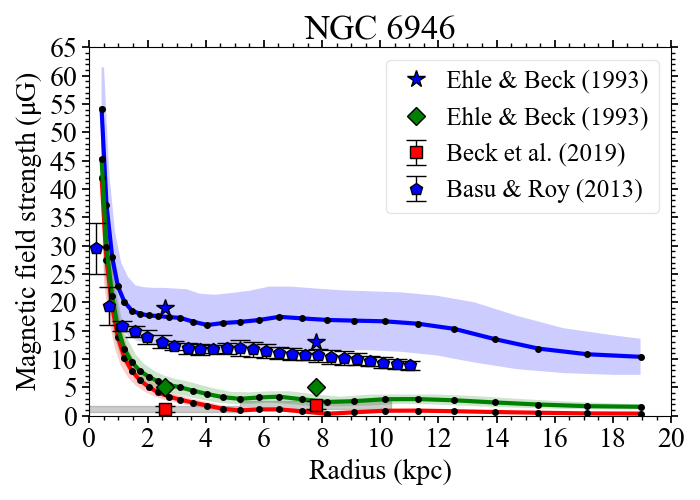}
    \caption{Strengths of the total, regular, and ordered magnetic field components from the model (blue, red and green lines respectively) compared to observational estimates from the literature. Shaded regions show associated errors. The radial coordinates for the observational data have been 
    rescaled using the distances in Table~\ref{tab:dist_inc}. An error of 20\% is assumed in $B\tot$ and $B\ord$ of M31 according to \cite{Fletcher+04}. All $B\reg$ data are computed in \cite{Beck+19}. For M51, $B\reg$ values are slightly different from those calculated in \cite{Fletcher+11}, but we adopt the \citet{Fletcher+11} error bars. 40\% error is assumed for $B\reg$ of M31, M33 and NGC 6946 according to \cite{Beck+19}. Uncertainties for the remaining data were not provided. The magnetic field model uses the thin-disk approximation ($h\ll R$) where $R$ is the radius of the galaxy. This assumption does not seem to hold true at distances less than 1 kpc from the galactic center according to Figure \ref{fig:h}. Thus, our magnetic field results are less reliable for $r<1\kpc$.
    \label{fig:B}}
\end{figure*}

\begin{figure*}
    \includegraphics[width=8.3cm,keepaspectratio]{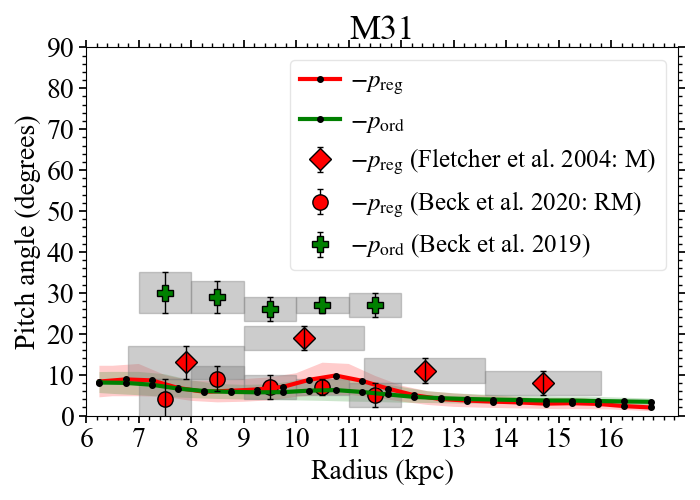}
    \includegraphics[width=8.3cm,keepaspectratio]{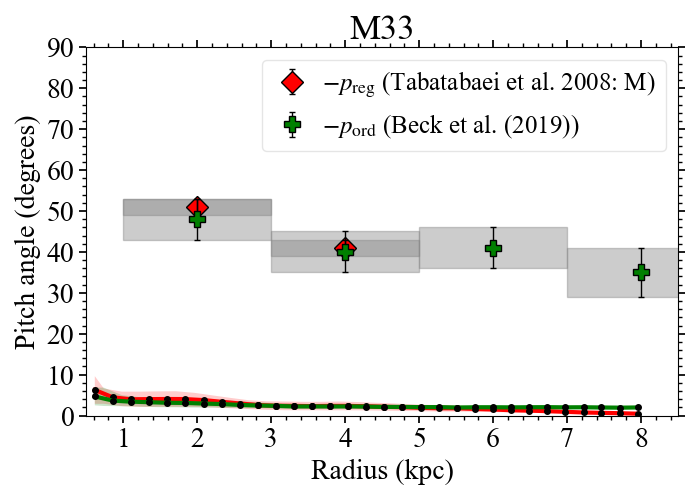}\\
    \includegraphics[width=8.3cm,keepaspectratio]{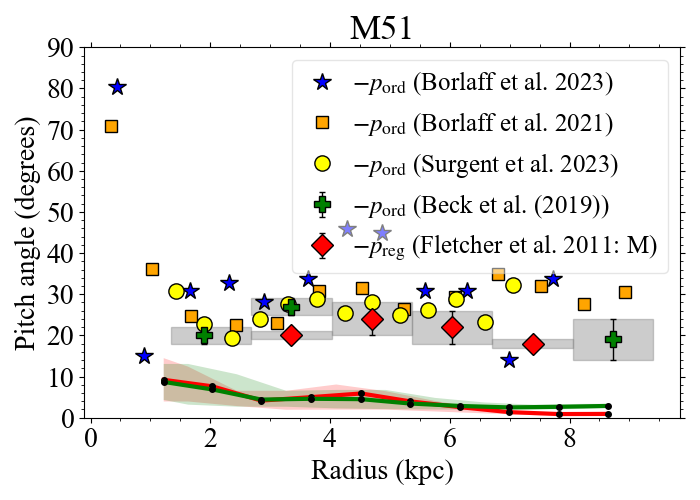}
    \includegraphics[width=8.3cm,keepaspectratio]{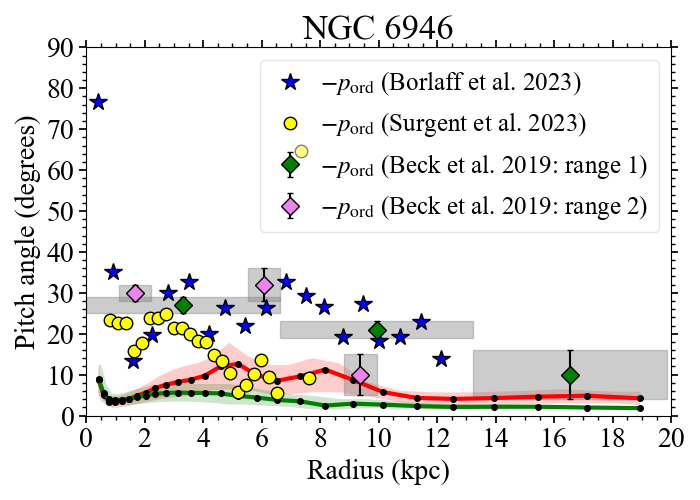}
    \caption{Pitch angles $-p\ord$ and $-p\reg$ estimated from the model and comparison with observations. Shaded regions show associated errors (Note that we do not include the dispersion in our estimated errors for the random isotropic pitch angles. The shaded region can thus be thought of as the lower bound for the errors). The `M' and `RM' denote mode analysis and rotation measure methods, respectively \citep{Beck+19}. M33 and M51 $p\reg$ data use mode analysis, while no data are available for NGC~6946. \cite{Beck+19} calculates $p\ord$ from Stokes Q and U maps, including a correction for Faraday rotation. (For M51, \cite{Borlaff+21} and \cite{Borlaff+23} attributes the larger values of their $p\ord$ at the innermost radius to resolution effects caused by the smaller number of polarization measurements available in the inner region. For NGC~6946, note that one point at $r\approx 7.2$ kpc, $-p\ord\approx -78^\circ$ of \citet{Surgent+23} does not fall within the plotting range used.) 
    }
    \label{fig:p}
\end{figure*}

\begin{figure*}
    \includegraphics[width=8.3cm,keepaspectratio]{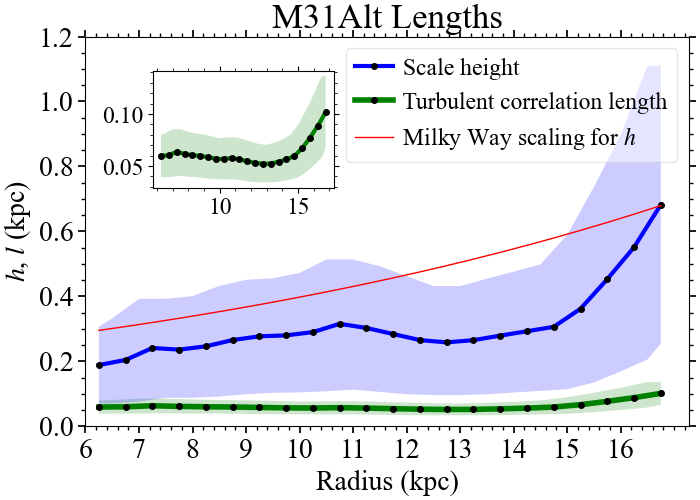}
    \includegraphics[width=8.3cm,keepaspectratio]{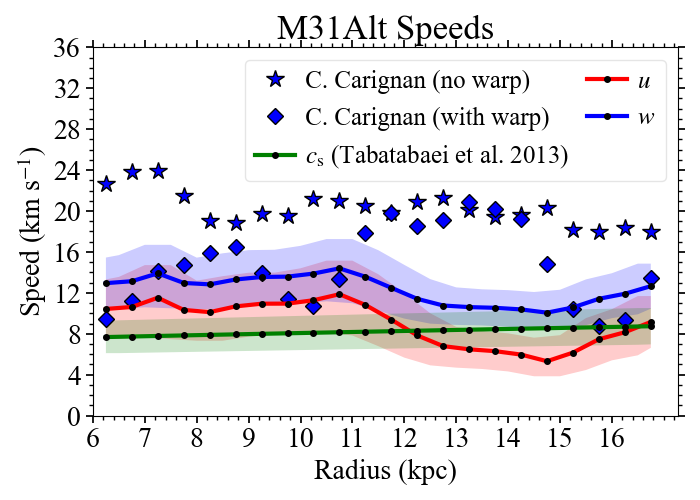}\\
    \includegraphics[width=8.3cm,keepaspectratio]{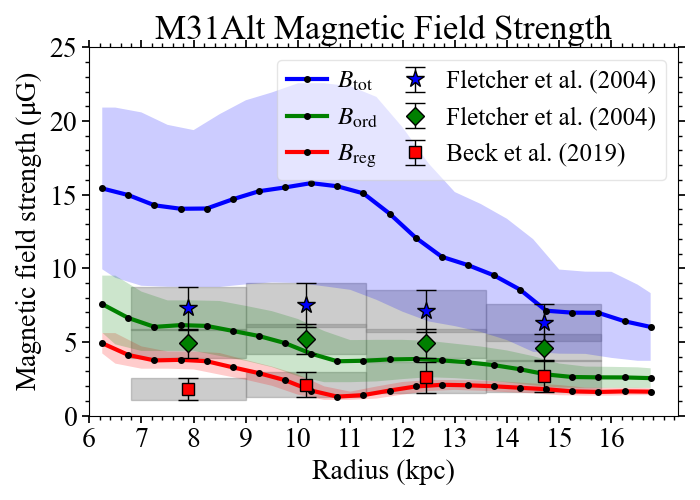}
    \includegraphics[width=8.3cm,keepaspectratio]{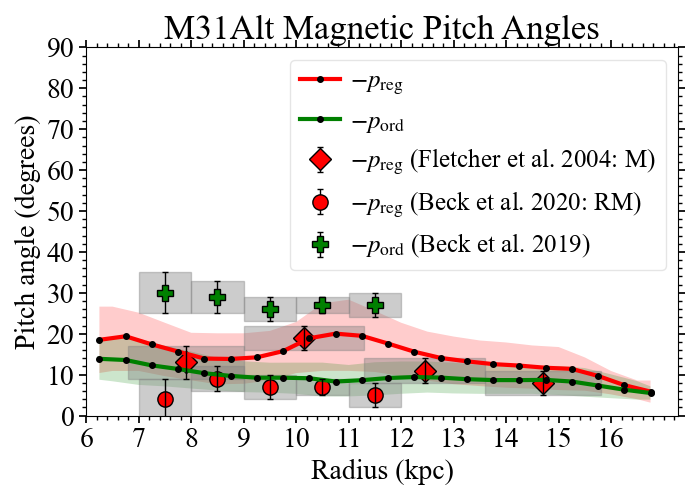}
    \caption{Alternative model for M31. This model uses $\psi=2$ instead of $\psi=1$, 
    which implies a twice larger value of the maximum radius attained by SNRs ($l\SN$) 
    as compared to the fiducial model.
    }
    \label{fig:M31_alt}
\end{figure*}

\begin{figure*}[h]
    \includegraphics[width=4.4cm,keepaspectratio]{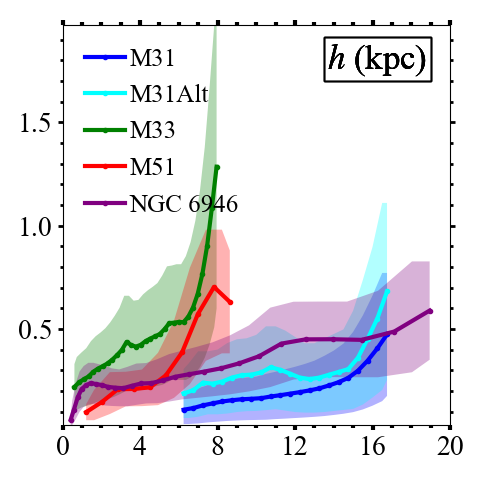}
    \includegraphics[width=4.4cm,keepaspectratio]{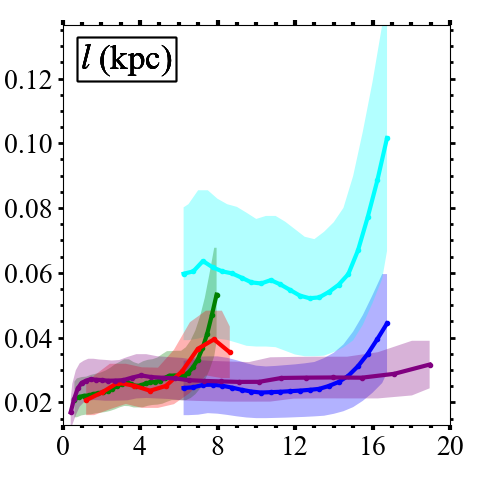}
    \includegraphics[width=4.4cm,keepaspectratio]{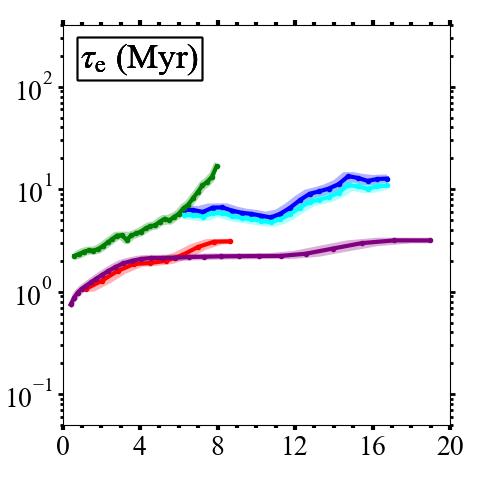}
    \includegraphics[width=4.4cm,keepaspectratio]{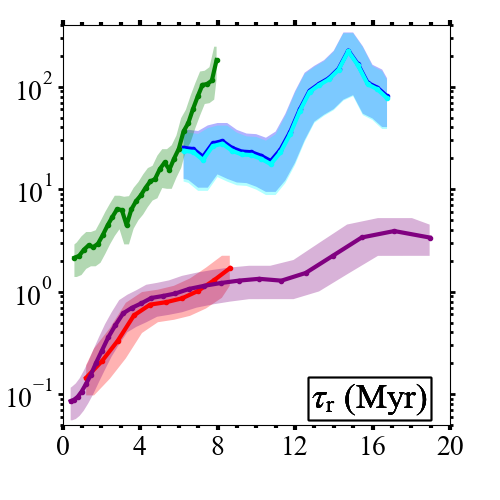}
    \\

    \vspace{-20pt}
    \includegraphics[width=4.45cm,keepaspectratio]{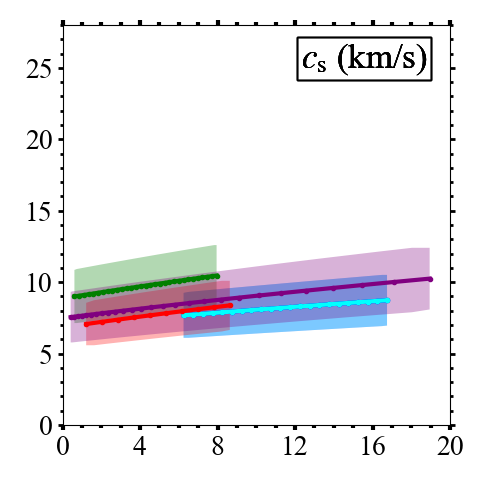}
    \includegraphics[width=4.45cm,keepaspectratio]{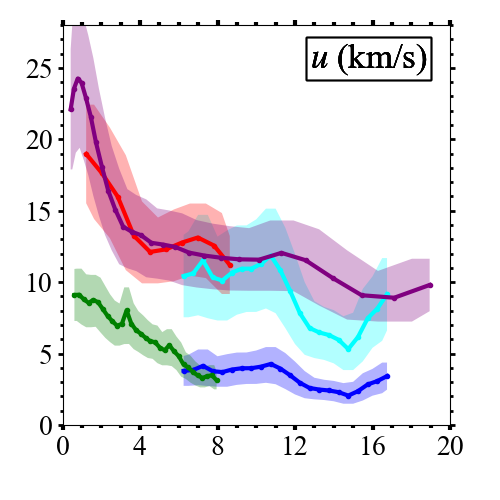}
    \includegraphics[width=4.45cm,keepaspectratio]{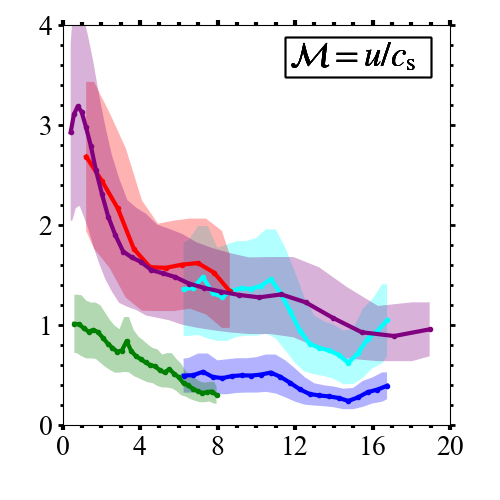}
    \includegraphics[width=4.45cm,keepaspectratio]{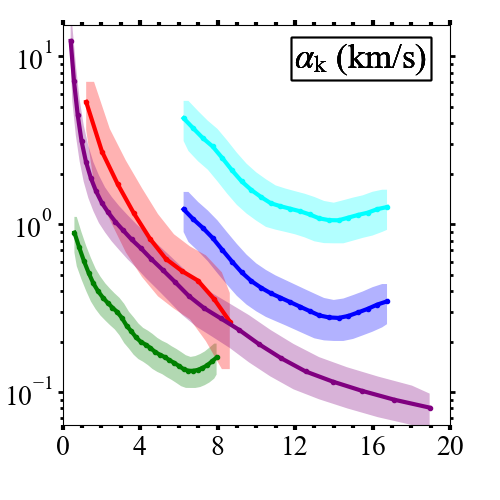}\\

    \vspace{-20pt}
    \includegraphics[width=4.45cm,keepaspectratio]{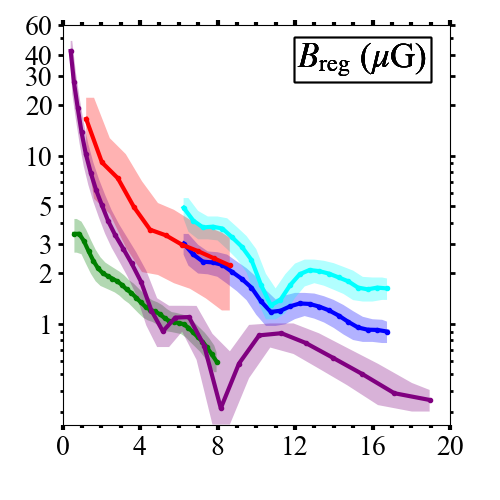}
    \includegraphics[width=4.45cm,keepaspectratio]{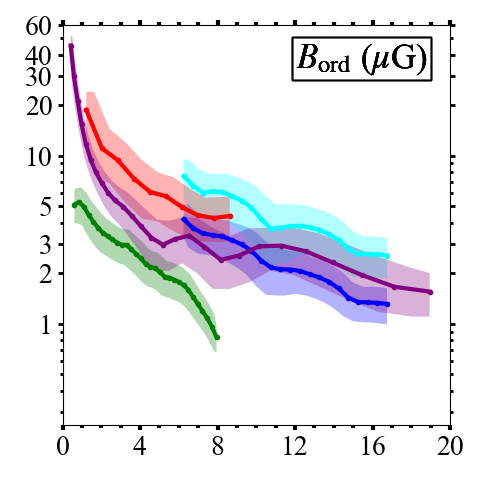}
    \includegraphics[width=4.45cm,keepaspectratio]{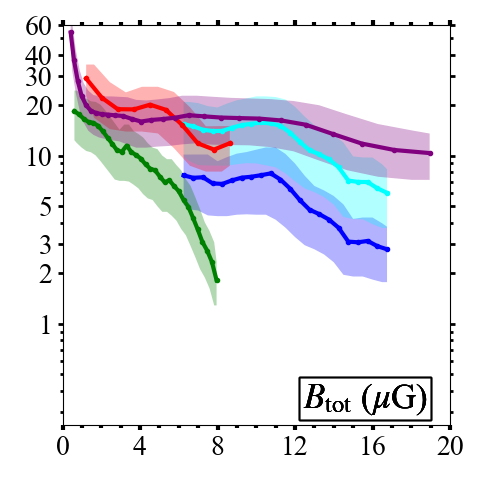}
    \includegraphics[width=4.45cm,keepaspectratio]{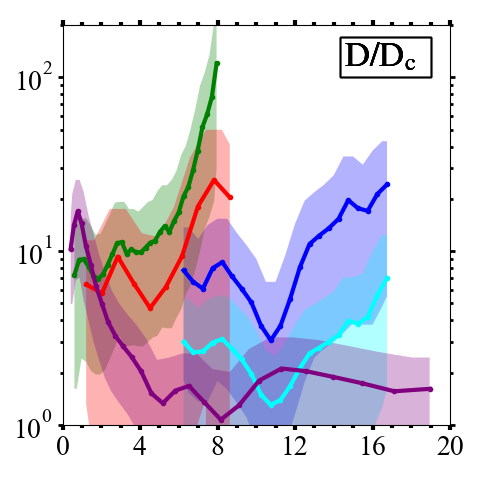}\\
    
    \vspace{-20pt}

    \includegraphics[width=4.45cm,keepaspectratio]{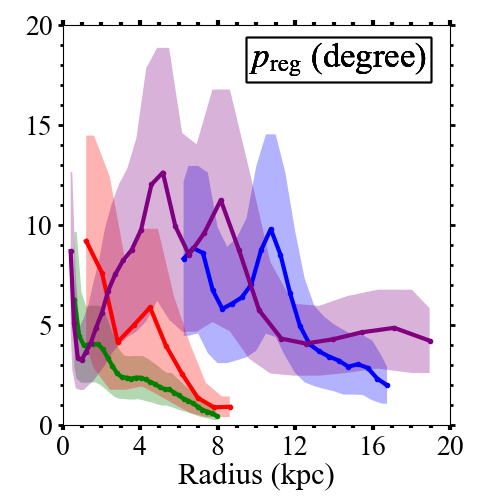}
    \includegraphics[width=4.45cm,keepaspectratio]{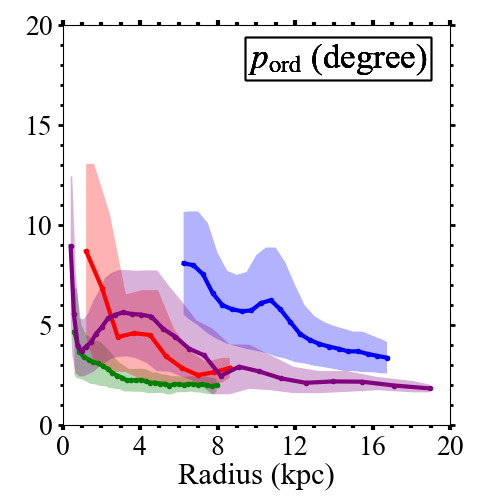}
    \includegraphics[width=4.45cm,keepaspectratio]{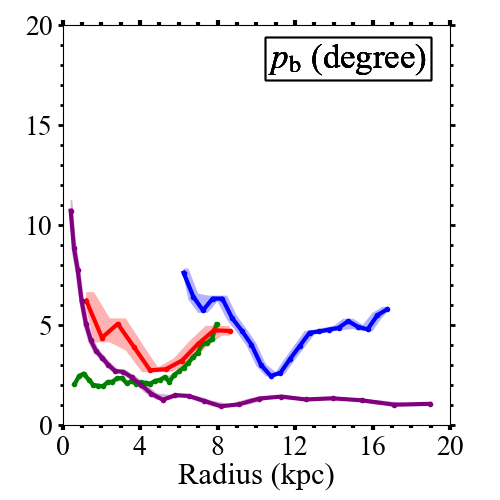}
    \includegraphics[width=4.45cm,keepaspectratio]{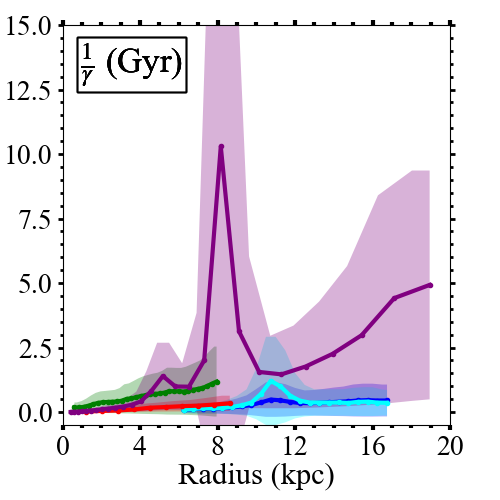}
    
    \caption{Radial variation of model outputs for all galaxies, along with associated errors (M31: blue, M31Alt (alternate model for M31 with $\psi=2$): cyan, M33: green, M51: red, NGC~6946: purple). The abscissa denotes radius in kpc and the ordinate (with unit) is provided inside each panel.
    The expression for the renovation time $\tau\renov$ is given in \citetalias{Chamandy+24}. 
    }
    \label{fig:combined_plot_r}
\end{figure*}

\section{Results and Discussion}
\label{sec:results}
Figures~\ref{fig:h}, \ref{fig:u}, \ref{fig:B}, and \ref{fig:p}
show the length scales, speeds, magnetic field strengths, and magnetic field pitch angles, respectively,
for our fiducial model.
Panels show model outputs plotted against radius, 
including shading for the estimated standard deviation
(Appendix~\ref{app_sec:erro_eq}).
Data points show the values determined from observations, along with their uncertainties (if available; 
see figure legends and captions for the references).
The model and data are seen to broadly agree, 
aside from exceptions that are discussed below.

\begin{table}
  \caption{
  Adjustable parameters of the model, with chosen values.
  The value of $\psi$ is chosen as $1$ in all the models (so the parameter $\psi$ is in effect omitted),
  but we also consider an alternative model for M31 where $\psi=2$ but all other parameter values are unchanged.
  \label{tab:params}}
  \hspace{-1.6cm}
  \begin{tabular}{@{}ccccccc@{}}
    \hline 
    Para- &Quantity  &Reference &\multicolumn{4}{c}{Value} \\
    meter &affected  &  &M31 &M33 &M51 &NGC~6946   \\
    \hline   
    $\psi$ &$l\SN(r)$ &Eq.~\eqref{l_SN} &1 (2)&1&1&1	\\ 
    $\zeta$ &$h(r)$ &Eq.~\eqref{h} &10&10&15&20	\\ 
    $\beta$ &$B\eq(r)$ &Eq.~\eqref{Beq} &7&7&7&7	\\ 
    $C_\alpha$ &$\alpha\kin(r)$ &Eq.~\eqref{alpha_k} &6&6&6&6	\\ 
    $K$ &$\mean{B}(r)$ &Eq.~\eqref{Bbar} &0.3&0.3&0.3&0.3	 \\ 
    \hline
  \end{tabular}
\end{table}

\subsection{Values of the adjustable parameters}
\label{sec:adjustable}
The best fit parameter values are listed in Table~\ref{tab:params}.
We assume that $K$ and $C_\alpha$ do not depend on other galaxy properties 
and take them to be adjustable universal constants.
Their best fit values are $K=0.3$ and $C_\alpha=6$.
We choose to make $\beta$ universal as well, finding a good fit for $\beta=7$. 
In principle, $\beta$ could depend on specifics such as the angular resolution of the observations,
but we choose to minimize the number of adjustable parameters in the model by setting it constant. 
The parameter $\psi$ is set equal to $1$ in our fiducial model, 
which means that we simply \textit{exclude} this parameter from the model,
which reduces the number of adjustable parameters.
However, we try an alternative model for M31 in Section~\eqref{sec:M31_alt} that relaxes this constraint.
Finally, the parameter $\zeta$ is allowed to vary between galaxies,
and we end up varying it between $10$ and $20$ to get good fits to the data.
Let us now consider each adjustable parameter in turn.

The best-fit value of $\beta=7$ is much larger than unity,
which suggests that either the overall magnetic field strength 
is overestimated by observational inference or underestimated by dynamo models.
Recently, \citet{Dacunha+24} compared global simulations and observations and found
that assuming that cosmic rays and magnetic fields are in energy equipartition 
tends to lead to overestimation of the total magnetic field strength from synchrotron observations
by typically an order of magnitude.
Our finding that $\beta=7$ is thus quite consistent with their finding.

On the other hand, the parameter $\zeta$ is allowed to vary between galaxies since it 
depends on quantities that are outside the scope of the model and difficult to observe,
e.g.~the stellar velocity dispersion.
We find $\zeta$ values in the range $10$--$20$.
These values are somewhat higher than other estimates (see \citealt{Forbes+12}, 
\citealt{Krumholz+18} and the discussion in \citetalias{Chamandy+24}).
A possible reason is that magnetic and cosmic ray pressure gradients, not included in the model, 
may be contributing significantly \citep[for a recent discussion of this topic in the context of the Parker instability, see][]{Tharakkal+23b,Tharakkal+23c}.

The parameter $C_\alpha$ controls the dynamo number $D$ ($D\propto C_\alpha$).
The ratio $D/D\crit$, where $D\crit$ is the critical dynamo number, 
must exceed unity for mean-field dynamo action,
and, from equation~\eqref{Bbar}, 
the strength of the mean magnetic field in the saturated state is proportional to $(D/D\crit-1)^{1/2}$.
In principle, the mean magnetic field could still grow even if this condition is not met locally,
due to the radial propagation of magnetic fronts from dynamo-active regions 
(see $\S11.6$ of \citealt{Shukurov+Subramanian21}).
However, this effect is not included in the simple mean-field dynamo model considered here.
We choose to make $C_\alpha$ as large as necessary in order to ensure 
that the mean-field dynamo is comfortably supercritical everywhere 
in all galaxies,
which leads us to adopt $C_\alpha=6$.
This suggests that the strength of the $\alpha$ effect in galaxies 
may be underestimated by using the standard formula~\eqref{alpha_k} with $C_\alpha$ set to unity.
This may simply be a consequence of a lack of precision in the theory, 
but it could be a hint that the mean-field dynamo is stronger than predicted owing to 
additional effects not included in the modeling.

The final adjustable parameter is $K$ (see equation~\ref{Bbar}),
which, like $C_\alpha$, affects the strength of the mean magnetic field in the saturated state 
($\mean{B}\propto K$).
But, unlike $C_\alpha$, 
$K$ does not affect the dynamo in the kinematic stage.
We note, however, that neither $C_\alpha$ nor $K$ affect the pitch angle of the mean magnetic field, $p_B$,
which is given by equation~\eqref{p_B}.
To obtain a reasonable match between $\mean{B}$ and $B\reg$, 
we adopt the universal value $K=0.3$.
Hence, $K$ is consistent with the approximate range of $0.1$--$1$ suggested in \citetalias{Chamandy+24}. 
Recall, however, that $K$ and $R_\kappa$ are degenerate, 
so the value of $K$ is partly a consequence of our choice $R_\kappa=0.3$ (Section~\ref{sec:mean}).

\subsection{Scale height, correlation length, and driving scale}\label{sec:h_results}
Figure~\ref{fig:h} shows the gas scale height (blue line) and turbulent correlation length (green line).
In all four galaxies, we find that the disk flares, i.e., $h$ increases with $r$, as expected
\citep[e.g.][]{Bacchini+19}.
For M33 and NGC~6946, we compare our results with scale height models from the literature. For M31, \cite{Braun91} models the spatial distribution of HI data using a simple geometric model. They obtain a linear fit with slope $16\pm3$ pc per $\!\kpc$, while we obtain a slope of $25.8\pm2.9$ pc per $\!\kpc$ for our data. 
\cite{Patra+20} assumes a three-component disc consisting of HI, $\mathrm{H}_{2}$ and stars. They numerically solve the Poisson-Boltzmann equation to obtain the density distribution of each component, and estimate the scale height as the half width at half maximum (HWHM) of the density distribution. This is different from our simpler model in which the disc is assumed to be uniform. A linear fit to the modeled scale height gives a slope of $21.2\pm1.2$ pc per $\!\kpc$, which agrees with the value of $23.9\pm0.6$ pc per $\!\kpc$ of \cite{Patra+20}. As \cite{Patra+20} points out, the difference in scale heights of NGC 6946 between \cite{Patra+20} and \cite{Bacchini+19} is due to the difference in the velocity dispersion $\sigma_{\mathrm{HI}}$ used. \cite{Bacchini+19} uses a tilted-ring model on the THINGS HI data cube while \cite{Patra+20} derives $\sigma_{\mathrm{HI}}$ from the THINGS moment map of NGC~6946. In our work, we model velocity dispersion with two components, turbulent (equation~\ref{u_noSBs}) and sound speed (equation~\ref{cs}).
For reference, we also show the Milky Way (MW) scale height scaled radially by the relative value of $r_{25}$,
taken from \citet{Chamandy+16}, which makes use of the MW model of \citet{Kalberla+Dedes08}. 

The turbulent correlation length is found to be in the range 
$17\pc\lesssim l\lesssim53\pc$ for all galaxies with a mean value close to $30\pc$.
These values are comparable to typical estimates \citep[e.g.][]{Hollins+17}.
The value of $l$ tends to increase with radius in a given galaxy,
though more gently than the scale height increases. This behavior is expected as lower gas density at larger radii allows supernova remnants to expand to larger sizes. Both the radial trends in density and temperature are included in equation~\eqref{l_SN}, but the density effect dominates for the galaxies we consider, as it decreases more rapidly than linearly, while the temperature increases only linearly. In our first paper~(\citetalias{Chamandy+24}), we also showed that the correlation length $l$ depends more strongly on gas surface density $\Sigma$ than on temperature $T$ or $\Sf$, with the exponent of $\Sigma$ ranging from –0.37 to –0.92. 

The driving scale of turbulence $l\SN$, 
which in our model is equal to the maximum radius attained by an SNR before it merges with the ISM, 
is given by equation~\eqref{l_SN}.
It is a factor $10/3$ larger than $l$, according to equation~\eqref{l};
hence 
$57\pc\lesssim l\SN\lesssim177\pc$, with a mean value of about $100\pc$.
These estimates can be compared with observational data for the diameters
of the largest 
SNRs measured in a given galaxy.
Only rough agreement should be expected because the value in the model is a theoretical limit
and the observed population is finite.
Moreover, our model predicts the sizes of SNRs that are in the process of merging with the ISM,
which would be harder to distinguish observationally.
These arguments suggest that model estimates should be larger than the largest observed sizes.
On the other hand, our model does not take into account local variation in the ISM parameters, 
which would produce scatter in the SNR sizes that increases the upper limit.
Thus, we consider agreement within a factor of a few to be acceptable.

\citet{Lee+Lee14} find SNR diameters in the range $17\pc < D < 50\pc$ for M31 out to about $15\kpc$ in radius, 
whereas we obtain $D\ma=2\max(l\SN)\approx 2\times(10/3)\times26\pc\approx 173\pc$ for $r<15\kpc$,
so more than three times larger than their maximum.
\cite{Long+10} study SNR candidates in the region within $4.3\kpc$ from the center of M33, 
and find $8\pc < D < 179\pc$. 
Our model predicts $D\ma\approx170\pc$ for $r< 4.3\kpc$, 
which agrees with their maximum, 
although they mention that several of their SNR candidates may actually be superbubbles,
which are not included in our fiducial model.
\cite{White+19} presents a larger catalog of SNR candidates with sizes $12\pc<D<183\pc$. Note that a subset of these sources also appear in \cite{Long+10}, but their sizes are estimated to be slightly larger in \cite{White+19}. 
\cite{Winkler+21} obtain $7\pc < D < 147\pc$ for SNR candidates in M51, 
while our model predicts $D\ma\approx 230\pc$, 
slightly larger than their maximum. 
For NGC~6946, \cite{Long+20} estimate diameters of SNRs within $13.3\kpc$ of the center,
and find values from $12$ to $337\pc$. 
Our model predicts $D\ma\approx 187\pc$,
which is roughly half of their maximum. Table \ref{tab:l_value_obs_summary} shows the model estimate of $D\ma=2\max(l\SN)$ from our model and the range of SNR diamaters $D$ from literature. In summary, our model predicts maximum diameters of SNRs that are usually consistent with the data to within a factor of two or three.%
\footnote{
Note that these differences are not caused by the choice of galactic distance,
as the distances used in those works are similar to those used in our model.
}

\begin{table*}
\begin{center}
  \def\arraystretch{1.1}
  \caption{Comparison of SNR sizes obtained from the fiducial model and literature values.}
  \label{tab:l_value_obs_summary}
  \hspace{-1.6cm}
  \begin{tabular}{@{}ccccc@{}}
    \hline 
    Galaxy & Model output ($D_{\mathrm{max}}$)  & Radial range & Literature value &Reference \\
    & [pc] & [kpc] & [pc] & \\
    \hline   
    M31 & 173 & $r<15$ & 17-50 & \cite{Lee+Lee14} \\
    \multirow{2}{*}{M33} & \multirow{2}{*}{170} & $r<4.3$ & 8-179 & \cite{Long+10} \\
     
     &     &  $r\approx 8$  &    12-183 & \cite{White+19} \\
    M51 & 230 & all $r$ & 7-147 & \cite{Winkler+21}	\\
    NGC~6946 & 187 & all $r$ & 12-337 & \cite{Long+20}\\
    \hline
  \end{tabular}
\end{center}
\end{table*}

\subsection{Turbulent and sound speeds}\label{sec:u}
In Figure~\ref{fig:u}, we plot the rms turbulent speed $u$ (red line), 
the sound speed $c\sound$ (green line),
and $(u^2+c\sound^2)^{1/2}$ (blue line).
Note that $c\sound$ is obtained from equation~\eqref{cs} 
after fitting temperature vs radius data with a straight line. 

In M31, the turbulence is predicted to be subsonic ($u<c\sound$), 
and the modeled 3D velocity dispersion is lower than observations (star symbols).
The data is taken from an observational model that was constructed using raw data from \citet{Chemin+09}
(C.~Carignan, priv.~comm.).
A version of this observational model that attempts to take disk warping into account is also shown (diamonds).
Our model predictions are generally lower than the data.
This discrepancy may be caused by effects other than turbulence and thermal motions 
contributing to the line width \citep[e.g.][]{Mogotsi+16}.
Even so, the low values obtained for the turbulent speed may not be realistic,
and in Section~\ref{sec:M31_alt} we explore an alternative model that addresses this apparent problem.

For M33, turbulence is again predicted by the model to be subsonic over most of the disk ($u<c\sound$), 
and the agreement between model and data is quite reasonable,
though the model values are slightly lower than the data.
For M51, the turbulence is predicted to be supersonic ($u>c\sound$),
and the level of agreement between model and data is excellent.
For NGC~6946, 
the velocity dispersion data sets by \citet{Boomsma+08} and \citet{Bacchini+19} are discrepant at small radius
but agree fairly well at other radii.
\luke{
\citet{Bacchini+19} themselves reject the innermost $14\%$ of their data points, 
stating that the drop in velocity dispersion toward smaller radius 
is likely an artefact caused by the low S/N of their data,
which have higher angular resolution as compared to those of \citet{Boomsma+08}.
This approximately corresponds to the innermost $9$ data points in our resampled \citet{Bacchini+19} data set,
which we show with open diamonds to emphasize that those data are less reliable.
}
\luke{Thus,} our model predictions agree rather well with the \luke{velocity dispersion data for NGC~6946. 
Note that our model predicts that} the turbulence is supersonic for most of the disk, 
becoming transonic ($u\approx c\sound$) for $r\ge 14\kpc$.

It should be noted that we also find that the modeled turbulent speed 
generally tends to decrease with radius, though in M31 it is flatter as compared to the other galaxies.

We also tried an alternative version of our model for which the turbulent speed 
is directly set to the velocity dispersion data (with multiplicative factor $\sqrt{3}$).
The results were qualitatively similar to those of our fiducial model, 
but using the velocity dispersion data directly leads to the mean-field dynamo being subcritical 
($D/D\crit$<1) in M31 and M33 unless the parameter $C_\alpha$ is increased to significantly larger values ($\ge10$)
that we consider to be somewhat unrealistic. 

\subsection{Magnetic field strengths and scale lengths}\label{sec:B}
Model fits for the strengths of the total (blue line), 
ordered (green line), and regular (red line) magnetic field components are shown in Figure~\ref{fig:B}.
The overall level of agreement between model and data is acceptable (after scaling the model by $\beta=7$).

The model predicts field strengths that generally decrease with radius. 
For the total field strength, the data do show a decrease with radius,
but data for the regular field strength tends to show an increase with radius, 
while for the ordered field strength there is no clear trend visible in Figure~\ref{fig:B}.
In Table~\ref{tab:scale_lengths}, we show the exponential scale lengths 
calculated by fitting exponential functions to the model results (see Appendix~\ref{sec:scale_length_calc}).
Most of the scale lengths are found to be roughly equal to $r_{25}/2$.

\citet{Beck07} plots magnetic energy density with radius for NGC~6946,
and finds a radial scale length for the total magnetic energy density ($B^2/8\pi$) 
of $7.0\pm0.1\kpc$ for $r>3\kpc$ (using that author's assumed distance to NGC~6946 of $5.5\Mpc$). 
This implies a radial scale length of $14.0\pm0.2\kpc$ for the total magnetic field strength.
\citet{Basu+Roy13} find the scale length for the total magnetic field of NGC~6946 to be $17.7\pm1.0\kpc$,
which is close to the \citet{Beck07} value.
We obtain $23\pm 4$ for the radial scale length of the total field in NGC~6946,
or $28\pm4$ when considering the same radial range used by \citet{Beck07}
(after multiplying by the ratio of assumed distances to NGC~6946, 
this radial range becomes $4.2$--$18.9\kpc$).
The scale length obtained is roughly twice the observational value, 
and the discrepancy arises because of the flatness of our $B\tot$ profile between about $5\kpc$ and $12\kpc$.
However, there is a large uncertainty in our model for $B\tot$ 
(blue shaded region in the lower-right panel of Figure~\ref{fig:B}),  
which is not factored into our error on the scale length.
Therefore, we are not overly concerned with the lack of close agreement in this case.

For the energy density of the ordered field, \citet{Beck07} obtains a scale length of $8.2\pm0.8\kpc$ for $r>6\kpc$
(using the distance assumed in that paper),
which translates to $16.4\pm1.6\kpc$ for the scale length of the ordered field strength.
We obtain a scale length of $7.8\pm 0.5$ for the entire profile of $B\ord$,
and $15\pm3\kpc$ in the radial range $8.4$--$18.9\kpc$, 
which is chosen to correspond with that used by \citet{Beck07}, 
after correcting for the choice of distance.
Thus, the model and data are in excellent agreement for the scale length of $B\ord$ in this radial range.

\begin{table*}
\begin{center}
  \def\arraystretch{1.1}
  \caption{Exponential scale length $r\tot$, $r\reg$, $r\ord$ (in kpc) and the ratio $r/r_{\mathrm{25}}$, corresponding to $B\tot$, $B\reg$ and $B\ord$ respectively for the galaxies used in the study. 
  \label{tab:scale_lengths}}
  \begin{tabular}{@{}ccccccc@{}}
    \hline 
    Galaxy & $r\tot$ & $r\tot$/$r_\mathrm{25}$ &  $r\reg$ & $r\reg$/$r_\mathrm{25}$ &$r\ord$ & $r\ord$/$r_\mathrm{25}$  \\
    \hline 
M31 & 
9.2 $\pm$ 0.9 & 
0.46 $\pm$ 0.05 &
9.2 $\pm$ 0.6 & 
0.46 $\pm$ 0.03 & 
8.8 $\pm$ 0.8  & 
0.44 $\pm$ 0.04 \\

M33 &
3.7 $\pm$ 0.2 & 
0.49 $\pm$ 0.03& 
4.7 $\pm$ 0.1 & 
0.63 $\pm$ 0.02 & 
4.5 $\pm$ 0.3  & 
0.60 $\pm$ 0.04 \\

M51 &
8.5 $\pm$ 1.1 & 
0.50 $\pm$ 0.07 & 
4.0 $\pm$ 0.5 & 
0.24 $\pm$ 0.03 & 
5.5 $\pm$ 0.4 & 
0.32 $\pm$ 0.03 \\

NGC~6946 & 
22.6 $\pm$ 4.3 & 
1.8 $\pm$ 0.3 & 
4.7 $\pm$ 0.6 & 
0.36 $\pm$ 0.05 & 
7.8 $\pm$ 0.5  & 
0.60 $\pm$ 0.04 \\ 
    \hline  
  \end{tabular}
\end{center}
\end{table*}

\subsection{Magnetic field pitch angles}\label{sec:p}
The modeled pitch angles associated with the ordered (green line), anisotropic (blue line),
and regular (red line) field components 
are plotted in Figure~\ref{fig:p}.
All pitch angles (both model and data) are negative (as expected), so we plot their absolute values. 
Aside from $|p\reg|$ in M31, where the model and data are in approximate agreement,
and $|p\ord|$ in NGC~6946, where there is partial agreement 
(note that there are no data for $|p\reg|$ for NGC~6946),
all of the model pitch angle values underpredict the values inferred from observation.
No reasonable tuning of the model parameters could significantly improve the level of agreement.

We can think of a few possible reasons for the overall poor level of agreement between modeled and observed pitch angles.
Firstly, 
we note that while the mode analysis method \citep{Fletcher+04,Tabatabaei+08,Fletcher+11} 
purports to measure the large-scale (mean) field,
an inherent assumption of the method is that the anisotropic small-scale (random) field is negligible,
which may not be true.
\citet{Beck+20} measured $|p\reg|$ using the azimuthal variation of the rotation measure (RM) in M31 
rather than the mode-analysis method (red circles in the top-left panel of Figure~\ref{fig:p}).
An advantage of the RM method is that it is not sensitive to the small-scale field.
The values found by the RM method are somewhat smaller than those inferred from the mode analysis method (red diamonds),
and much smaller than the values of $|p\ord|$, 
determined from the intrinsic orientations of the polarization angles.
Thus, perhaps pitch angles obtained using the mode analysis method are actually a kind of weighted average
between $p\reg$ and $p\ord$.
This interpretation is consistent with the pitch angles derived by mode analysis for M51 
(red diamonds in the lower left panel of Figure~\ref{fig:p}),
where the values are consistent with the $|p\ord|$ data and slightly lower than the average values of $|p\ord|$.
The larger ratio of $B\ord/B\reg$ for M51 compared to M31 (Figure~\ref{fig:B}) 
would suggest that the pitch angles found
by mode analysis should be closer to $|p\ord|$ as compared to those in M31, which they are.
However, M33 does not quite fit this interpretation because the pitch angles from mode analysis 
(red diamonds in the upper-right panel) are essentially equal to $|p\ord|$, 
despite $B\ord/B\reg$ being smaller than for M51.
On the other hand, observational estimates of $B\ord$ and $B\reg$ rely on different sets of assumptions, 
so the values of $B\ord/B\reg$ may lack accuracy.
In any case, it seems likely that the mode analysis method tends to overpredict the value of $|p\reg|$,
owing to the presence of an anisotropic random component of the magnetic field.

The above reasoning may help to explain why the data and model do not generally agree for $|p\reg|$,
but a different reason is needed to explain the lack of agreement for $|p\ord|$,
where the theoretical predictions are smaller than the observations.
Having said that, there is quite a bit of scatter in the $p\ord$ data, 
and for NGC~6946 the theoretical prediction (green line in the lower-right panel of Figure~\ref{fig:p}) 
does marginally agree with some of the data points.
Nevertheless, it seems likely that the theory developed in this work to determine $p\ord$ is inadequate.
One possibility
is that $p\ord$ is influenced by the spiral arms and their associated non-axisymmetric flows.
Indeed, 
$p\ord$ and $p\arm$ data are correlated, as are $p\reg$ and $p\arm$ data,
as shown in Appendix~\ref{sec:pcorrel}, which suggests that important effects stemming from the spiral structure of the galaxy
are missing from our simple axisymmetric model, 
and influence both mean and random components of the magnetic field.
\luke{
Future work to include the effects of compression and shear associated with spiral arms
is needed to address this limitation,
along the lines outlined in $\S13.10$ of \citet{Shukurov+Subramanian21} (see also~\citealt{Beck+05}).
}

We have also not attempted to include the generation of small-scale random magnetic field 
by the turbulent tangling of the mean field, 
which could influence both the strength and pitch angle of the random field.
Simulations by \citet{Gent+24} find that tangling can amplify the random field to energy densities over two orders of magnitude higher than those of the mean field.

To summarize, the shortcomings of our model in reproducing the pitch angle data 
suggest directions for improving both the data analysis and theoretical models in the future.



\subsection{Alternative model for M31}\label{sec:M31_alt}
As discussed in Section~\ref{sec:u}, our modeled turbulent speed is quite low in M31, being only $2$--$4\kms$, 
which is about a factor of two less than the modeled sound speed, and a factor of several less 
than the 3D velocity dispersion data.
Thus, for this galaxy, we deemed it worthwhile to explore an alternative model, M31Alt,
for which the parameter $\psi$ is introduced (equation~\ref{l_SN})
and $\psi=2$ is chosen, based on the goodness of fit with the various data.
The values of the other parameters are kept the same as in the fiducial model.

The key outputs already presented for our fiducial model 
are presented for Model~M31Alt in Figure~\ref{fig:M31_alt}.
It can be seen that the turbulent speed (red line, top-right panel) is now comparable to the sound speed,
and that turbulence is mildly supersonic for $r<12\kpc$.
On the other hand, the scale height (top-left panel) is no longer fully monotonic.
A more serious concern is that doubling $\psi$ causes $l\SN$ to double, 
and thus the prediction for $D\ma$, which was already too high (Section~\ref{sec:h_results}), 
to also double. 
This may suggest that turbulence is also driven at larger scales than those attained by individual SNRs,
perhaps by superbubbles.

Turning to the field strengths (bottom-left panel of Figure~\ref{fig:M31_alt}), 
the level of agreement is slightly superior to that obtained in the fiducial model 
(top-left panel of Figure~\ref{fig:B}).
Likewise, the modeled $p\reg$ values (bottom-right panel of Figure~\ref{fig:M31_alt}) 
are now seen to agree more closely with the data obtained from mode analysis
and less well with that obtained from the RM fitting as compared to the fiducial model
(top-left panel of Figure~\ref{fig:p}), though the predictions for $|p\ord|$ are still too small.

We see then that Model~M31Alt produces somewhat better agreement with most of the data than the fiducial M31 model,
with the important exception of the maximum SNR diameters, 
and with the cost that an extra adjustable parameter ($\psi$) must be introduced.
In any case, these results for M31Alt demonstrate that by adjusting the driving scale of turbulence
($l\SN\propto\psi$)
by only a factor of two (which is certainly plausible) it becomes possible to alleviate
the discrepancy in the predicted and observed 3D velocity dispersions seen in the fiducial model.
We note that we also tried adjusting $\psi$ from unity for M33, M51, and NGC~6946,
but found that any improvements in the level of agreement with data that this allows are rather marginal,
so we decided to retain $\psi=1$ for those galaxies.
Since our goal in introducing the free parameters was to account for uncertainty, setting $\psi = 1$ for the other three galaxies is preferred, as it effectively removes one free parameter.

\subsection{Comparison between galaxies}\label{sec:comparison_galaxies}
In Figure~\ref{fig:combined_plot_r} we plot various modeled quantities for all four galaxies.
Our fiducial model is shown, as well as Model~M31Alt, which is presented using cyan color. 
A similar figure in Appendix~\ref{app_sec:output_vs_r_r25} plots output quantities against $r/r_{25}$.
In most cases, values of quantities are fairly similar from galaxy to galaxy,
and also show similar trends with radius.
This provides confidence that the model gives results that are reasonably consistent between galaxies.

The correlation time $\tau$ varies between about $1\Myr$ and $10\Myr$, 
and tends to increase with radius.
Note that the renovation time $\tau\renov$ is shown for comparison but is not used in the 
fiducial model presented in this work, 
where we adopt $\tau=\tau\eddy$ rather than $\tau=\min(\tau\renov,\tau\eddy)$
as in \citetalias{Chamandy+24}.
Adopting instead the definition of \citetalias{Chamandy+24}
affects significantly the value of $\tau$ in some cases. 
We find that $\tau\renov<\tau\eddy$ (and hence 
$\tau=\tau\renov$ according to the prescription of \citetalias{Chamandy+24}) 
for all radii in M51 and for $r\lesssim15\kpc$ in NGC~6946.
In these galaxies, $\tau$ is reduced by up to an order of magnitude at $r=0$ 
relative to the fiducial model, and by a factor of $2$--$3$ for most other radii.
Such small values of $\tau$ cannot be ruled out by observations and simulations.
Simulations, in particular, tend to use input parameter values similar to the Solar neighborhood,
so it is possible that for some galaxies $\tau$ could be significantly smaller 
than $1\Myr$ near the centers of galaxies.
On the other hand, in our fiducial model where $\tau=\tau\eddy$, 
we obtain greater consistency in the range of $\tau$ between galaxies.
Moreover, the value of $\tau$ does not significantly affect most other quantities in our model.
For M51 and NGC~6946, the values of the pitch angles $|p\reg|$ and $p\ord$ are somewhat smaller at small radii when the original \citetalias{Chamandy+24} model where $\tau=\min(\tau\renov,\tau\eddy)$ is adopted
rather than $\tau=\tau\eddy$,
while for M31 and M33 the results for the two models are the same since $\tau\eddy<\tau\renov$ at all radii.
A comparison between results for the magnetic field strength and pitch angle for M51 and NGC~6946 
for the two models for $\tau$ is shown in Appendix~\ref{sec:tau_models}. 

Figure~\ref{fig:combined_plot_r}
also allows one to see clear trends with radius that are rather consistent between galaxies.
The turbulent speed $u$, sonic Mach number $\mathcal{M}$,
and kinetic $\alpha$ effect of the mean-field dynamo all tend to decrease with radius,
as do the strengths of the magnetic field components (Section~\ref{sec:B}).
The scale height, turbulent correlation length, turbulent correlation time, 
and local mean-field dynamo growth rate $\gamma$ tend to increase with radius.

\begin{figure}
    \includegraphics[width=8.3cm,keepaspectratio]{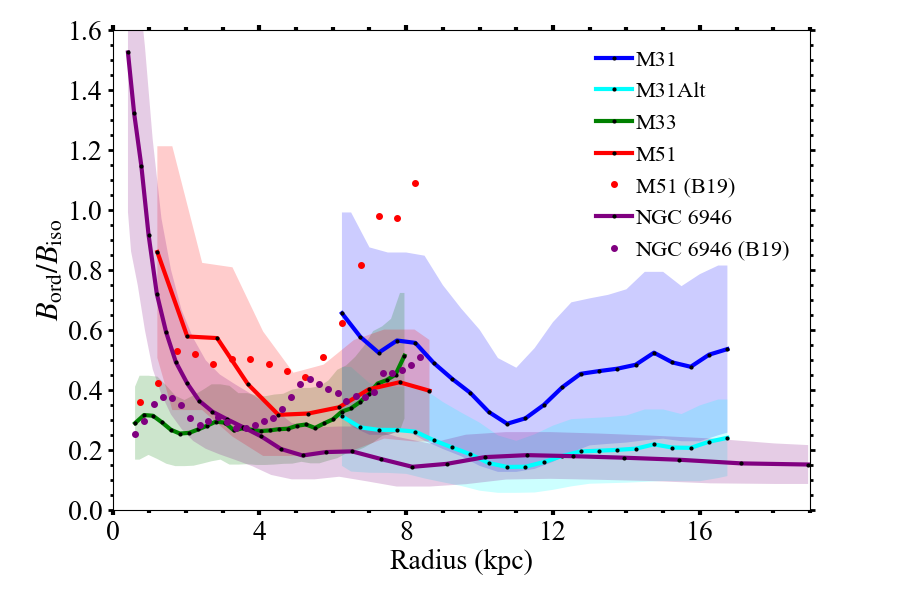}
    \caption{Ratio of ordered to isotropic components of magnetic field for all the galaxies, and comparison with values from \cite{Beck+19} (B19) for NGC 6946 (purple dots) and M51 (red dots).
    }
    \label{fig:bord_biso_ratio}
\end{figure}

In Figure~\ref{fig:bord_biso_ratio}, we plot the ratio $B\ord/B\iso$, 
and include data for M51 and NGC~6946 from Figure~2 of \citet{Beck+19}.
For these two galaxies, the model and data do not agree in detail,
but do agree fairly well for part of the radial range.

\subsection{Two-phase interstellar medium model}\label{sec:molecular}
Our model effectively averages over the phases of the interstellar gas.
We do not attempt to include input data for the hot gas ($T\sim10^6\K$). 
While this phase may have a large fractional volume, 
its contribution to the gas mass in the disk is likely to be small
\citep[e.g.][]{Gent+13a}.
Moreover, partly owing to its transient nature \citep{Evirgen+17}, 
the hot phase has been found to play a subdominant role 
in the galactic dynamo compared to the warm phase.

The cold gas of temperature $\lesssim100\K$ is
concentrated in molecular clouds and has a low fractional volume,
which suggests that it may be unimportant for modeling properties of the mean and random magnetic field components
averaged over large scales.
\luke{
Furthermore, compression during gravitational collapse 
likely leads to an enhancement of the saturated magnetic field strength in molecular clouds, 
an effect not included in our model \citep{Sur+12,Muhammed+25}.
For these reasons, we choose to exclude molecular gas from our fiducial model.
}

On the other hand, \luke{the molecular} phase can contribute a large fraction of the total gas mass, 
so\luke{, despite these caveats,} 
we decided to \luke{also} try a version of our model that includes molecular gas data as input,
in addition to HI data. 
As the surface density of molecular gas is subdominant in M31 and M33 (see supplementary material)
the solutions are only slightly affected by including the molecular gas for those galaxies,
and the fits remain reasonable using the same parameter values.
However, for M51 and NGC~6946 molecular gas tends to dominate at inner radii,
and including it has a large effect on solutions.
To obtain a reasonable estimate for the sound speed, 
we adopted a mass-weighted average of $c\sound^2$, 
assuming a molecular gas temperature equal to one tenth 
of the temperature used in the fiducial model.
We obtain scale heights more than an order of magnitude smaller than our fiducial model
at small radius \luke{for these two galaxies}.
This causes the dynamo number to be subcritical, which quenches mean-field dynamo action.%
\footnote{This problem is still present, though to a lesser extent, 
if the sound speed is not modified to account for the molecular gas.}
Increasing the parameter $\zeta$ in equation~\eqref{h} only helps so much~--
above a certain value of $\zeta$, 
the term involving $\zeta$ becomes subdominant and has no effect on $h$.
Solutions can only be found for very large, unrealistic values of $C_\alpha$ 
(or $C'_\alpha$; see \citetalias{Chamandy+24})
and even then, the fits to magnetic field data are poor.
Solutions do not improve significantly if we replace the definition of $\tau$ 
with the more general definition used in \citetalias{Chamandy+24}.

In summary,
we do not find reasonable solutions for M51 and NGC~6946 when molecular gas is included.
This seems to suggest that the molecular gas plays a subdominant role in shaping the magnetic field
probed by current instruments
\luke{or that our dynamo model cannot reliably be used to model molecular clouds because
important physical processes like compression are not included. A}nother possibility is that equation~\eqref{h} underpredicts the scale height
at small radius when molecular gas is included, and thus needs to be modified.
Improving the scale height model should be a priority of future work.
It may also be possible to develop a \luke{more complete} two-phase ISM model, 
with different scale heights and magnetic field properties for each phase.

\section{Summary and Conclusions}
\label{sec:conclusions}
In \citetalias{Chamandy+24} we presented a dynamo model for the magnetic fields of spiral galaxies
that incorporates a model for the ISM turbulence parameters \citep{Chamandy+Shukurov20}.
The non-linear mean-field dynamo equations can be solved semi-analytically 
(after making various approximations) to calculate the large-scale magnetic field,
and we use a simple fluctuation dynamo model motivated by direct numerical simulations 
to calculate the small-scale magnetic field.
Our radially dependent axially symmetric model takes as input a handful of observables, 
namely the galaxy rotation curve, $\Vc(r)$, the gas temperature $T(r)$, 
the gas surface density $\Sg(r)$, the star formation rate surface density $\Sf(r)$, 
and the stellar surface density $\Ss(r)$,
and derives profiles for the rms turbulent speed $u(r)$, gas scale height $h(r)$,
turbulent correlation length $l(r)$ and time $\tau(r)$, total magnetic field strength $B\tot(r)$,
ordered magnetic field strength $B\ord(r)$ and pitch angle $p\ord(r)$,
and regular magnetic field strength $B\reg(r)$ and pitch angle $p\reg(r)$.
The key points stemming from our study are as follows:
\begin{itemize}
  \item The main goal of this work is to assess whether
  our interstellar turbulence--galactic dynamo model can explain quantitatively
  the vertically and azimuthally averaged properties 
  of the magnetic fields and interstellar turbulence in the disks of nearby spiral galaxies
  inferred from observational data.
  The short answer is broadly yes, but not entirely.
  To perform this test, we apply our model to the galaxies M31, M33, M51, and NGC~6946,
  and compare model outputs to the most recent observationally derived values from the literature.
  
  \item The model contains five adjustable parameters; 
  two of these are allowed to vary between galaxies while the other three are treated as adjustable universal constants.
  However, our fiducial model drops one of these ($\psi$) by setting it to unity,
  and thus in our fiducial model there is only one adjustable parameter ($\zeta$)
  that is allowed to vary between galaxies,
  in addition to three adjustable universal constants ($\beta$, $C_\alpha$, and $K$).
  This choice was made with the intention of keeping the model as simple (and predictive) as possible.
  However, for the galaxy M31, 
  we found that agreement with the data improves significantly if we set $\psi=2$ (Model~M31Alt).
  
  \item The level of agreement between model and data is reasonable in most cases,
  and there is a remarkable level of consistency in the values 
  and profiles of the model outputs across our small galaxy sample.

  \item For the correlation time $\tau$ of interstellar turbulence, 
  we chose to set $\tau$ equal to the eddy turnover time $\tau_\mathrm{e}$
  even when the renovation time $\tau\renov$ is shorter than $\tau_\mathrm{e}$.
  This choice simplifies the model of \citetalias{Chamandy+24}
  (where $\tau$ was set to the minimum of these two quantities)
  and has very little impact on the results except that it leads to $\tau$ values that are more similar between galaxies.
  We obtain values in the approximate range $1$--$10\Myr$, typically increasing with radius in a given galaxy
  (Figure~\ref{fig:combined_plot_r}).
  For the correlation length, we obtain $20\pc\lesssim l\lesssim50\pc$ for our fiducial model
  and about twice as large for our alternative M31 model that uses $\psi=2$,
  and $l$ typically increases with radius (Figure~\ref{fig:h} inset and Figure~\ref{fig:combined_plot_r}).
  The disk scale height also tends to increase with radius, as expected (Figure~\ref{fig:h}).

  \item The scale lengths of the modeled magnetic field components tend to be of order $r_{25}/2$,
  with a few exceptions (Table~\ref{tab:scale_lengths}).
  
  \item The best-fit values of the model parameters (Table~\ref{tab:params})
  suggest a mismatch in normalization between classical dynamo theory (fluctuation and mean-field dynamos)
  and observational inference insofar as the overall strength of the magnetic field is concerned. 
  In particular, we need to set the overall scaling factor for the magnetic field to $\beta=7$ 
  to achieve good agreement with the latest observationally derived values.
  This could suggest that dynamo models tend to underestimate the saturation strength of the magnetic field
  or that the magnetic field strength tends to be overestimated by observers,
  who assume energy equipartition between total magnetic field and cosmic rays,
  which in turn requires assumptions about the nature of the cosmic ray electrons.
  
  \item There is disagreement between the modeled pitch angles of the 
  so-called ``ordered'' magnetic field (observable through its polarized emission)
  and those inferred from observations 
  The model values of $|p\ord|$ are generally lower than observationally derived values. 
  This may point to deficiencies in the modeling of the anisotropic small-scale field.
  In Section~\ref{sec:new_pitch} we present new theory to predict $p\ord$ as a function of the global radial shear.
  The mismatch suggests that more physical effects need to be taken into account~--
  perhaps effects stemming from the galactic spiral structure, which is not included in our axisymmetric model.

  \item There is also a disagreement between the value of the pitch angle for the large-scale regular field
  and $|p\reg|$ is found to be larger in the data than in the theory 
  for two out of three galaxies (M33 and M51) for which it has been measured,
  though for M31 the model predictions and observations are in excellent agreement.
  We suggest that the discrepancy may be at least partly due to a questionable assumption
  in the mode analysis technique used to model the magnetic field structure from the data 
  \citep{Fletcher+04,Tabatabaei+08,Fletcher+11},
  namely that the anisotropy of the random small-scale field is negligible.
  This assumption seems to be contradicted by the significant differences between $p\ord$ and $p\reg$.
  The pitch angle derived from the mode analysis method may be a kind of weighted average between $p\ord$ and $p\reg$,
  and given that $|p\ord|$ is found to be larger than $|p\reg|$ observationally, 
  this could help to explain why $|p\reg|$ data from mode analysis is larger than theoretical predictions.
\end{itemize}

Several possibilities exist for extending the theoretical model presented.
Firstly, the mean-field dynamo model could be made to be global as opposed to local in radius,
which would necessitate numerical solutions,
in which case multiple 1D (in radius) mean-field simulations could be performed to explore the parameter space.
This would significantly improve the accuracy of the solutions,
though we do not expect it to lead to drastically different conclusions,
given that the approximations used to derive the mean-field dynamo
solutions used in this paper have been found to be fairly accurate \citep{Chamandy16}.

Secondly, the dynamo model could be improved by incorporating additional physical effects, 
such as those related to the spiral structure, as mentioned above, or the
interaction between the mean and random components of the magnetic field
\citep[e.g.][]{Rogachevskii+Kleeorin07,Subramanian+Brandenburg14,Chamandy+Singh18,Bhat+19,Gopalakrishnan+Subramanian23,Gent+24},
but the nature of this interaction is still not very well understood.

Thirdly, the turbulence model could be improved, 
for example by including turbulence driving by superbubbles (and associated galactic outflows), 
in addition to that by isolated SNe \citep{Chamandy+Shukurov20}.
This functionality is already included, to some extent, in our numerical code, 
but it adds parameters to the model that are difficult to constrain,
so we did not attempt to apply this more general version of our model to the galaxies studied.

Fourthly, the model (or an improved version thereof)
could be applied to a larger collection of nearby spiral galaxies,
and a joint statistical fit to the data could be performed.
Working with a larger, statistical sample would help 
to identify areas of agreement and tension between models and data and lead to improvements to the models.
A further aim is to make true predictions 
for the turbulence and magnetic field properties of galaxies for which such quantities have not yet been measured.

\section*{Acknowledgements}
We thank Anvar Shukurov for comments on an early draft \luke{and subsequent discussion},
Claude Carignan for providing kinematic data, 
and Tuhin Ghosh, Nishikanta Khandai, Rainer Beck, Fatemeh Tabatabaei, 
and Łukasz Bratek for useful discussions. GS acknowledges support from INAF under the Large Grant 2022 funding scheme (project "MeerKAT and LOFAR Team up: a Unique Radio Window on Galaxy/AGN co-Evolution")

\section*{Data Availability}
The sources of data used in this paper are tabulated in Table~\ref{tab:datasource}. 
Data available only in graphical form were extracted 
using the \texttt{\href{https://automeris.io/}{WebPlotDigitizer}} software. 
Plots of these data, with the original inclination and distance used in the source paper, 
are provided in the supplementary material. 
The corrected data after interpolation are presented as .csv files,
which are available as part of the code linked above 
A summary of the sources of error is given in Table~\ref{tab:errors_combineddata}. 

\bibliographystyle{aasjournal}
\bibliography{refs}

\appendix

\section{Effect of shear on the random magnetic field}
\label{sec:shear}
\begin{figure}[h t]
\centering
  \includegraphics[width=\columnwidth, clip = true,trim={5 0 0 0}]{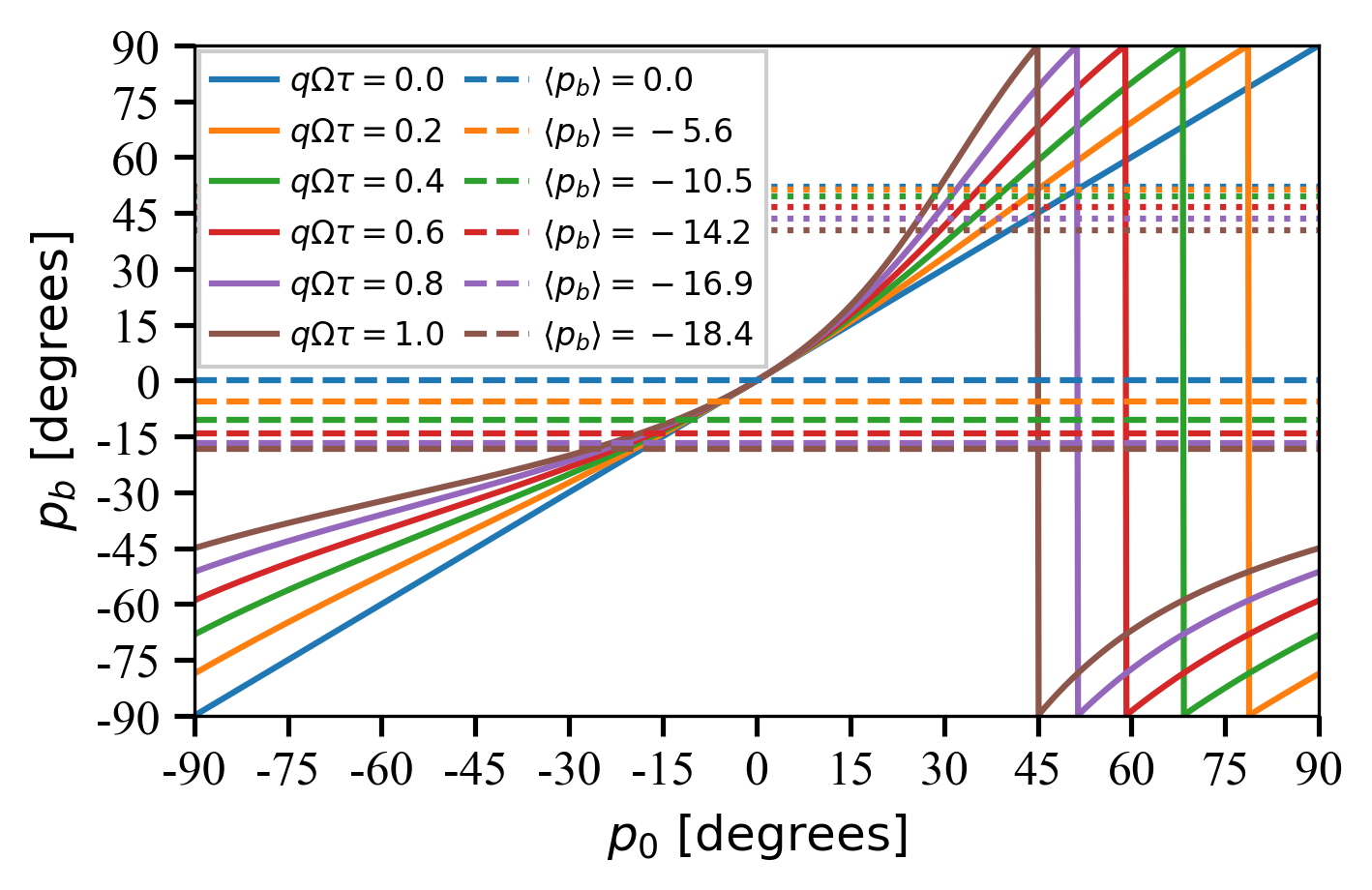}\\
  \includegraphics[width=\columnwidth, clip = true,trim={5 0 0 0}]{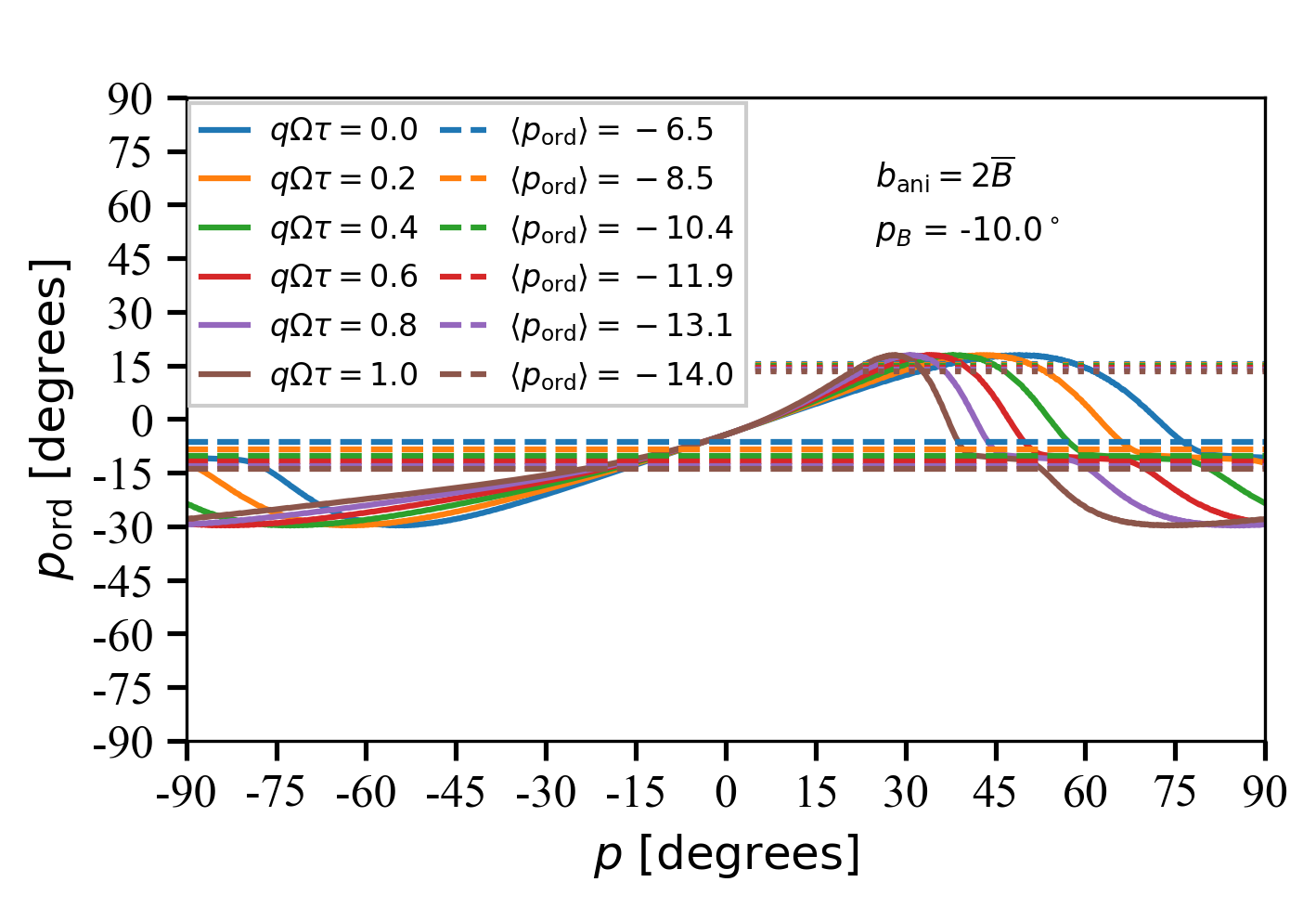}
  \caption{
  The pitch angle $p_b$ (upper panel) of the random field, as given by equation~\eqref{p_b}
  and $p\ord$ (lower panel) of the ``ordered'' field detected in polarized emission, 
  as given by equation~\eqref{pord_appendix},
  plotted for different values of $q\Omega\tau$.
  As $p_0$ is a uniformly distributed random variable, 
  the mean value (dashed) is equal to the expectation value.
  Also shown is the standard deviation (dotted).
  In the lower panel, we choose $p_B=-10^\circ$ and $b\ani/\Bbar=2$ for illustration.
  The mean value $\langle p\ord\rangle$ is what is used in our analysis 
  (formally, we define $p'\ord = \langle p\ord\rangle$ and then redefine $p\ord=p'\ord$).
  }
  \label{fig:p_sol}
\end{figure}

For a compressionless fluid with negligible microscopic diffusivity, the induction equation is 
\begin{equation}\label{ind}
  \frac{\del\bfB}{\del t} = -(\bfV\cdot\bfDel)\bfB + (\bfB\cdot\bfDel)\bfV.
\end{equation}
For $\bfV=r\Omega\bfphihat$ and assuming $\del\Omega/\del\phi=0$, $\del\Omega/\del z=0$, 
and $\del B_i/\del\phi=0$, where $i=r,\phi,z$ (cylindrical polar coordinates),
the advective term becomes
\begin{equation}\label{a}
  - (\bfV\cdot\bfDel)\bfB = - r\Omega\frac{1}{r}\frac{\del\bfB}{\del\phi}
  = \Omega B_\phi \bfrhat - \Omega B_r\bfphihat.
\end{equation}
For the stretching term, we have
\begin{equation}\label{s}
  (\bfB\cdot\bfDel)\bfV 
  = -\Omega B_\phi \bfrhat + \left(\Omega + r\frac{\del\Omega}{\del r}\right)B_r\bfphihat .
\end{equation}
Substituting expressions~\eqref{a} and \eqref{s} into equation~\eqref{ind}, 
multiple terms cancel and we finally obtain
\begin{equation}
  \frac{\del \bfB}{\del t} = r\frac{\dd \Omega}{\dd r} B_r\bfphihat,
\end{equation}
which has the solution \citep[e.g.~\S2.1.2 of][]{Shukurov+Subramanian21}
\begin{equation}\label{B_shear}
  B_r = \const, \qquad B_\phi = B_\phi|_{t=0} - q\Omega t B_r,
\end{equation}
where $q$ is defined in equation~\eqref{q}.
The flow becomes uncorrelated after a time $\tau$, 
so we consider the effect of the shear over this timescale,
setting $t=\tau$ in equation~\eqref{B_shear}. 
This approach is similar but not identical to that of \citet{Stepanov+14} and \citet{Hollins+17},
who derived equations for the standard deviations of the field components rather than the components themselves.
This then leads to the expression
\begin{equation}\label{p_b}
  \tan p_b = \frac{b_r}{b_\phi} 
  = \frac{b_{r,0}}{b_{\phi,0} - q\Omega\tau b_{r,0}}   
  = \frac{\tan p_0}{ 1 - q\Omega\tau \tan p_0 },
\end{equation}
where $p_0 = \arctan (b_{r,0}/b_{\phi,0})$ 
is the pitch angle of the background isotropic field (equation~\ref{p0}). 
The value of $p_b$ for a given value of $p_0$ is shown in the upper panel of Figure~\ref{fig:p_sol},
for different choices of $q\Omega\tau$.
Recall that $p_0$ has a uniform probability distribution function between $-\pi/2$ and $\pi/2$.
The mean values of $p_b$, listed in the legend, are shown by the dashed lines.
The standard deviation of $p_b$, $\sigma_{p_b}$, is shown with dotted lines.
Note that the effect of the large-scale radial shear is to reduce the magnitude of the pitch angle if $p_0<0$,
and to increase the magnitude of the pitch angle if $p_0>0$,
though in the latter case, the pitch angle can jump from $\pi/2$ to $-\pi/2$.
For this reason, shear tends to make the mean pitch angle negative.

\section{Ordered magnetic field pitch angle}
\label{sec:pord}
Here we model the pitch angle of the ordered magnetic field $p\ord$ 
in terms of the strengths and pitch angles of the mean and random fields.
We define the ordered field as 
\begin{equation}
\label{vec_Bord}
  \bmB\ord = \meanv{B} + \bmbtilde,
\end{equation}
where $\bmbtilde$ is the part of the random field that contributes to the polarized synchrotron emission,
which is assumed to have rms value $b\ani$ and mean pitch angle $p_b$, 
defined in equations~\eqref{b_ani} and \eqref{p_b}, respectively.
Note that with this definition ``ordered'' is not the same as mean,
which is consistent with the terminology used in some, but not all, of the literature.
To calculate $p\ord$, 
we need to model the distributions of the magnitude and direction of $\bmbtilde$.

A simple but not very realistic choice is to assume that $\bmbtilde$ 
has constant magnitude $b\ani$ and constant pitch angle $p_b$,
and that its direction reverses randomly, 
such that $\bmbtilde$ has either a component aligned with $\meanv{B}$ (so that $\meanv{B}\cdot\bmbtilde>0$)
or anti-aligned with $\meanv{B}$ (so that $\meanv{B}\cdot\bmbtilde<0$).
In the first case, $B\ordr=\mean{B}\sin p_B + b\ani\sin p_b$ and $B\ordp=\mean{B}\cos p_B + b\ani\cos p_b$,
whereas in the second case $B\ordr=\mean{B}\sin p_B - b\ani\sin p_b$ and $B\ordp=\mean{B}\cos p_B - b\ani\cos p_b$.
Also, the polarized intensity is proportional to the square of the ordered field,%
\footnote{Here we neglect any possible dependence of the cosmic ray electron density on the magnetic field, for simplicity.}
which is given by
\begin{equation}
  \bmB\ord^2 =
  \begin{dcases}
    \mean{B}^2 + b\ani^2 + 2\mean{B}b\ani \cos(p_b - p_B) \quad &\mbox{if  } \meanv{B}\cdot\bmbtilde>0;\\
    \mean{B}^2 + b\ani^2 - 2\mean{B}b\ani \cos(p_b - p_B) \quad &\mbox{if  } \meanv{B}\cdot\bmbtilde<0.
  \end{dcases}
\label{vec_Bord2}
\end{equation}
Its mean value is thus given by
\begin{equation}
\label{meanBord2}
  \mean{\bmB\ord^2} = \mean{B}^2 + b\ani^2.
\end{equation}
We now take a weighted average with normalization factor $\mean{\bmB\ord^2}$, which leads to the expression
\begin{equation}
\begin{split}
  p\ord &= \frac{1}{2}\left\{\left[1+\frac{2\mean{B}b\ani}{\mean{B}^2+b\ani^2}\cos\left(p_b-p_B\right)\right]\right.\\
  &\qquad\times \arctan\left(\frac{\mean{B}\sin p_B + b\ani\sin p_b}{\mean{B}\cos p_B + b\ani\cos p_b}\right)\\
  &+\left[1-\frac{2\mean{B}b\ani}{\mean{B}^2+b\ani^2}\cos\left(p_b-p_B\right)\right]\\
  &\left.\qquad\times\arctan\left(\frac{\mean{B}\sin p_B - b\ani\sin p_b}{\mean{B}\cos p_B - b\ani\cos p_b}\right)\right\}.
\end{split}
\end{equation}

Next, we try a slightly more realistic model where the pitch angle of $\bmbtilde$ remains fixed
but the component along the axis corresponding to this pitch angle in the $r$-$\phi$ plane is normally distributed with mean zero and standard deviation $b\ani$.
In this case, we find
\begin{equation}
\label{pord_appendix}
\begin{split}
 p\ord = &\frac{1}{(2\pi)^{1/2}b\ani}\displaystyle\int_{-\infty}^\infty
    e^{-\frac{\btilde^2}{2b\ani^2}}\left[1 + \frac{2\mean{B}\btilde}{\mean{B}^2+\btilde^2}\cos\left(p_b-p_B\right)\right]\\
    &\times\arctan\left(\frac{\mean{B}\sin p_B + \btilde\sin p_b}{\mean{B}\cos p_B + \btilde\cos p_b}\right)\,d\btilde,
\end{split}
\end{equation}
which is the same as equation~\eqref{pord}, used in our analysis.
This could be made more realistic by factoring in the dispersion of $p_b$,
but we choose to leave such complications for future study. 

In the lower panel of Figure~\ref{fig:p_sol} 
we plot $p\ord$ as a function of the pitch angle of the isotropic background field $p_0$
(which is uniformly distributed),
for different choices of $q\Omega\tau$ and adopting $b\ani=2\mean{B}$ and $p_B=-10^\circ$, 
for illustration.
The dashed lines show the mean value $\langle p\ord\rangle$.
The mean values are the values used in our analysis (e.g.~Figure~\ref{fig:p_sol});
thus, formally, we define $p'\ord=\langle p\ord \rangle$ and then set $p\ord=p'\ord$.

\section{Re-calibration of data to the chosen distance and inclination values}\label{sec:di}
The references in Table~\ref{tab:datasource} use different values of the distance $d$
and inclination $i$ for each galaxy. 
Therefore, we re-calibrated the data to the chosen values of $d$ and $i$,
listed in Table~\ref{tab:dist_inc}.
The plots of original data as obtained from the references in Table \ref{tab:datasource} can be found in the supplementary materials, with the corresponding inclination and distance in the legend.

The conversion of line-of-sight component of the rotation curve ($V_\mathrm{c} \,\sin i$) 
to total velocity is done according to 
\begin{equation}
    V_{\mathrm{c,{\text{corrected}}}}= V_\mathrm{c}\times \frac{\sin{i_0}}{\sin{i}}
    \label{eqn:rotcurvecorr}
\end{equation}
Distance correction is done by rescaling all distances as per 
\begin{equation}
    r_{\text{corrected}}=r\times\frac{d}{d_0}
    \label{eqn:distcorr}
\end{equation}
All quantities in Table~\ref{tab:datasource}, except 
$\sigma_\mathrm{HI}$ and $T$, 
are corrected for inclination according to 
\begin{equation}
    \Sigma_{\text{corrected}}= \Sigma\times \frac{\cos{i}}{\cos{i_0}}.
    \label{eqn:obscorr}
\end{equation}
In all cases, subscript $0$ refers to the values of distance and inclination 
used in the original reference from which the data is sourced.

\section{Error Analysis \label{app_sec:erro_eq}}
\subsection{Model inputs}
Each observable presented in Table \ref{tab:datasource}, after correction (equations \ref{eqn:rotcurvecorr}, \ref{eqn:obscorr}, \ref{eqn:distcorr}) has an associated error that depends on the 
uncorrected value of the observable and the inclinations and/or distances used in the source paper and our study. These are derived using Equation \ref{eqn:rotcurvecorr} and are presented below. 
For the circular speed, we have

\begin{equation}
\begin{split}
    \sigma_{V_\mathrm{c}}^2 =
    &\left( \sigma_v \frac{ d_0\sin{i_0}}{r d\sin{i}  }\right)^2 + \left( \sigma_{d_0} \frac{ V_0\sin{i_0}}{r d\sin{i}  }\right)^2+ \left( \sigma_{i_0}\cos{i_0} \frac{ V_0 d_0}{r d\sin{i}  }\right)^2 \\
    &+ \left( \frac{\sigma_i \cos{i}}{\sin^2 {i}} \frac{ V_0 d_0\sin{i_0}}{r d}\right)^2 
    + \left( \frac{\sigma_d}{d^2} \frac{ V_0 d_0\sin{i_0}}{r \sin{i}  }\right)^2
    \label{eqn:vcircerror}
\end{split}
\end{equation}
For X = HI, $\mathrm{H_2}$ or SFR, we can express the error in the surface densities $\Sigma_\mathrm{X}$ as:
\begin{equation}
\sigma_{\Sigma_\mathrm{X}}^2 = \left(\sigma_{\Sigma_{X_0}} \frac{\cos{i_0}}{\cos{i}}\right)^2 + \left(\sigma_{i_0} \frac{\Sigma_{X_0} \sin{i_0}}{\cos{i}}\right)^2   
+ \left(\sigma_{i} \frac{\Sigma_{X_0}\sin{i} \cos{i_0}}
{\cos^2{i}}\right)^2. \\
\label{eqn:sigmaerrors}
\end{equation}
However, 
if we calculate molecular gas surface density ($\Sigma_{\mathrm{H_2}}$) using molecular fraction data (as for M31), 
using the relation $\Sigma_{\mathrm{H_2}} = \Sigma_{\mathrm{HI}} \left(n/n-1\right)$, 
then $\sigma_{\Sigma_{\mathrm{H_2}}}$ 
accounts for the error in $n$ as well. 
So, the equation becomes
\begin{equation}
    \sigma_{\Sigma_{\mathrm{H_2}}}^2 = \left(\sigma_{\Sigma_{\mathrm{HI}}} \frac{n}{n-1} \right)^2 + \left(\sigma_{n} \frac{\Sigma_{\mathrm{HI}}}{(n-1)^2} \right)^2.
    \label{eqn:sigmaerrors_molfrac}
\end{equation}
For the other galaxies in this work for which data for $\Sigma_{\mathrm{HI}}$ are directly available, the total gas surface density is 
$\Sg = 3\Sigma_{\mathrm{HI}}/\left(4-\mu\right) +\Sigma_{\mathrm{H_2}}/\left(4-\mu^\prime\right)$. 
Note that this equation is a more general version 
of equation~\eqref{eqn:sigmagas}; see Appendix~\ref{sec:include_mol_gas}. 
Here, we assume a 10\% value for both $\sigma_\mu$ and $\sigma_{\mu^\prime}$, 
which are the respective errors in the mean molecular masses $\mu$ and $\mu^\prime$. 
The error is then
\begin{equation}
\begin{split}
&    \sigma_{\Sg}^2 =  \left(\sigma_{\Sigma_{\mathrm{HI}}} \frac{3\mu}{4-\mu} \right)^2 + \left(\sigma_{\mu} \frac{12\Sigma_{\mathrm{HI}}}{\left({4-\mu}\right)^2} \right)^2 \\
& \hspace{16pt} + \left(\sigma_{\Sigma_{\mathrm{H_2}}} \frac{\mu^\prime}{4-\mu^\prime} \right)^2 + \left(\sigma_{\mu^\prime} \frac{4\Sigma_{\mathrm{H_2}}}{\left({4-\mu^\prime}\right)^2} \right)^2 \\
    \label{eqn:sigma_err_tot}
\end{split}
\end{equation}
with $\sigma_\mu/\mu$ and $\sigma_{\mu^\prime}/\mu^\prime$ set to $10\%$.

For $\Sigma_\star$, where data are available directly, 
we use  equation~\eqref{eqn:sigmaerrors} to calculate error $\sigma_{\Sigma_\star}$. 
However, for M33, we use the prescription from \cite{Kam+17}:
\begin{equation*}
     \St = \Upsilon \left[\mathrm{M_\odot\, pc^{-2}} \right] \times 10^{-0.4(\mu_{3.6}-C_{3.6})}.
\end{equation*}
Here, $\mu_{3.6}$ and $C_{3.6}$ are surface brightness and correction values at 3.6 $\mu \mathrm{m}$. We have used $\Upsilon=0.72$ and $0.52$, assuming an error of $\pm 0.1$ for both. So, for M33, the error $\sigma_{\St}$ includes error from $\Upsilon$ and the surface brightness. $\mu$ evolves as 
$\mu = \mu_0 + 1.10857 r/R$, 
where $\mu_0 = 18.01$, $R = 1.82$, and $r$ is the distance from the center.
The final equation for the error is: 
\begin{equation}
    \sigma_{\St}^2 = \mathrm{a}\sigma_{\Upsilon}^2 + \left(0.4\ln 10\,\Upsilon\sigma_\mu\right)^2
\end{equation}
Here, $\mathrm{a}=10^{-0.4(\mu_{3.6} - C_{3.6})}$ and
\begin{equation*}
    \sigma_{\mu} = \sigma_{\mu_0} + \frac{1.10857}{\mathrm{R}}\sqrt{\left(\frac{r\sigma_R}{R}\right)^2 + \sigma^2_r},
\end{equation*}
where $\sigma_r$ is found using error propagation based on equation~\eqref{eqn:distcorr}. Values of $\sigma_\mathrm{R}$ and $\sigma_{\mu_0}$ are 0.02 kpc and 0.03 mag\,arcsec$^{-2}$, respectively.

\subsection{Note of error sources and methods} \label{sec:error_prop_formulae}

\begin{table*}
\begin{center}
  \def\arraystretch{1.1}
  \caption{Summary of adopted uncertainty values.} 
  \begin{tabular}{@{}ccccc@{}}
\hline 
Quantity		            &M31		 &M33	   &M51		  &NGC 6946		\\
\hline   
$\Vc$		            &\cite{Chemin+09} &\cite{Kam+17}&\cite{Sofue+99}           &\cite{Sofue+99}	\\   
&&&(section 2.5)& (section 2.5)\\

$\Sigma_{\mathrm{HI}}$		    & 6\% based on M51 and  & 6\% based on M51 and &6\% based on \cite{Kumari+20} &	6\% based on \cite{Kumari+20} 	\\ 
&NGC~6946 estimates &NGC~6946 estimates &(Fig F.6)&(Fig F.8)\\

$\Sigma_{\mathrm{H_2}}$		    &Uses molecular fraction error  &	\cite{Gratier+10} 	   & 6\% estimate from &	6\% estimate from  	\\ 
&from \cite{Nieten+06} (Fig 10)&(Fig 8)&\cite{Kumari+20} (Fig F.6)&\cite{Kumari+20} (Fig F.8)\\
                  
$\Sf$		    &10\%  &	 10\%  &10\% &	10\% \\ 

$\St$  		&	10\%		 &\cite{Kam+17} &	10\%  &	10\%	\\ 
&&(propagate using Eq. 3)&&\\

$T$  		            &	Standard deviation		 &	Standard deviation &Standard deviation		  &Standard deviation		
\\
  		            &	\cite{Tabatabaei+13b}		 &	\cite{Lin+17} &\cite{Bresolin+04}		  &\cite{Gusev+13}		
\\
\hline
  \label{tab:errors_combineddata}
  \end{tabular}
  \end{center}
\end{table*}

Rotation curve errors for M31 and M33 are available for every data point used, while for M51 and NGC 6946, \cite{Sofue+99} mentions an error of $\pm 10$--$20\kms$. So, we choose the error in $\Vc$ for those galaxies to be $\pm 15\kms$.
For $\SHI$, $\SHmol$ and $\Sf$, we use the $\log\mathrm{\Sigma_{\mathrm{SFR}}}$ vs $\log\mathrm{\SHI}$ plots in the appendix of \cite{Kumari+20} 
for M51 and NGC 6946 to estimate a percent error by manually measuring the error bars and taking an average. 
Using this technique, we finalised on an error of 6\% for all galaxies. A different technique had to be used for M31 as we had error in molecular fraction from \cite{Nieten+06}. For $\St$, we use an arbitrarily chosen error of 10\%. The temperatures used are linear fits of available data, and error is taken to be the standard deviation of their values from the mean.

Errors in $\Omega$ and $\Sigma$ are calculated as described in Section~\ref{sec:error}. 
The error in $q$ is taken to be the standard deviation 
around the mean overall radii. 

\subsection{Model outputs}
\label{sec:error}
We make use of the standard error formula
\begin{equation}
    \label{errorf}
    \left(\frac{\sigma_f}{f}\right)^2 = a^2\left(\frac{\sigma_X}{X}\right)^2 + b^2\left(\frac{\sigma_Y}{Y}\right)^2
\end{equation}
where $f(X,Y) \propto X^aY^b$, 
with $X$ and $Y$ observables and $a$ and $b$ constants.
In \citetalias{Chamandy+24}, we calculated the scaling relations analytically for limiting cases where we assumed that the turbulence was either supersonic or subsonic. To approximate the error, we calculate the errors for both cases and choose the maximum of the two. This gives us a conservative upper limit for the error. Even though the error in $c_\mathrm{s}$ is found similarly, an extra contribution due to uncertainty in the adiabatic index $\gamma_\mathrm{ad}$, $\sigma_{\gamma_\mathrm{ad}} = 20\%$, was introduced manually into $\sigma_{c_\mathrm{s}}$.

The error in local growth rate $\gamma$ is found by applying the error propagation formula on Equation \ref{eqn:local_growth_rate_formula} to be

\begin{equation}
    \sigma_\gamma^2 = \left(\frac{\sigma_\tau}{\tau}\right)^2 +  \left(\frac{2\sigma_u}{u}\right)^2 +  \left(\frac{2\sigma_h}{h}\right)^2 +  \left(\frac{2\sigma_D}{D_\mathrm{c} \left[\frac{D}{D_\mathrm{c}} - \sqrt{\frac{D}{D_\mathrm{c}}}\right]}\right)^2 ,
\end{equation}
where the critical dynamo number $D_\mathrm{c} = -(\pi/2)^5$.

\section{Including molecular gas in the model}
\label{sec:include_mol_gas}
For a molecular cloud component (denoted by prime)
composed of molecular hydrogen and atomic helium, 
the particle surface number density $\ntilde^\prime$ is given by
\begin{equation} 
\ntilde^{\prime} \mu^{\prime}=2 \ntilde_{\mathrm{H_2}}+4\left(\ntilde^{\prime}-\ntilde_{\mathrm{H_2}}\right) 
\quad\Rightarrow \quad \ntilde^{\prime}=\frac{2 \ntilde_{\mathrm{H_2}}}{4-\mu^{\prime}}.
\end{equation}
Using $\Sigma^{\prime}=\ntilde^{\prime} \mu^{\prime} m_\mathrm{H}$ 
and $\Sigma_{\mathrm{H_2}}=\ntilde_{\mathrm{H_2}} \left(2 m_\mathrm{H}\right)$, 
we obtain
\begin{equation} 
  \Sigma^{\prime} 
  = \frac{2 \ntilde_{\mathrm{H_2}}}{4-\mu^{\prime}} \mu^{\prime} m_\mathrm{H}
  = \frac{\mu^{\prime}} {4-\mu^{\prime}} \Sigma_{\mathrm{H_2}}.
\label{eqn:ap_molgas_sigma}
\end{equation}
We thus obtain a general expression for the total surface density of gas excluding metals:
\begin{equation} 
\label{Sigma_gen}
  \Sigma
  = \frac{3 \mu}{4-\mu} \Sigma_{\mathrm{HI}} + \frac{\mu^{\prime}}{4-\mu^{\prime}} \Sigma_{\mathrm{H_2}}.
\end{equation}
The mean particle mass $\mu^\prime$ for a molecular, 
non-ionized disc is given by 
\begin{equation}
  \mu^\prime =  
  \left(\frac{X}{2} + \frac{Y}{4} + \frac{Z}{A_Z}\right)^{-1},
\end{equation}
where $Z$ and $A_Z$ are the mass fraction and mean atomic weight of metals.
With $X=0.7$, $Y=0.3$, and $Z=0$, we obtain $\mu^\prime \approx 2.4$.
The mass density is still given by $\Sigma/2h$ (equation~\ref{rho}) but the number density is given by
\begin{equation}
  n = \frac{1}{2h\mH}\left(\frac{3\Sigma_{\mathrm{HI}}}{4-\mu}  + \frac{\Sigma_{\mathrm{H_2}}}{4-\mu^{\prime}} \right).
\end{equation}

\section{Calculation of scale length}
\label{sec:scale_length_calc}

Assuming that the field strengths $B\tot$, $B\reg$ and $B\ord$ (denoted as $B$ in this section) 
follow an exponential profile $B = B_0 \exp{\left(r/r_0\right)}$, 
the scale length $r_0$ is calculated by a linear fit of $\log{B}$ vs $r$ graphs. 
The scale length is obtained from the slope $m$ of the fit using 
\begin{equation}
    r_0 = - \frac{\log{e}}{m}.
\end{equation}
The values of scale lengths, their ratios with $r_\mathrm{25}$, and errors associated with these quantities are presented in Table \ref{tab:scale_lengths}. 
The error in the scale length is given by 
$\sigma_{r} = r \sigma_m / m$. 
Similarly, the error in the ratio $r/r_\mathrm{25}$ is given by $\sigma_{r/r_\mathrm{25}} = \left[\left(\sigma_{r_\mathrm{25}}/r_\mathrm{25}\right)^2 + \left(\sigma_{r}/r\right)^2\right]^{1/2}$. 

\section{Variation of model outputs with $r/r_{25}$}\label{app_sec:output_vs_r_r25}

\begin{figure*}
    \includegraphics[width=4.4cm,keepaspectratio]{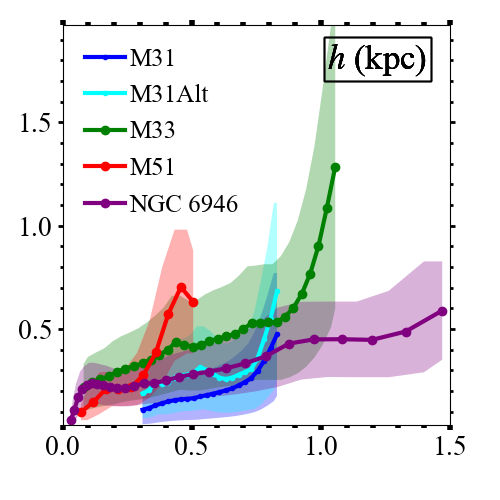}
    \includegraphics[width=4.4cm,keepaspectratio]{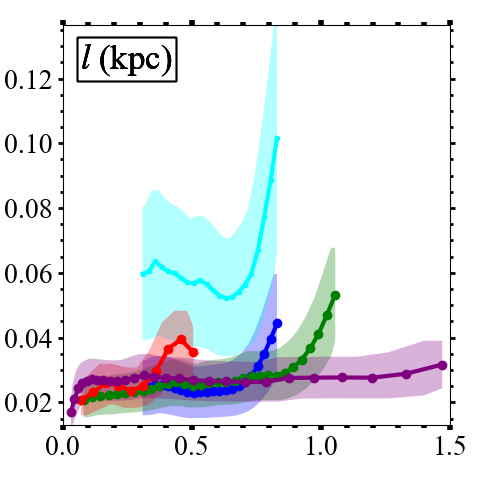}
    \includegraphics[width=4.4cm,keepaspectratio]{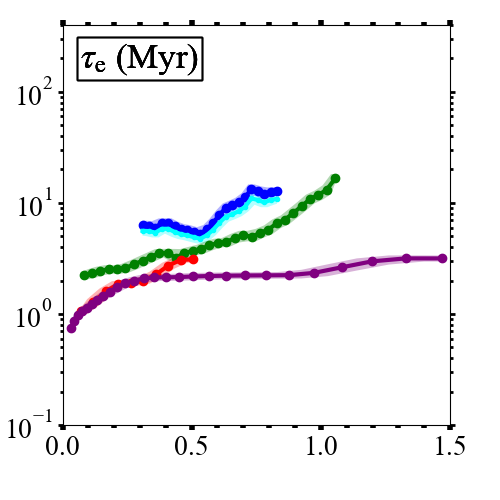}
    \includegraphics[width=4.4cm,keepaspectratio]{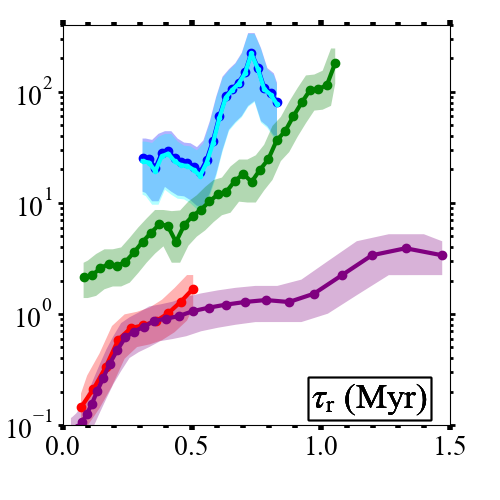}
    \\

    \vspace{-20pt}
    \includegraphics[width=4.45cm,keepaspectratio]{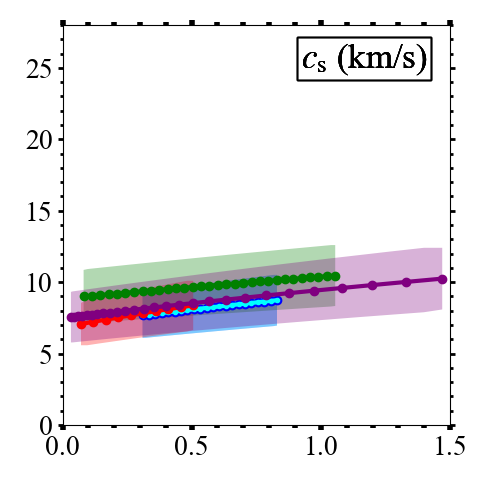}
    \includegraphics[width=4.45cm,keepaspectratio]{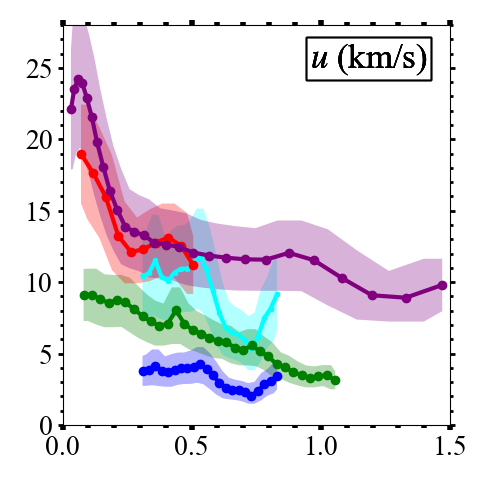}
    \includegraphics[width=4.45cm,keepaspectratio]{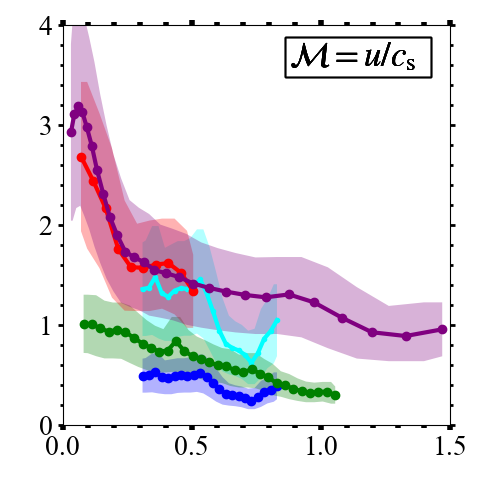}
    \includegraphics[width=4.45cm,keepaspectratio]{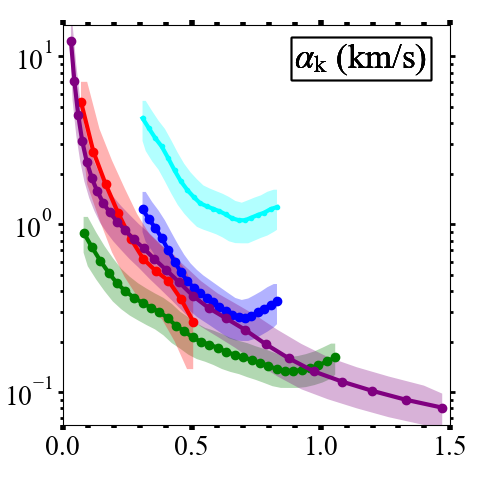}\\

    \vspace{-20pt}
    \includegraphics[width=4.45cm,keepaspectratio]{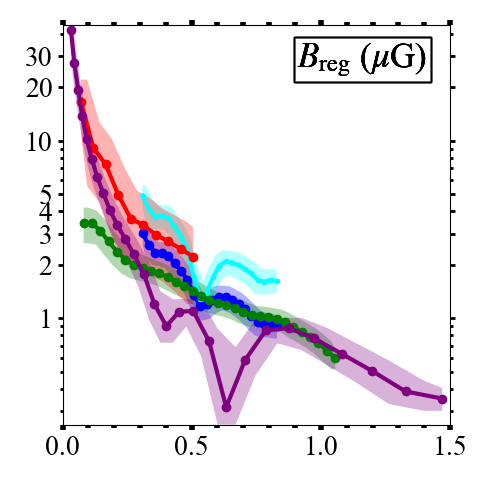}
    \includegraphics[width=4.45cm,keepaspectratio]{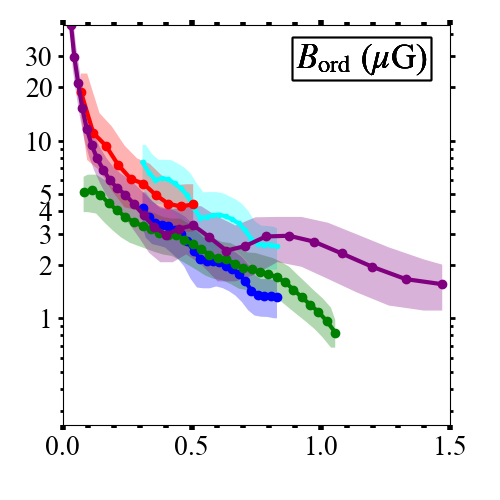}
    \includegraphics[width=4.45cm,keepaspectratio]{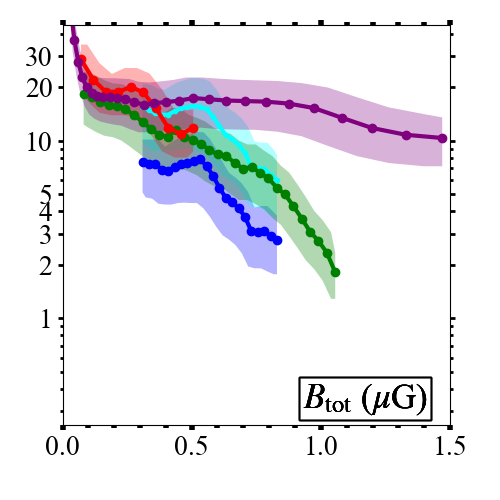}
    \includegraphics[width=4.45cm,keepaspectratio]{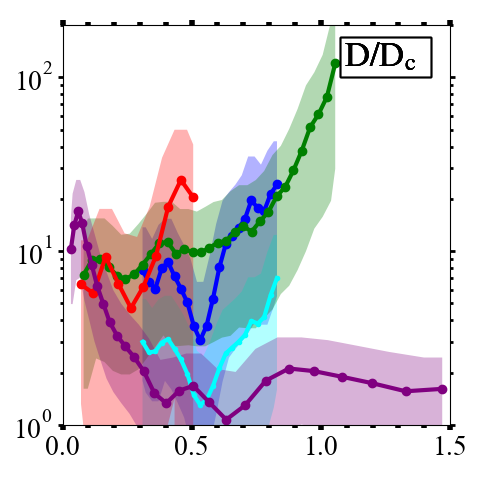}\\
    
    \vspace{-20pt}

    \includegraphics[width=4.45cm,keepaspectratio]{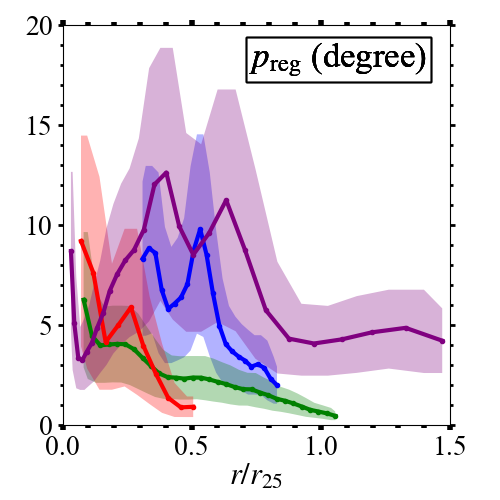}
    \includegraphics[width=4.45cm,keepaspectratio]{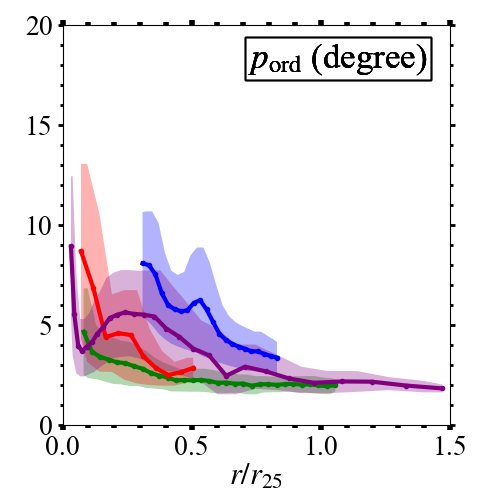}
    \includegraphics[width=4.45cm,keepaspectratio]{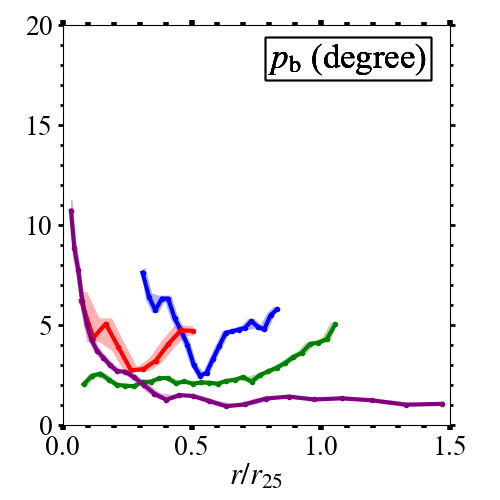}
    \includegraphics[width=4.45cm,keepaspectratio]{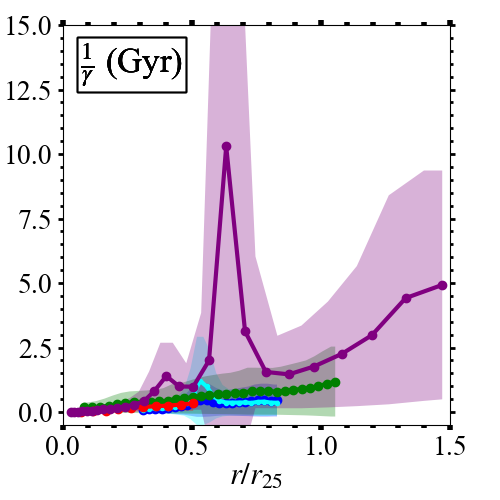}
    
    \caption{Same as Figure \ref{fig:combined_plot_r} but abscissa is $r/r_{25}$.
    }
    \label{fig:combined_plot_r_r25}
\end{figure*}

Figure \ref{fig:combined_plot_r_r25} shows plots of the various model outputs against the radius normalised by $r_\mathrm{25}$. 

\section{Different models for $\tau$}\label{sec:tau_models}
\begin{figure*}
    \includegraphics[width=8cm,keepaspectratio]{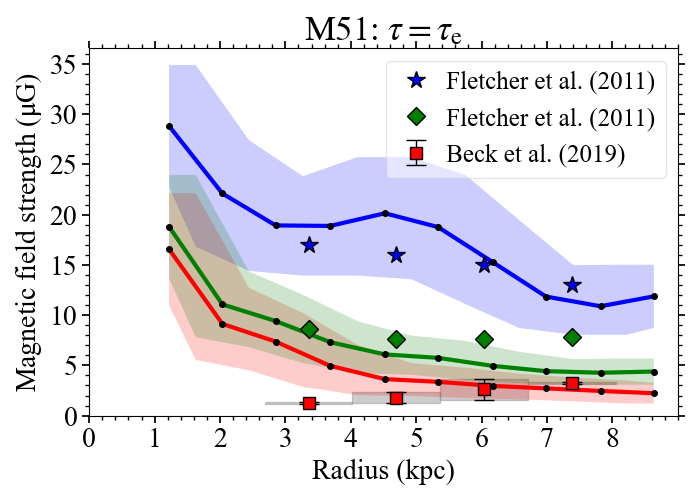}
    \includegraphics[width=8cm,keepaspectratio]{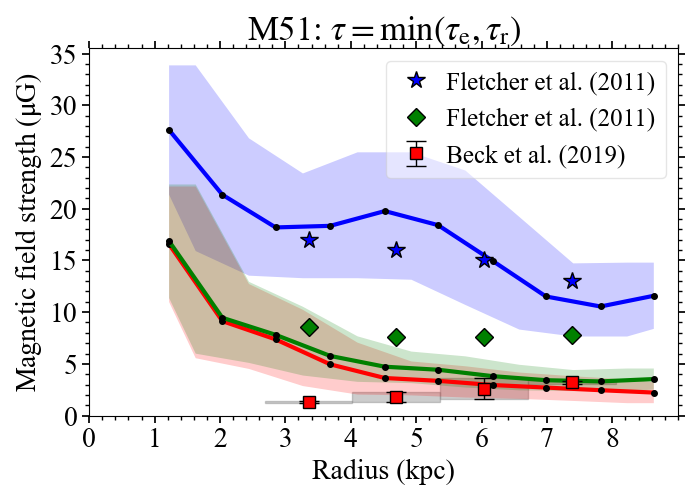}
    
    \includegraphics[width=8cm,keepaspectratio]{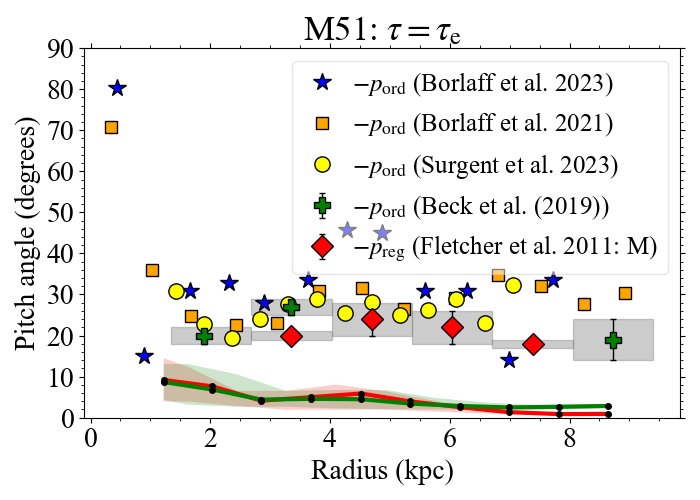}
    \includegraphics[width=8cm,keepaspectratio]{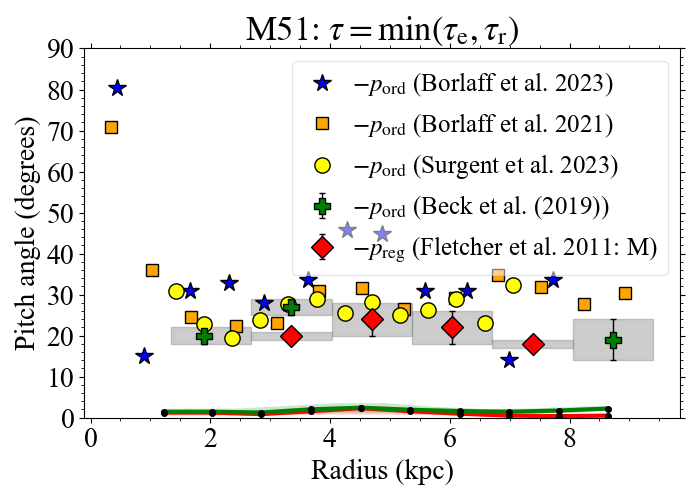}
    
\caption{Comparison of M51 field strengths (top) and pitch angles (bottom) for $\tau = \tau_\mathrm{e}$ (left panel) and $\tau = \mathrm{min}(\tau\renov, \tau_\mathrm{e})$ (right panel).
    }
    \label{fig:M51_tau_compare}
\end{figure*}

\begin{figure*}
    \includegraphics[width=8cm,keepaspectratio]{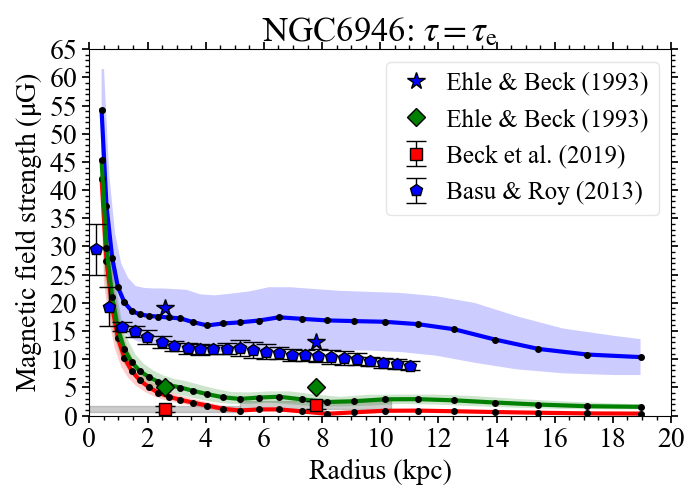}
    \includegraphics[width=8cm,keepaspectratio]{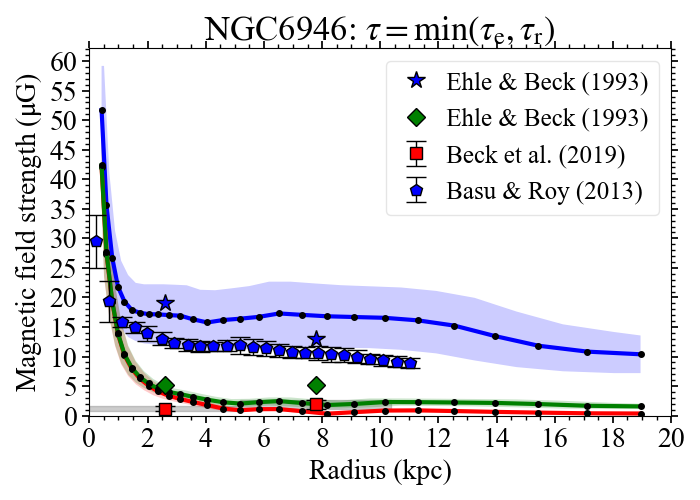}
    
    \includegraphics[width=8cm,keepaspectratio]{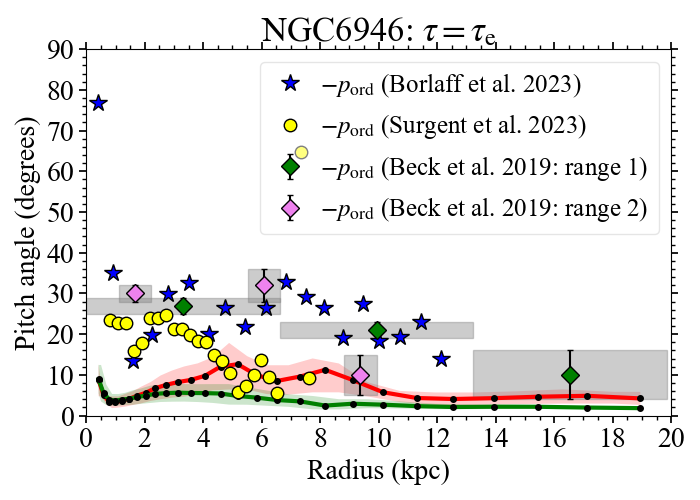}
    \includegraphics[width=8cm,keepaspectratio]{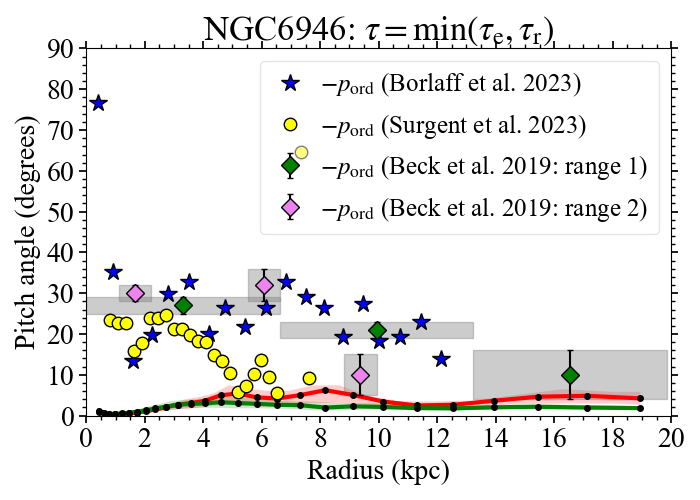}
    
\caption{Same as Figure \ref{fig:M51_tau_compare} but for NGC~6946.}
    \label{fig:NGC6946_tau_compare}
\end{figure*}

In this section, we make a comparison of the model outputs for field strengths and pitch angles of M51 (Figure \ref{fig:M51_tau_compare}) and NGC 6946 (Figure \ref{fig:NGC6946_tau_compare}) when two models for $\tau$ are used. The left panels show the results for the fiducial model used in this paper ($\tau = \tau_\mathrm{e}$), and the right panels show the results when we choose $\tau$ to be the least of eddy turnover time ($\tau_\mathrm{e}$) and renovation time ($\tau_\mathrm{r}$). 

As seen in Figure \ref{fig:combined_plot_r}, $\tau_\mathrm{r}$ is almost an order of magnitude smaller than $\tau_\mathrm{e}$ at all radii. This further reduces the values of pitch angles for both galaxies. So, we used $\tau = \tau_\mathrm{e}$ for all the galaxies studied. This choice does not affect the results for M31 and M33 for which $\tau_\mathrm{r} > \tau_\mathrm{e}$ always. 

\section{Empirical relationship between magnetic field and spiral arm pitch angles}
\label{sec:pcorrel}

A possible explanation for the discrepancy between predicted and observed pitch angles is the absence of spiral arm modeling in our theoretical framework. To investigate this idea, we examine the relationship between the pitch angles of the ordered and regular fields with those of the spiral arms. Using data from \citet{Beck+19} (Table 4), we find a positive correlation between the ordered field and spiral arm pitch angles (Pearson correlation coefficient $0.49$) 
as well as between the regular field and spiral arm pitch angles (Pearson correlation coefficient $0.86$). 
The magnitudes of the different types of pitch angle are also quite similar, though there is a lot of scatter
\citep[c.f.][]{Vaneck+15}.
New theoretical models that can explain these correlations are needed, 
but this is beyond the scope of the present work, which is limited to axisymmetric models.
\begin{figure*}[h b]
    \centering
    \includegraphics[width=\textwidth, clip = true,trim={5 0 0 0}]{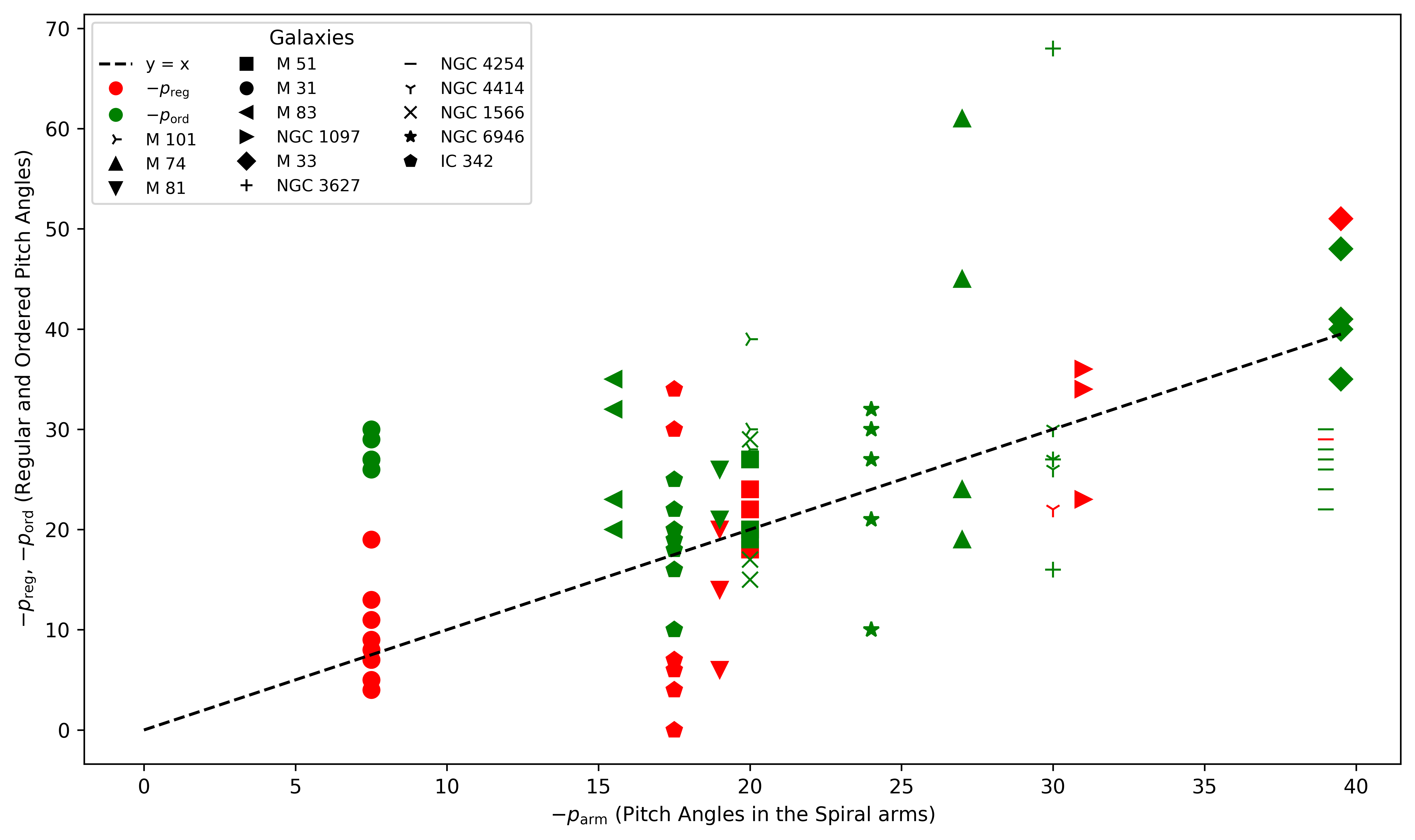}
    \caption{The pitch angles (with a negative sign to represent the magnitudes) for the ordered field $-p\ord$ and the regular field $-p\reg$ plotted against the pitch angles of the spiral arms $-p\arm$. Data is taken from \citet{Beck+19}. 
}
    \label{fig:pcorr}
\end{figure*}

\clearpage
\section{Supplementary Material}

\subsection{Data used as input for the model}
\begin{figure*}[h]
    \includegraphics[width=6.5cm,keepaspectratio]{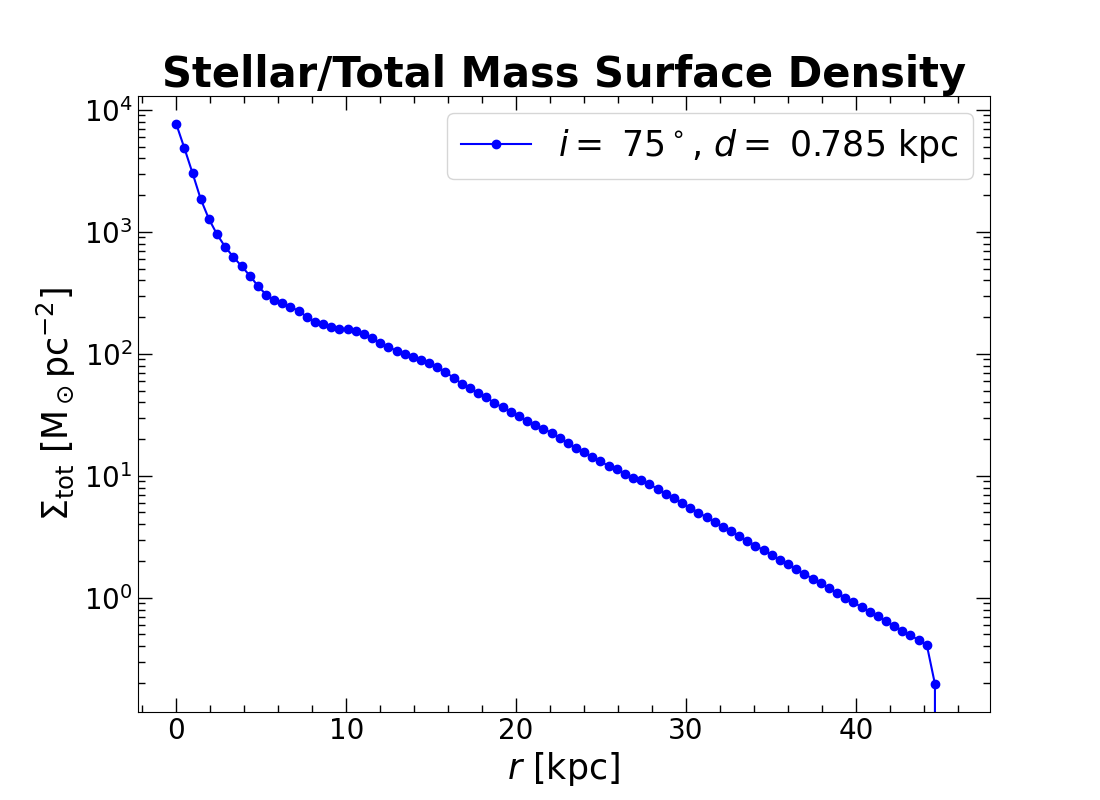}
    \includegraphics[width=6.5cm,keepaspectratio]{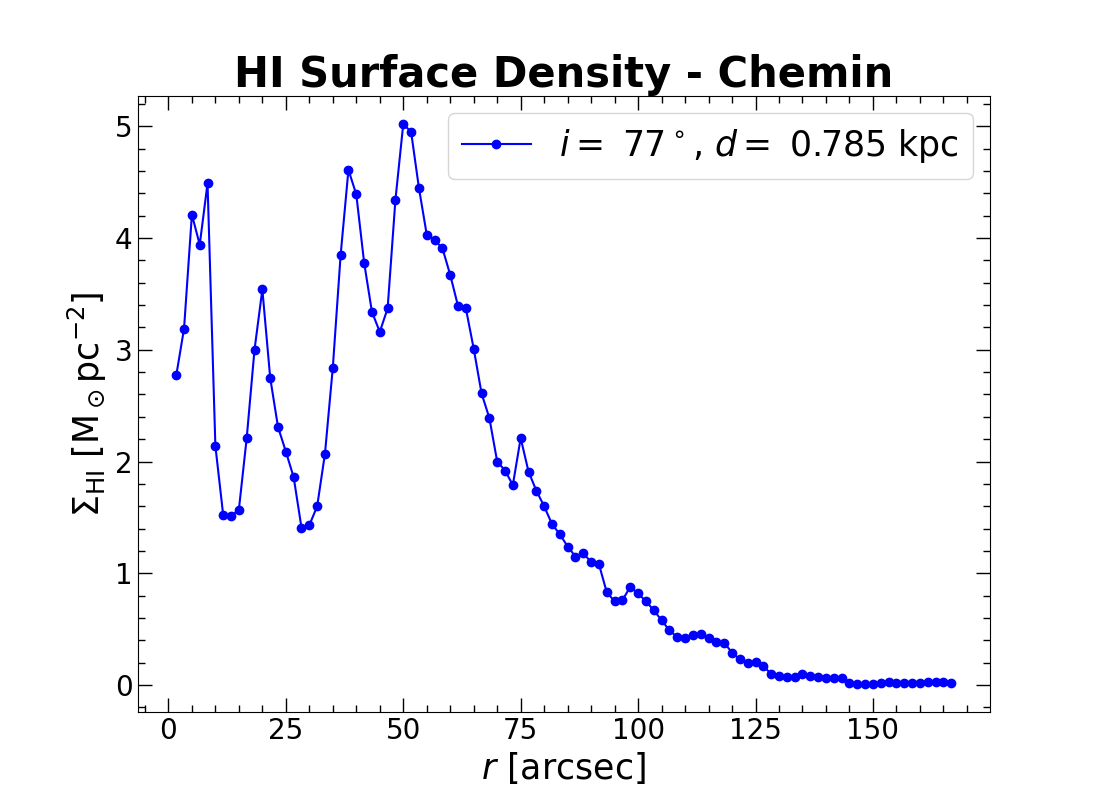}
    \includegraphics[width=6.5cm,keepaspectratio]{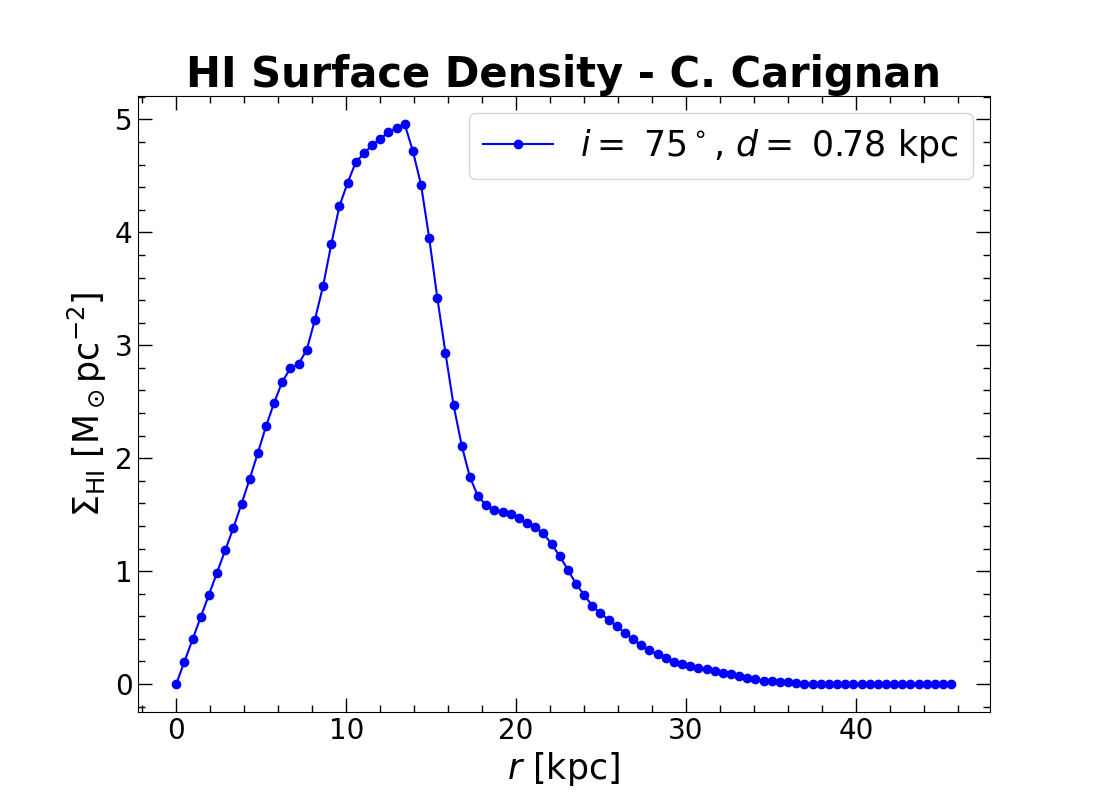}\\ 
    \includegraphics[width=6.5cm,keepaspectratio]{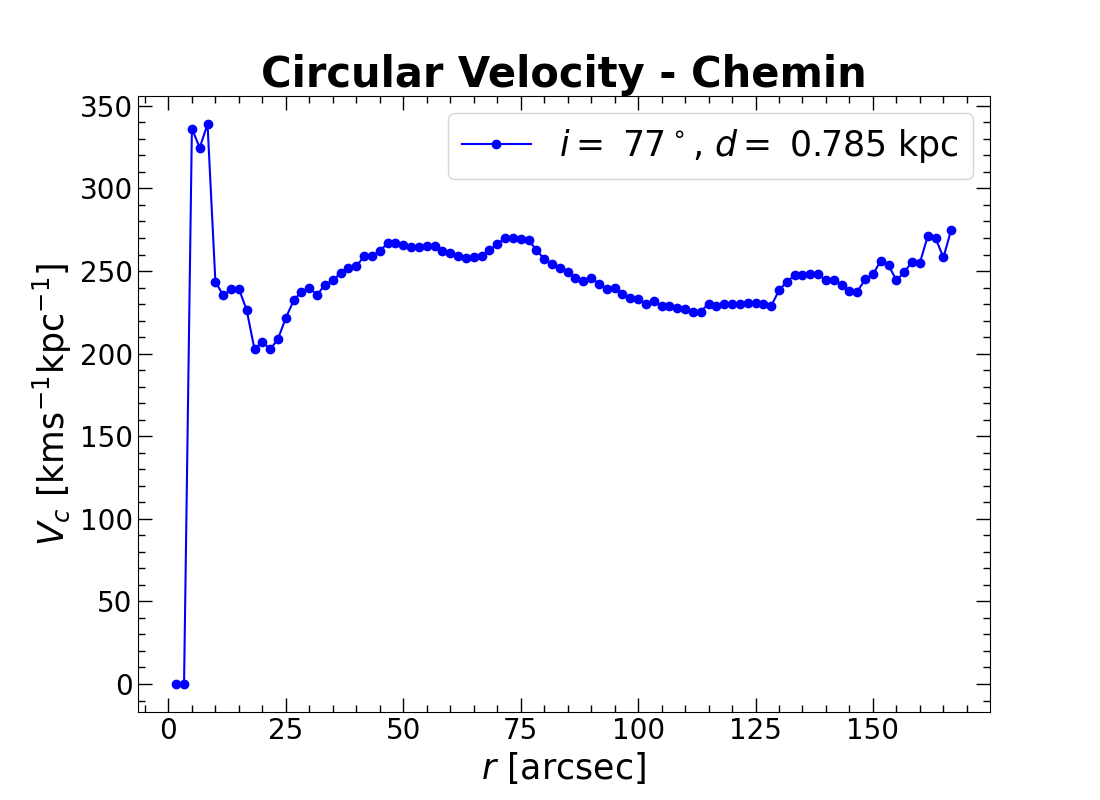}
    \includegraphics[width=6.5cm,keepaspectratio]{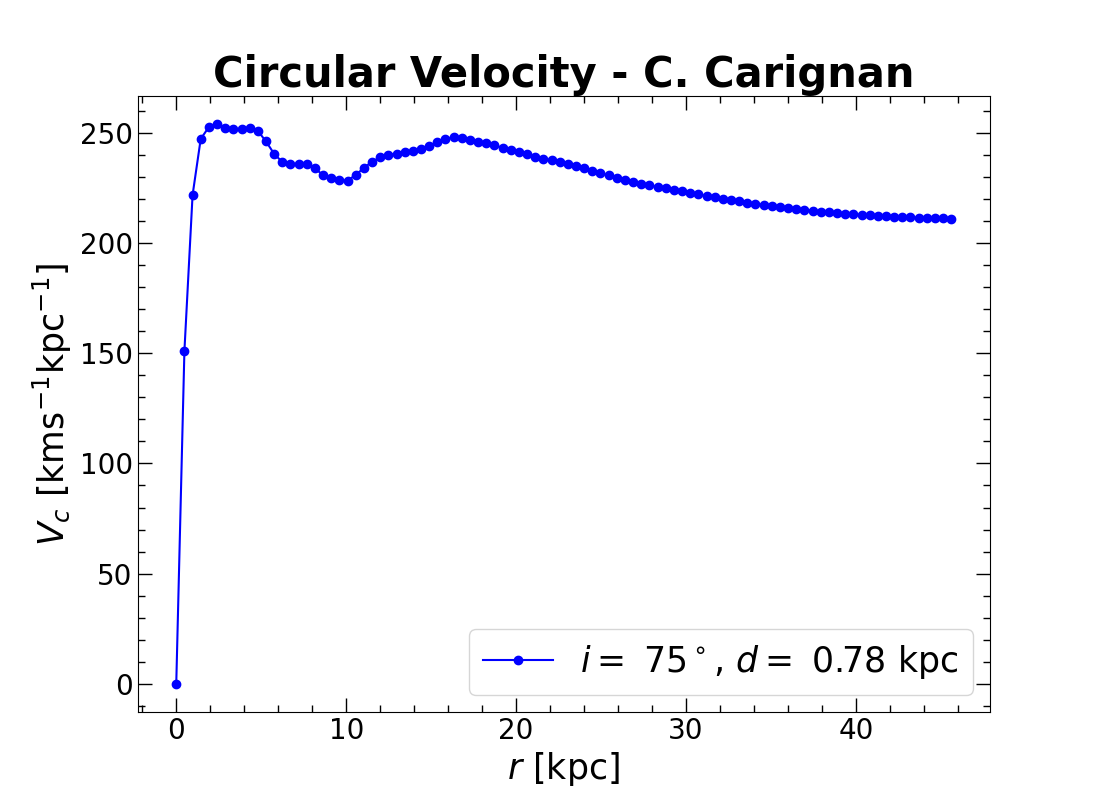}
    \includegraphics[width=6.5cm,keepaspectratio]{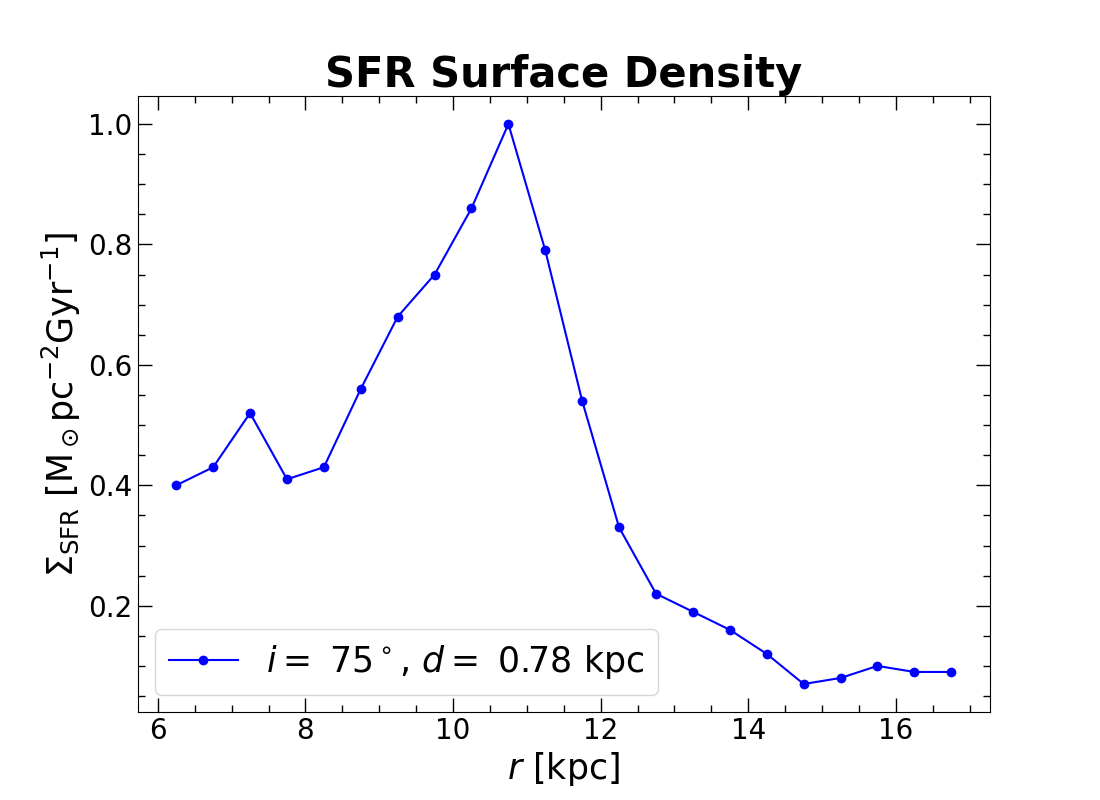}\\ 
    \includegraphics[width=6.5cm,keepaspectratio]{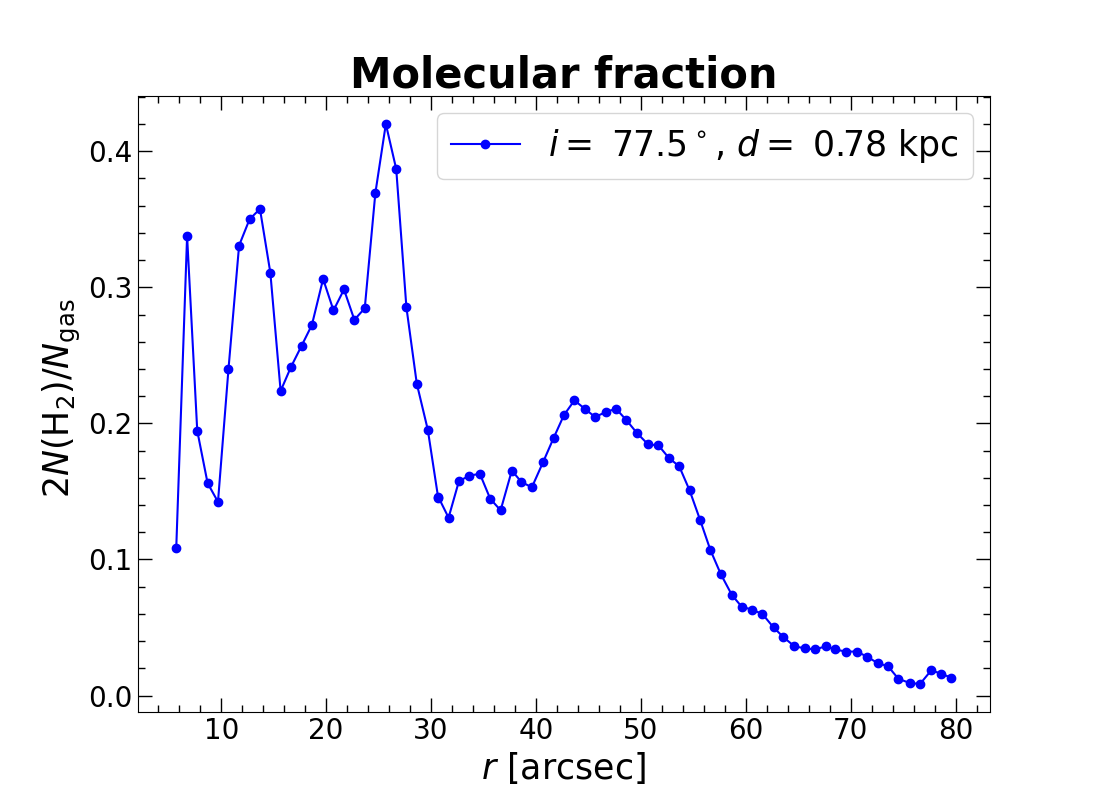} 
    \includegraphics[width=6.5cm,keepaspectratio]{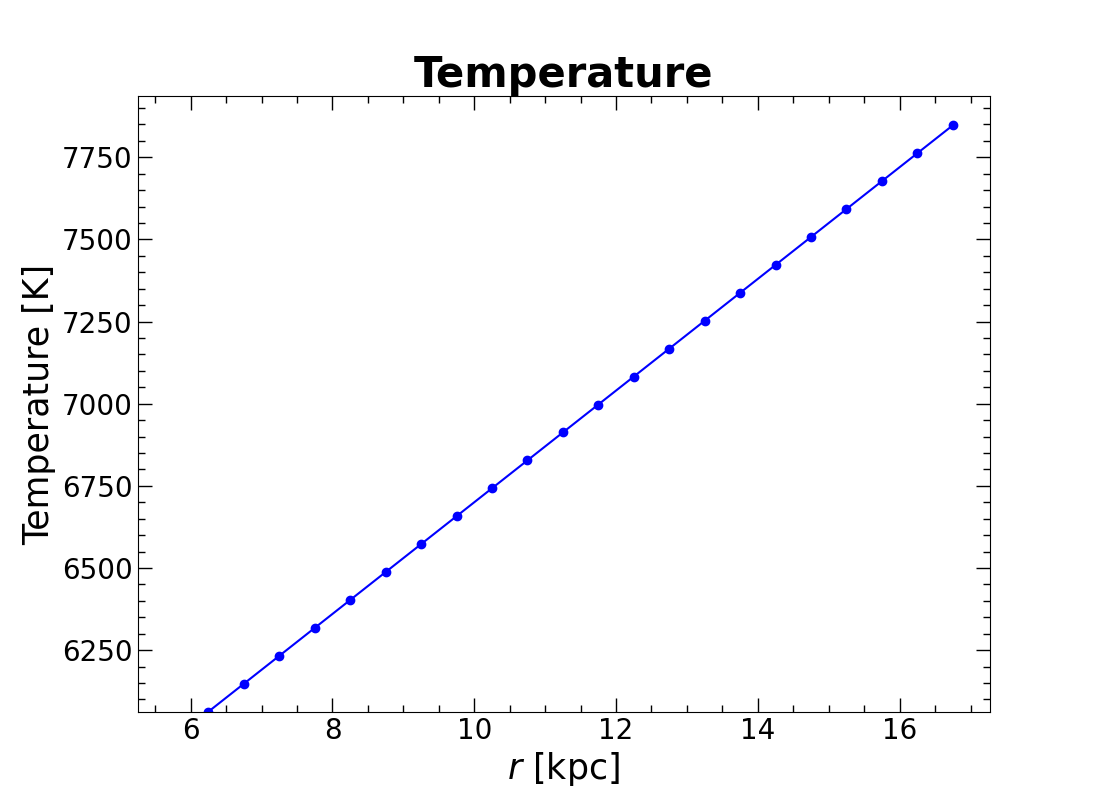} 
    \caption{Input quantities for M31 (Sources: C.~Carignan, priv.~comm., \cite{Chemin+09}, \cite{Nieten+06}, \cite{Tabatabaei+Berkhuijsen10}, \cite{Tabatabaei+13b}).
    }
    \label{fig:m31_input_qty}
\end{figure*}

\begin{figure*}[h]

    \includegraphics[width=6.5cm,keepaspectratio]{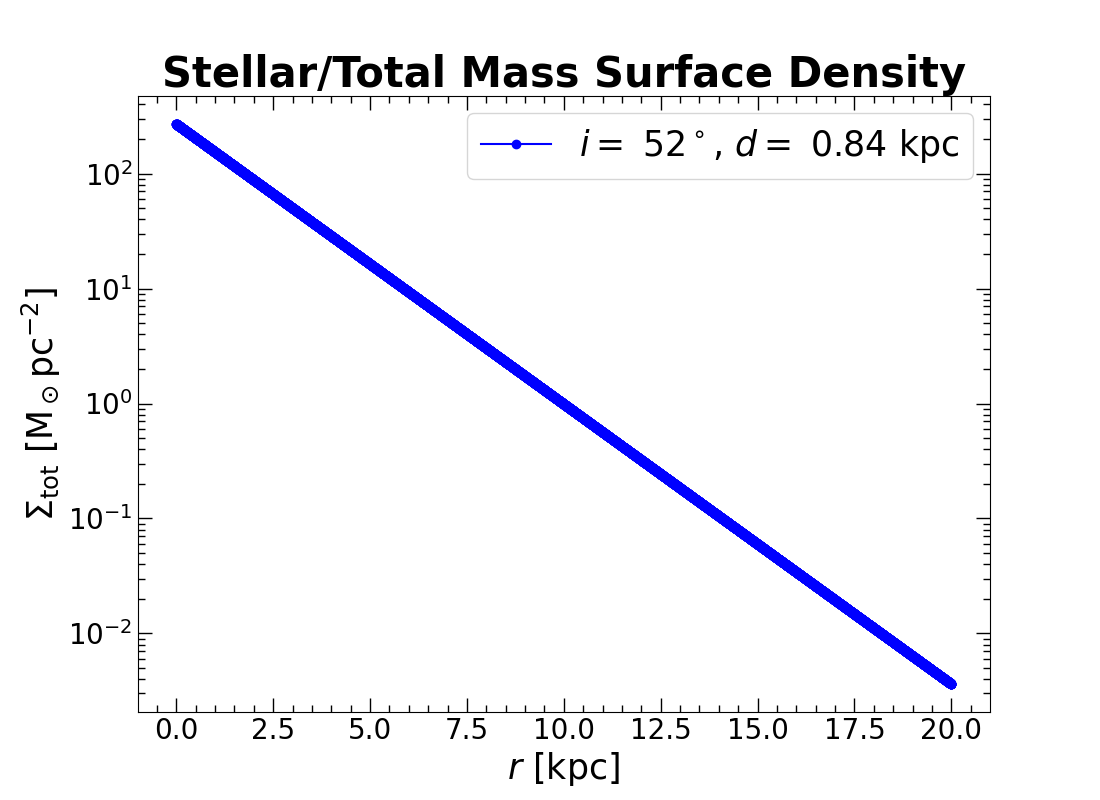}
    \includegraphics[width=6.5cm,keepaspectratio]{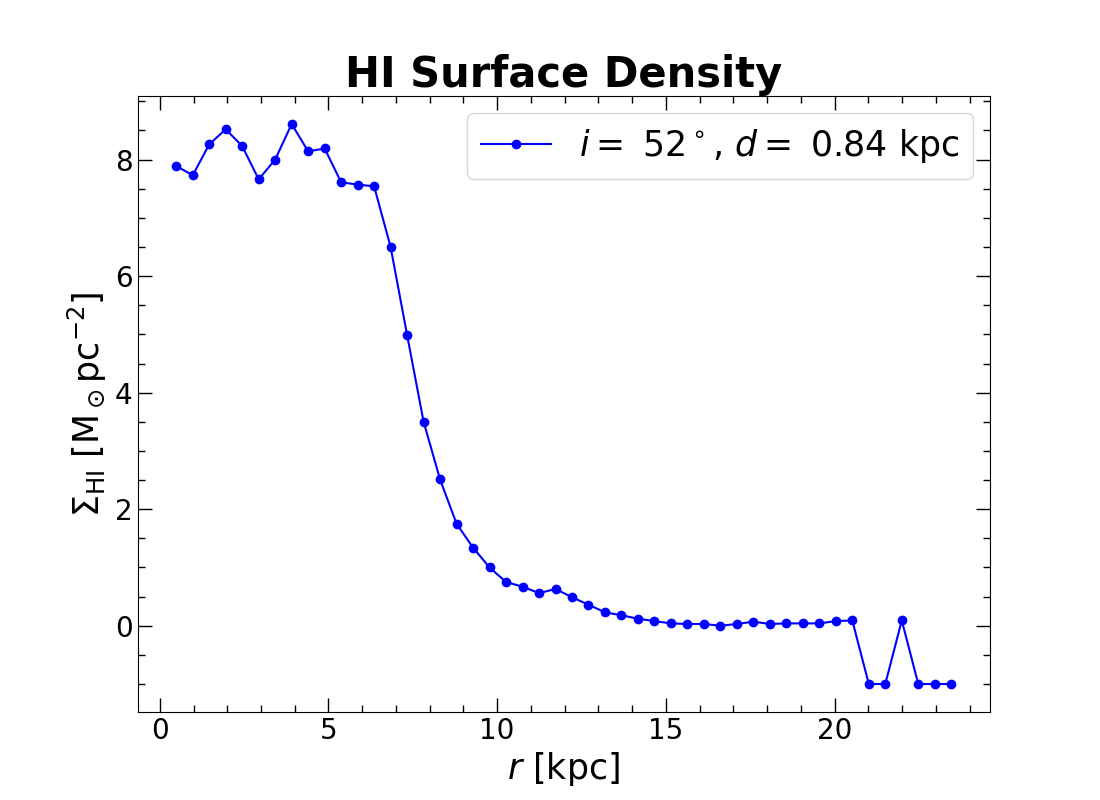}
    \includegraphics[width=6.5cm,keepaspectratio]{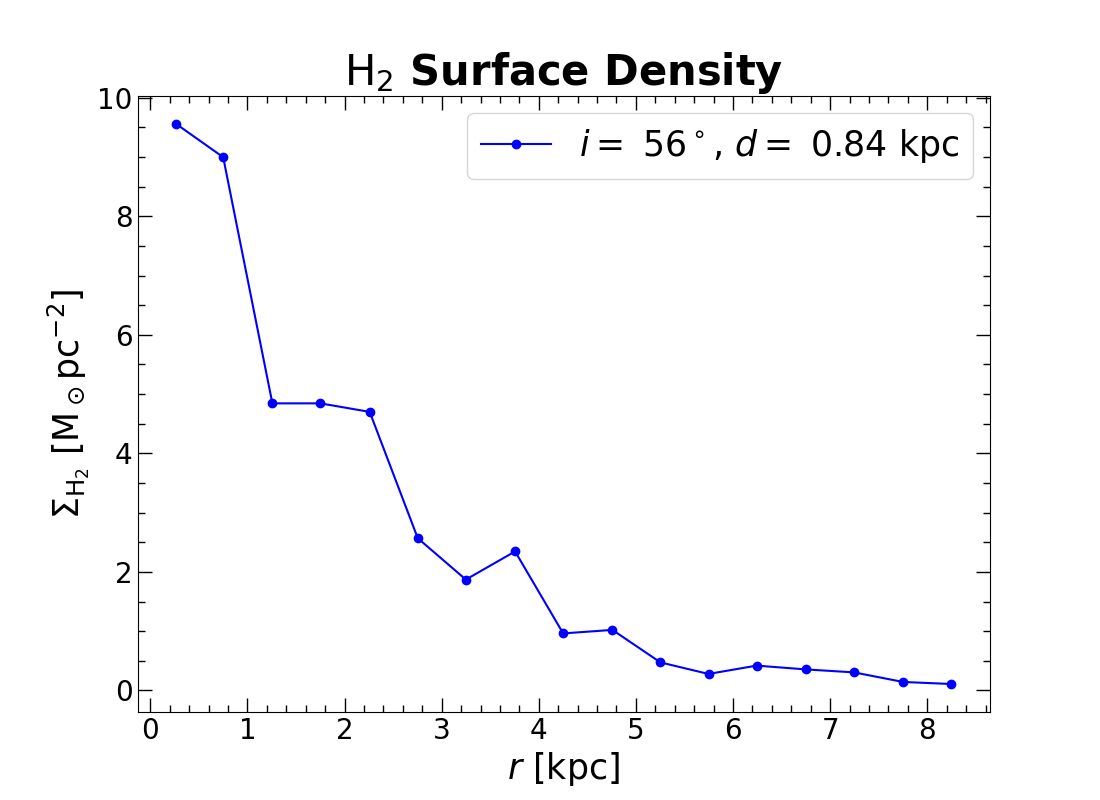}\\ 
    \includegraphics[width=6.5cm,keepaspectratio]{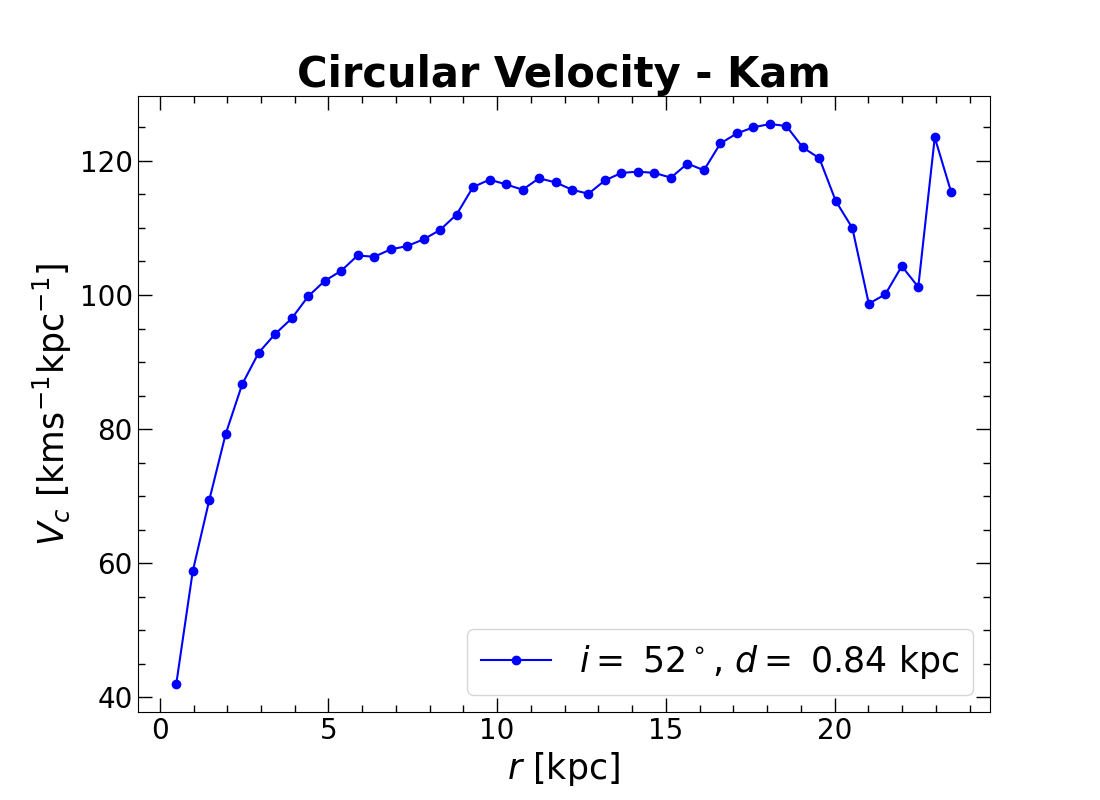}
    \includegraphics[width=6.5cm,keepaspectratio]{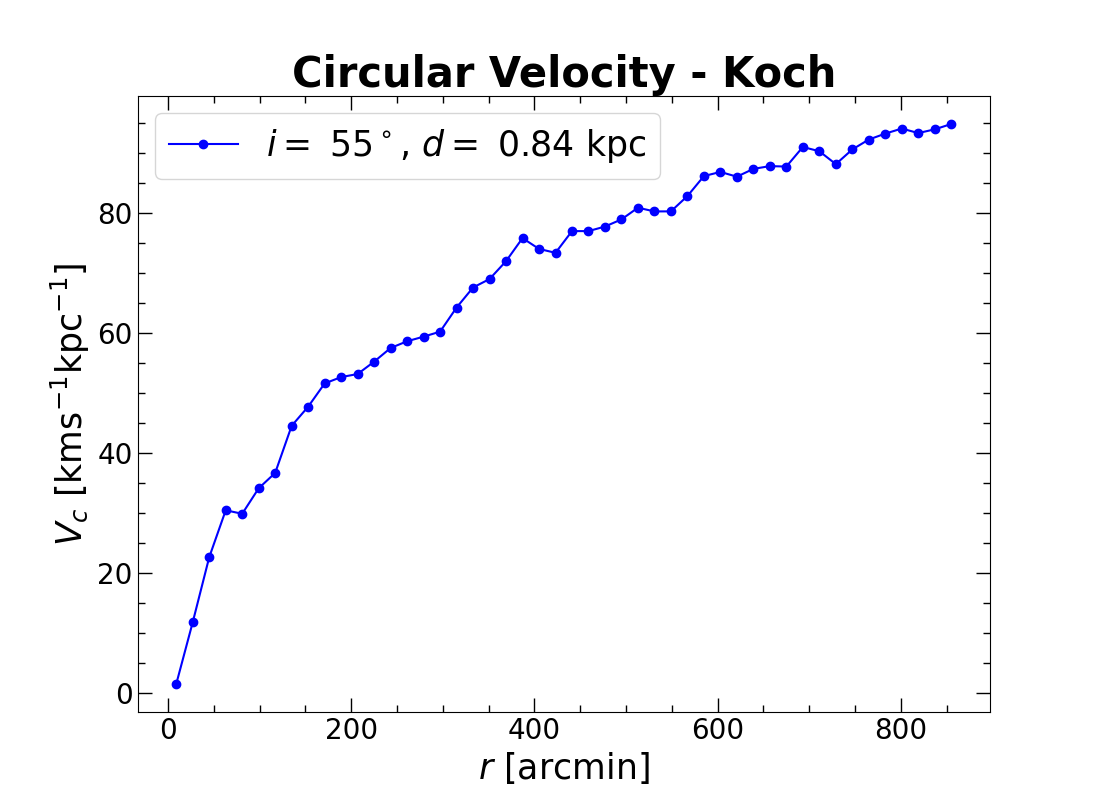}
    \includegraphics[width=6.5cm,keepaspectratio]{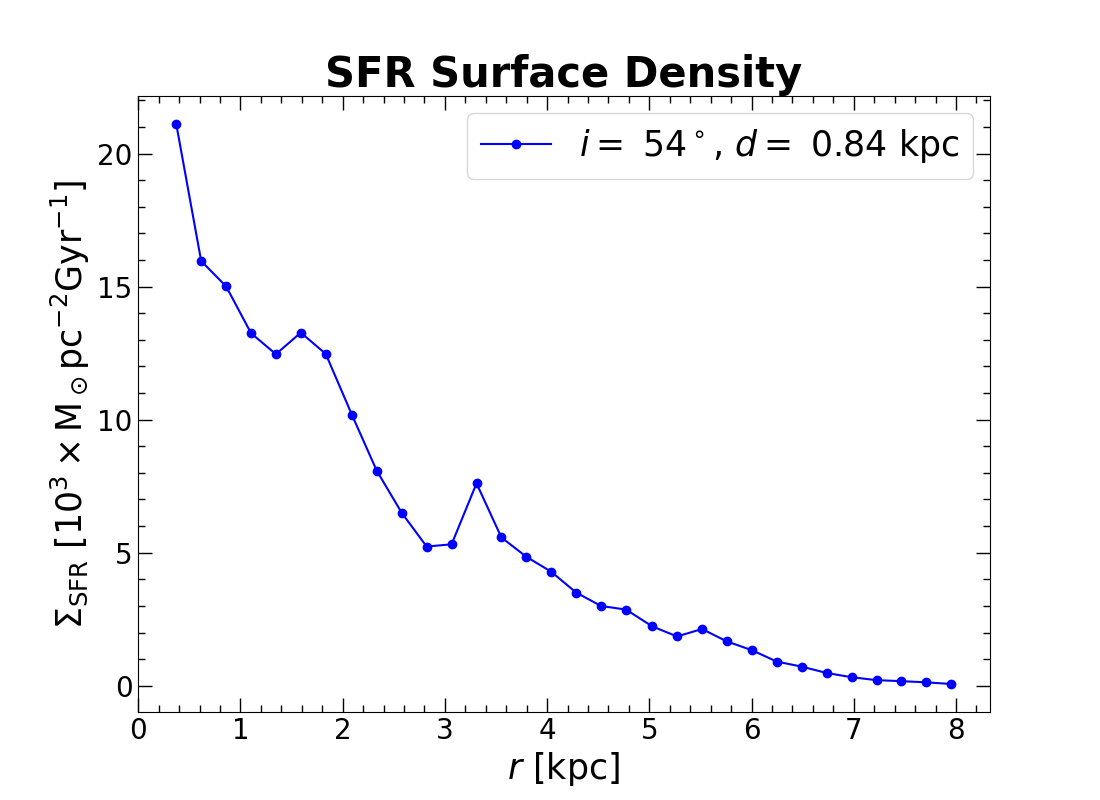}\\ 
    \includegraphics[width=6.5cm,keepaspectratio]{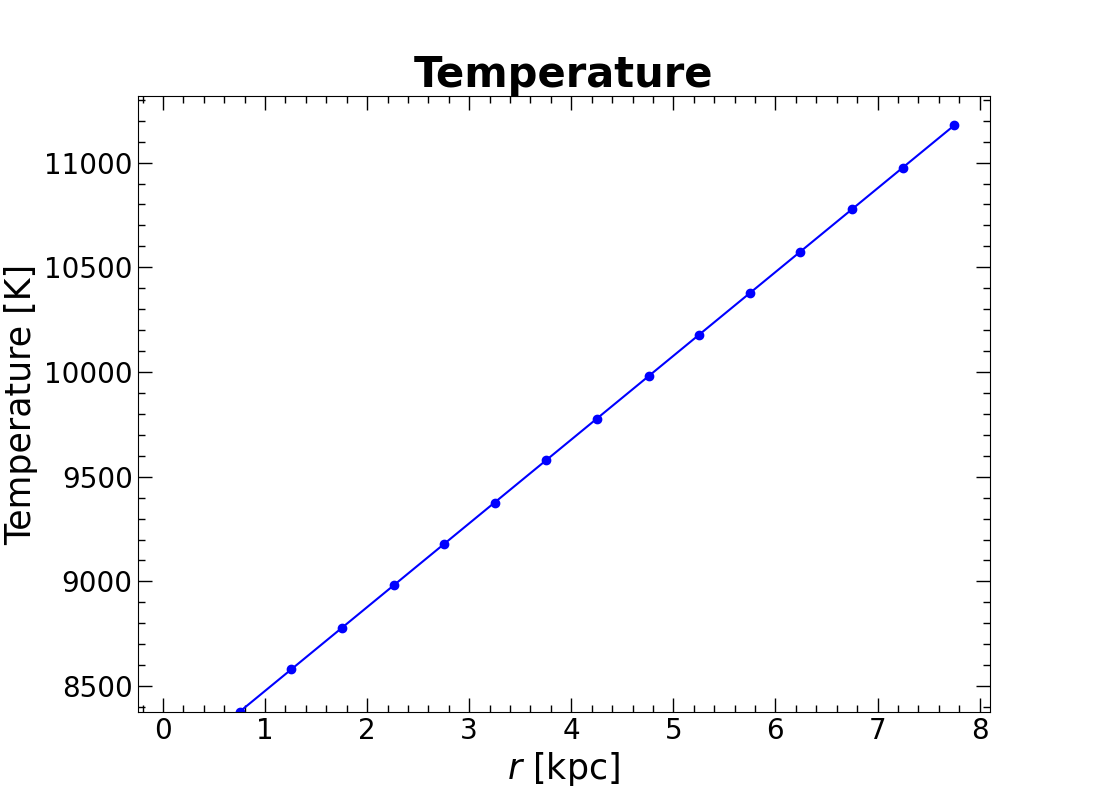} 
    
    \caption{Input quantities for M33 (Sources: \cite{Kam+17}, \cite{Koch+18a}, \cite{Gratier+10}, \cite{Verley+09}, \cite{Lin+17}).
    }
    \label{fig:m33_input_qty}
\end{figure*}

\begin{figure*}[h]

    \includegraphics[width=6.5cm,keepaspectratio]{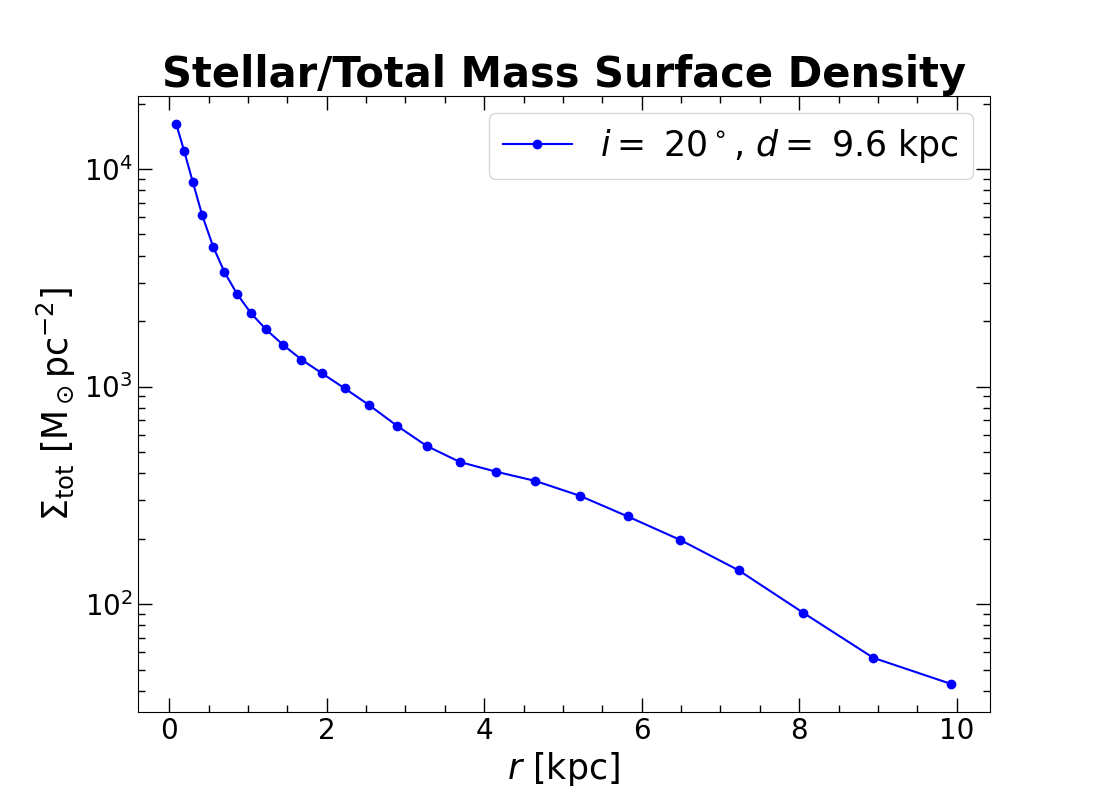}
    \includegraphics[width=6.5cm,keepaspectratio]{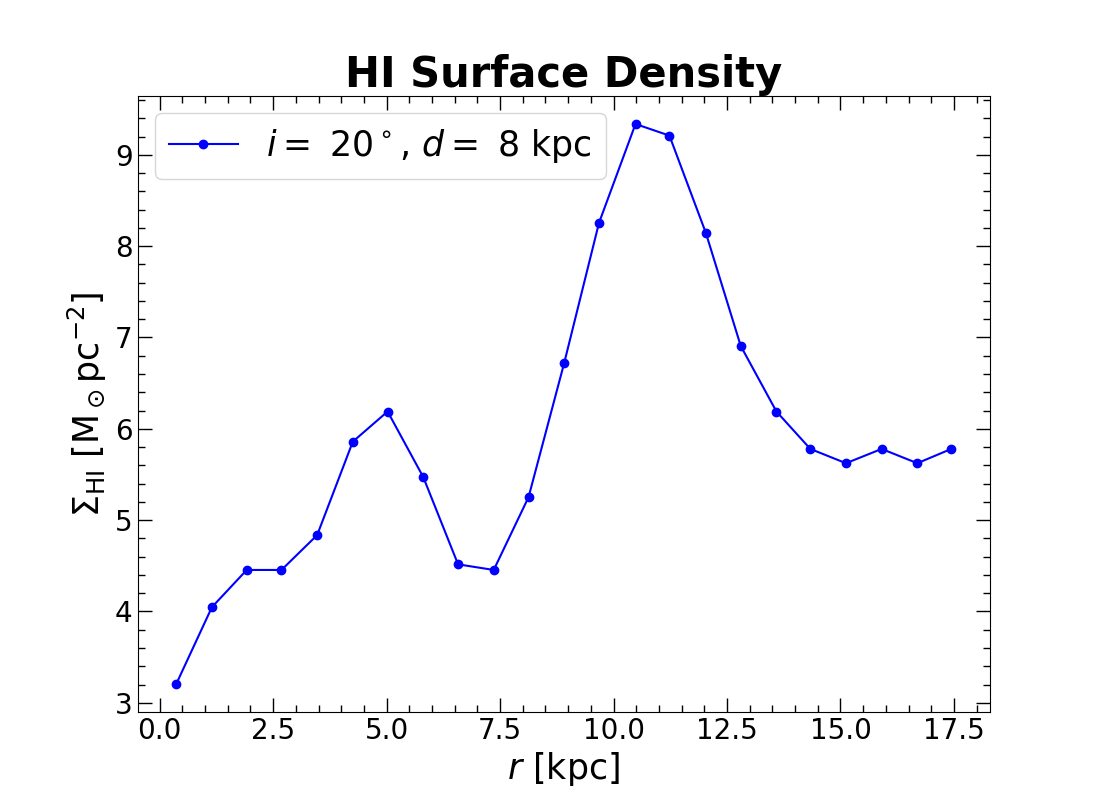}
    \includegraphics[width=6.5cm,keepaspectratio]{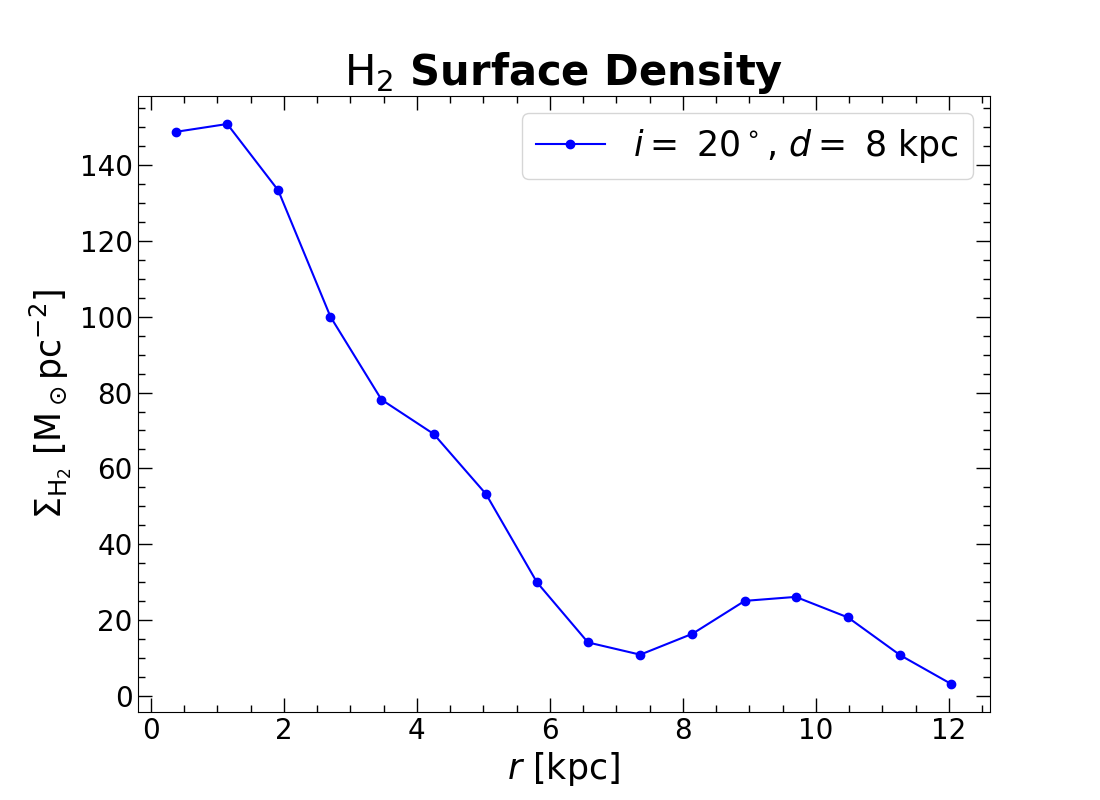}\\ 
    \includegraphics[width=6.5cm,keepaspectratio]{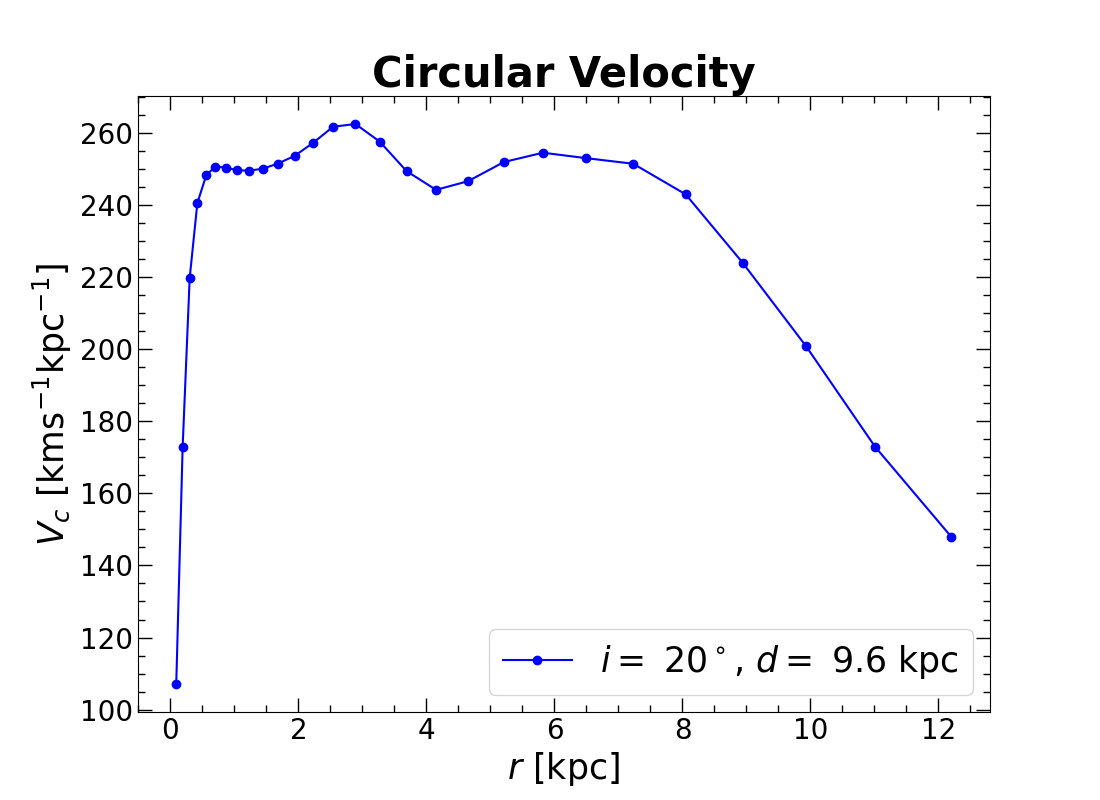}
    \includegraphics[width=6.5cm,keepaspectratio]{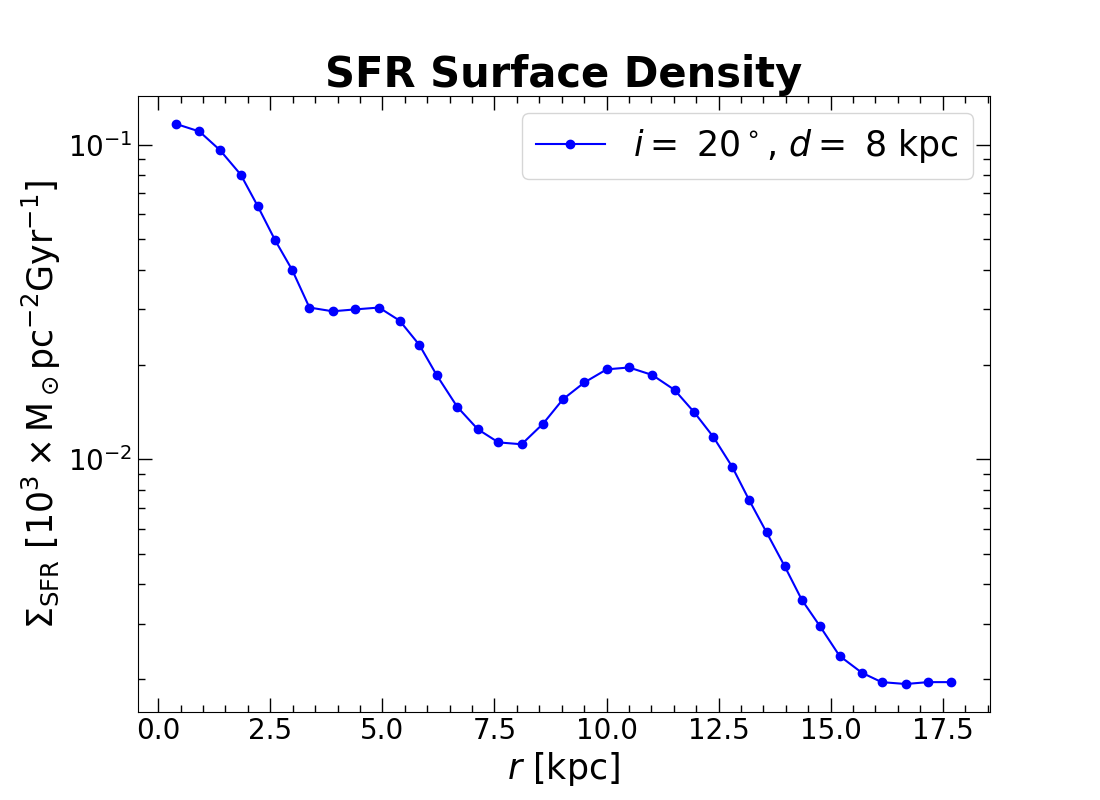}
    \includegraphics[width=6.5cm,keepaspectratio]{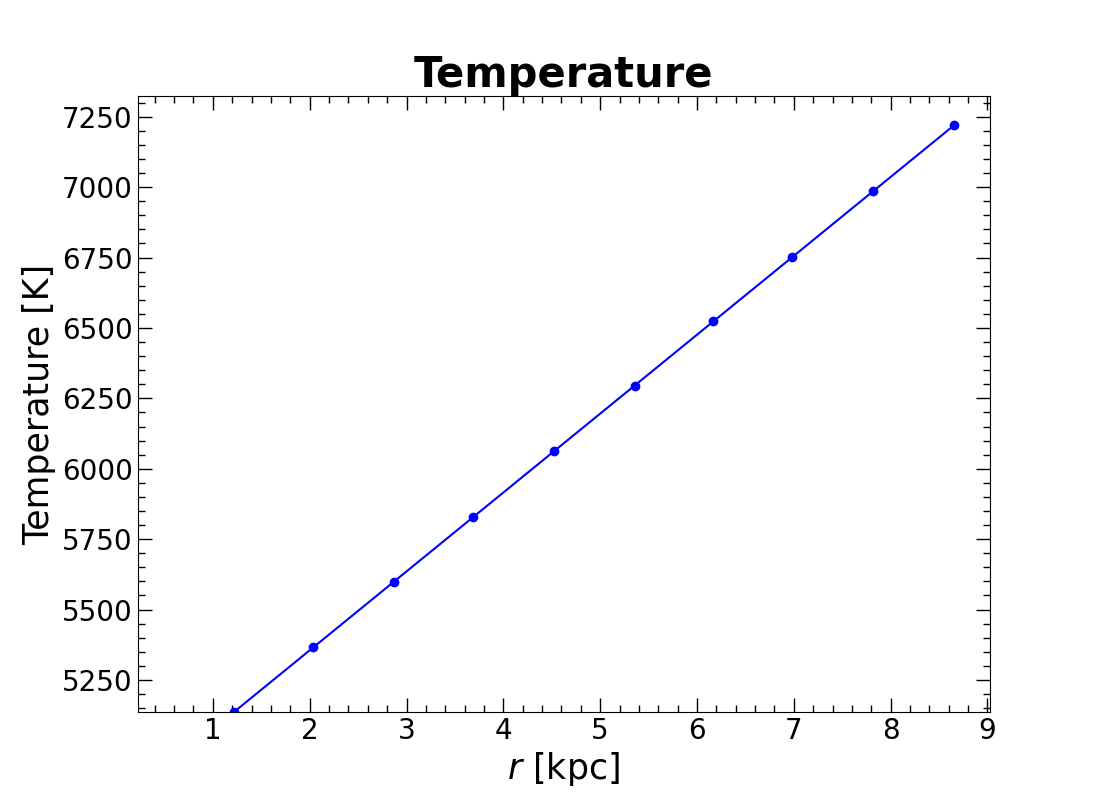} 
    
    \caption{Input quantities for M51 (Sources: \cite{Sofue+18}, \cite{Bigiel+08}, \cite{Kumari+20}, \cite{Bresolin+04}).
    }
    \label{fig:m51_input_qty}
\end{figure*}

\begin{figure*}[h]

    \includegraphics[width=6.5cm,keepaspectratio]{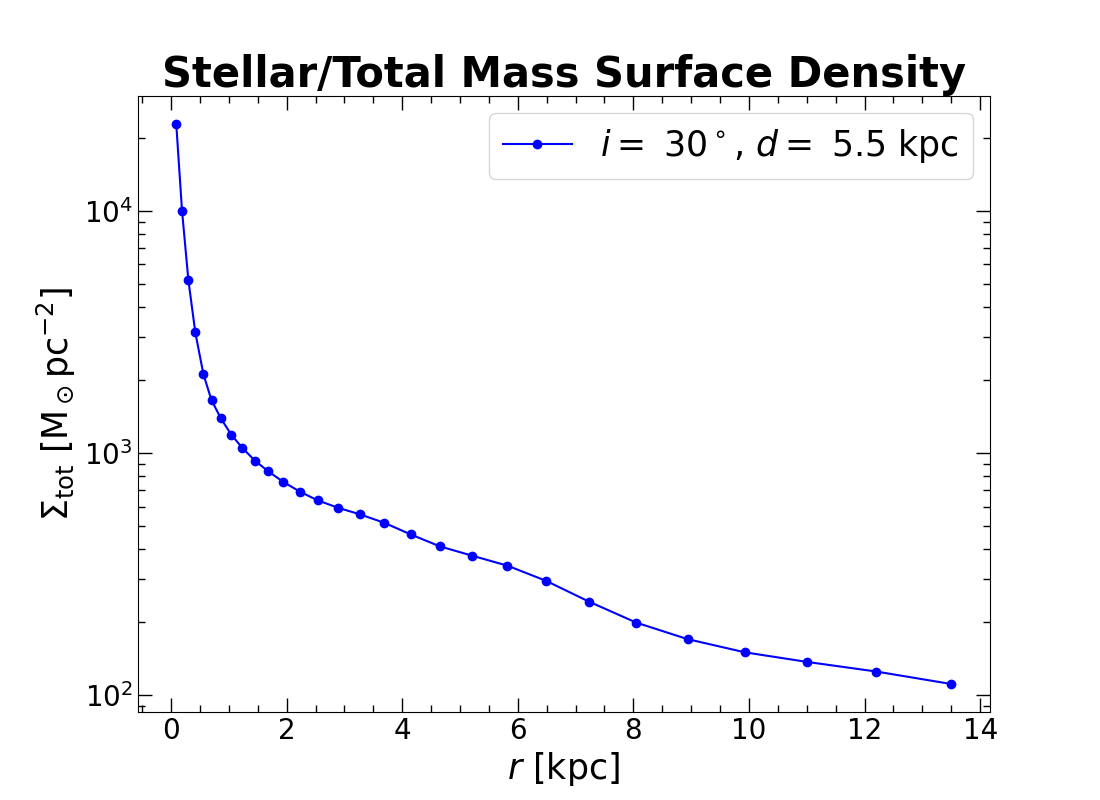}
    \includegraphics[width=6.5cm,keepaspectratio]{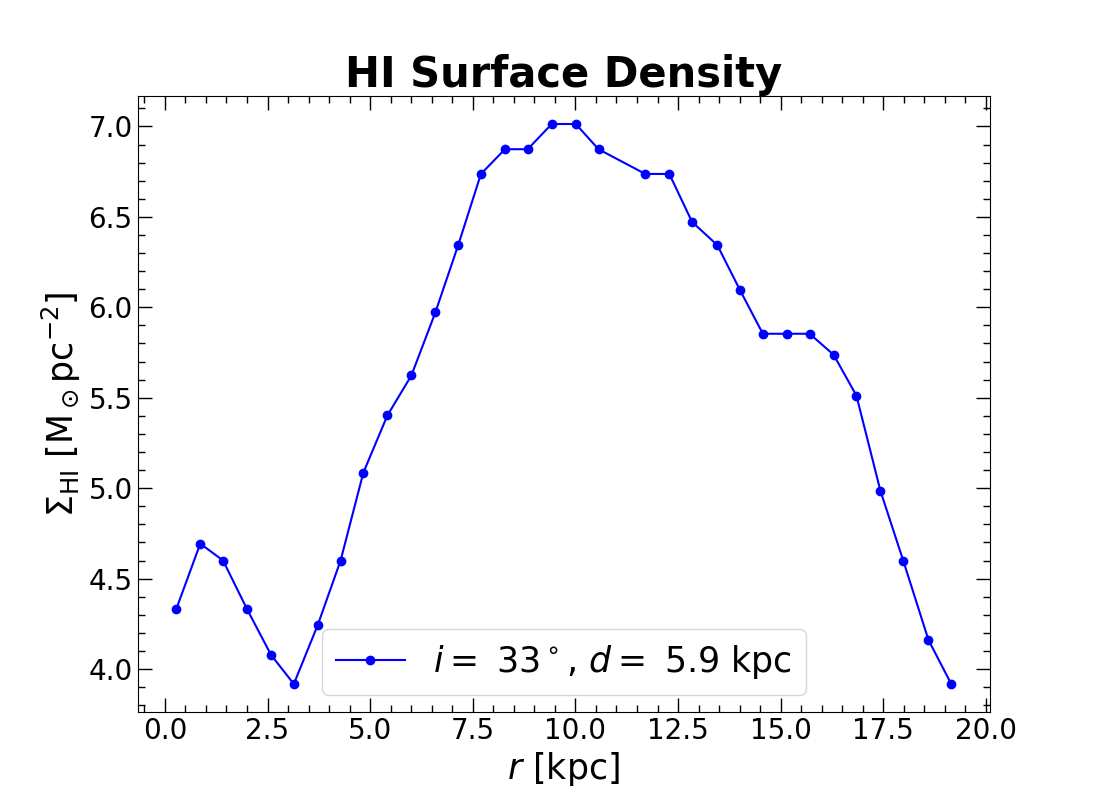}
    \includegraphics[width=6.5cm,keepaspectratio]{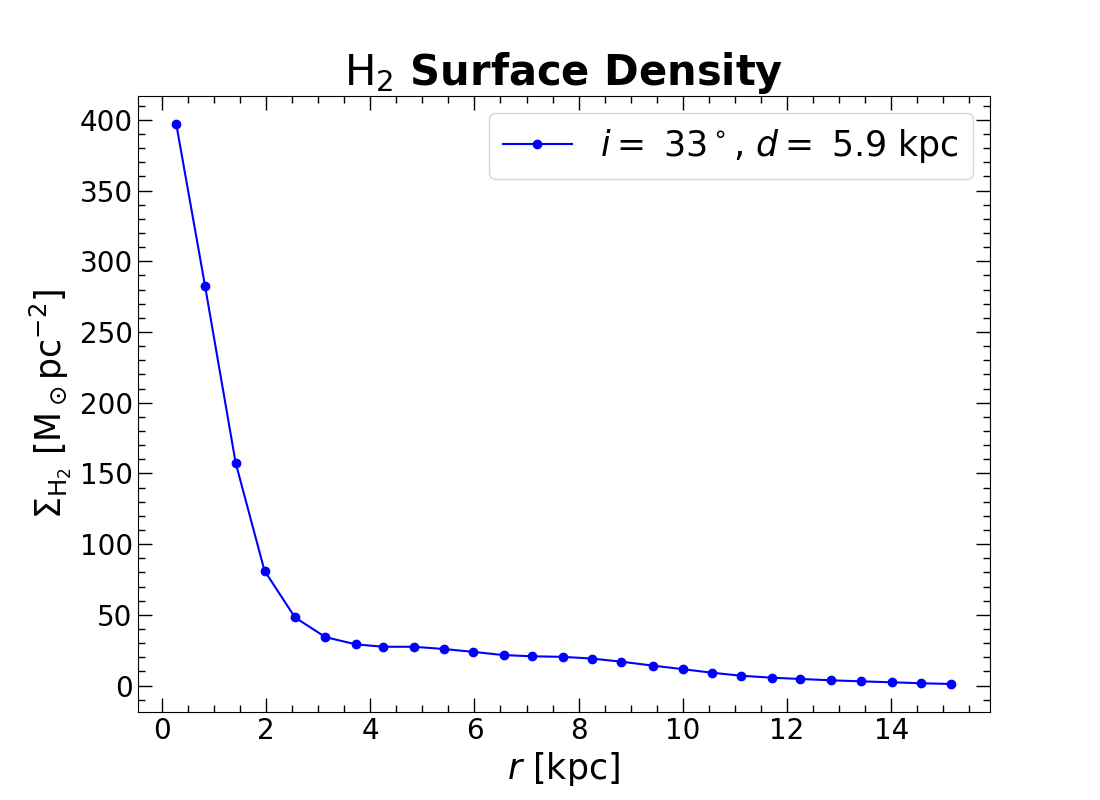}\\ 
    \includegraphics[width=6.5cm,keepaspectratio]{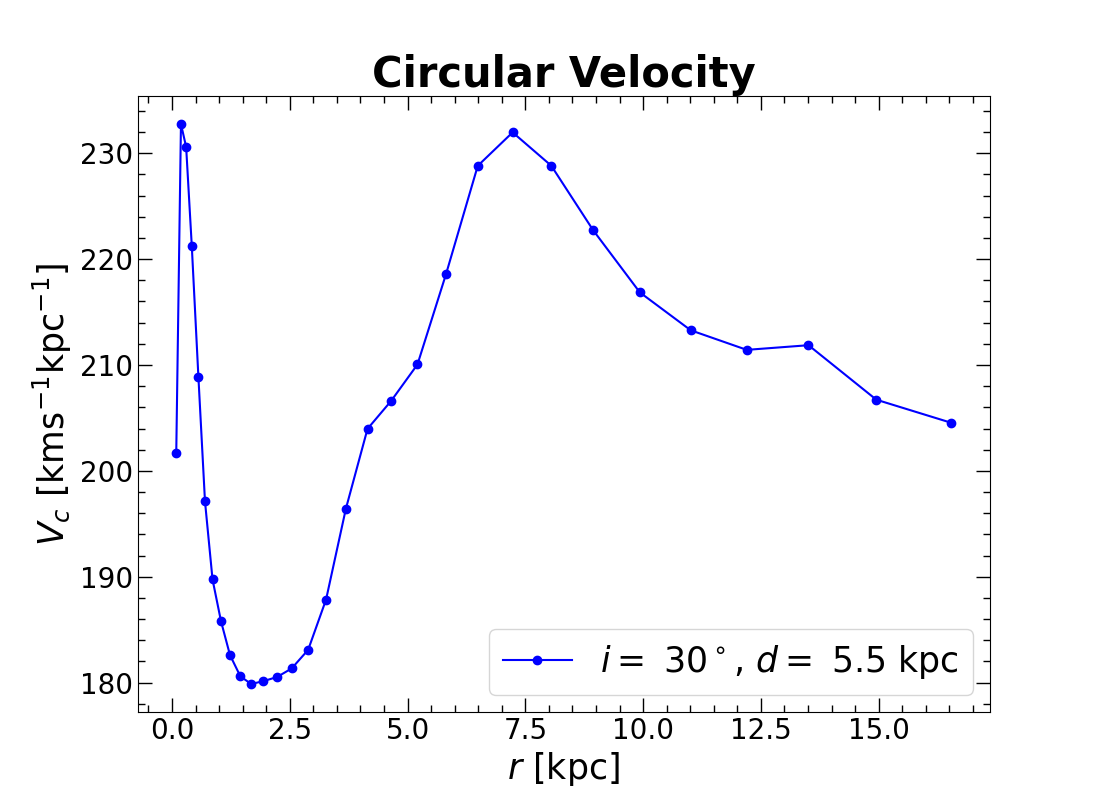}
    \includegraphics[width=6.5cm,keepaspectratio]{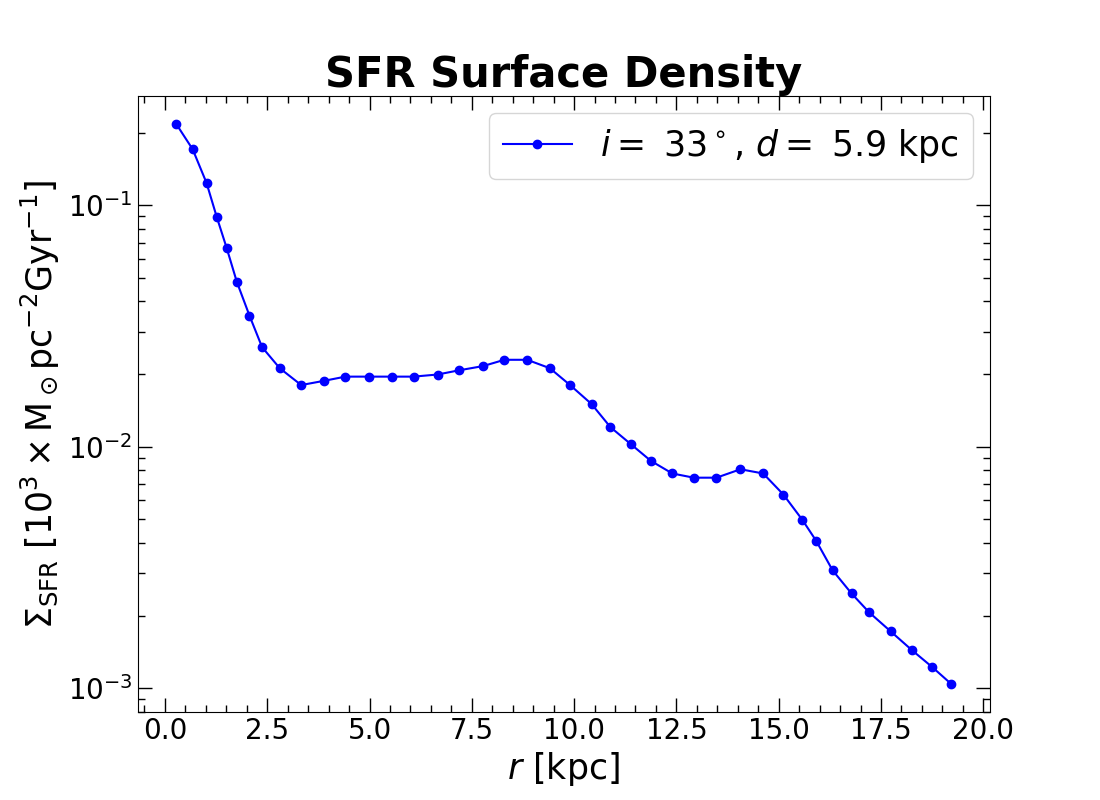}
    \includegraphics[width=6.5cm,keepaspectratio]{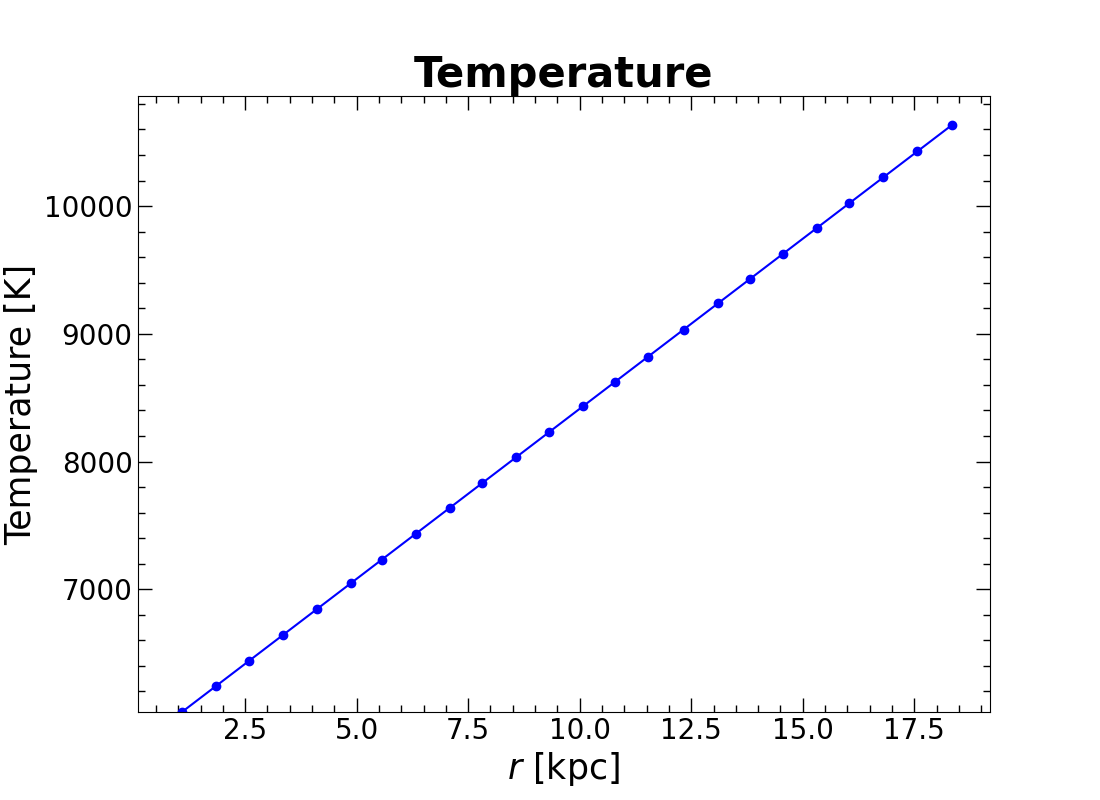} 
    \caption{Input quantities for NGC~6946 (Sources: \cite{Sofue+18}, \cite{Bigiel+08}, \cite{Kumari+20}, \cite{Gusev+13}).
    }
    \label{fig:ngc6946_input_qty}
\end{figure*}

Figures \ref{fig:m31_input_qty}, \ref{fig:m33_input_qty}, \ref{fig:m51_input_qty} and \ref{fig:ngc6946_input_qty} show the original data obtained for the study for M31, M33, M51 and NGC~6946 respectively. Note that these plots show the data for the entire radial extent as presented in the source papers, and have not been corrected for distance and inclination. The inclination and distance used in the source paper are indicated in the legend of each plot. Note that the temperature data plotted corresponds to the straight line fits of the data from the source papers. When we use these data in our model, they are first converted to the coarsest resolution and then corrected for inclination and distance. The data with the coarsest resolution are as follows for each of the galaxies (when molecular gas is not considered): M31- $\Sigma_{\mathrm{SFR}}$, M33- $\Sigma_{\mathrm{SFR}}$, M51- $\Sigma_{\mathrm{HI}}$ and NGC~6946- $\Sigma_{\star}$.  

\subsection{Comparison of ratio of gas to stellar surface density, with and without molecular gas}

\begin{center}
    
\begin{figure*}[h]

    \includegraphics[width=8cm,keepaspectratio]{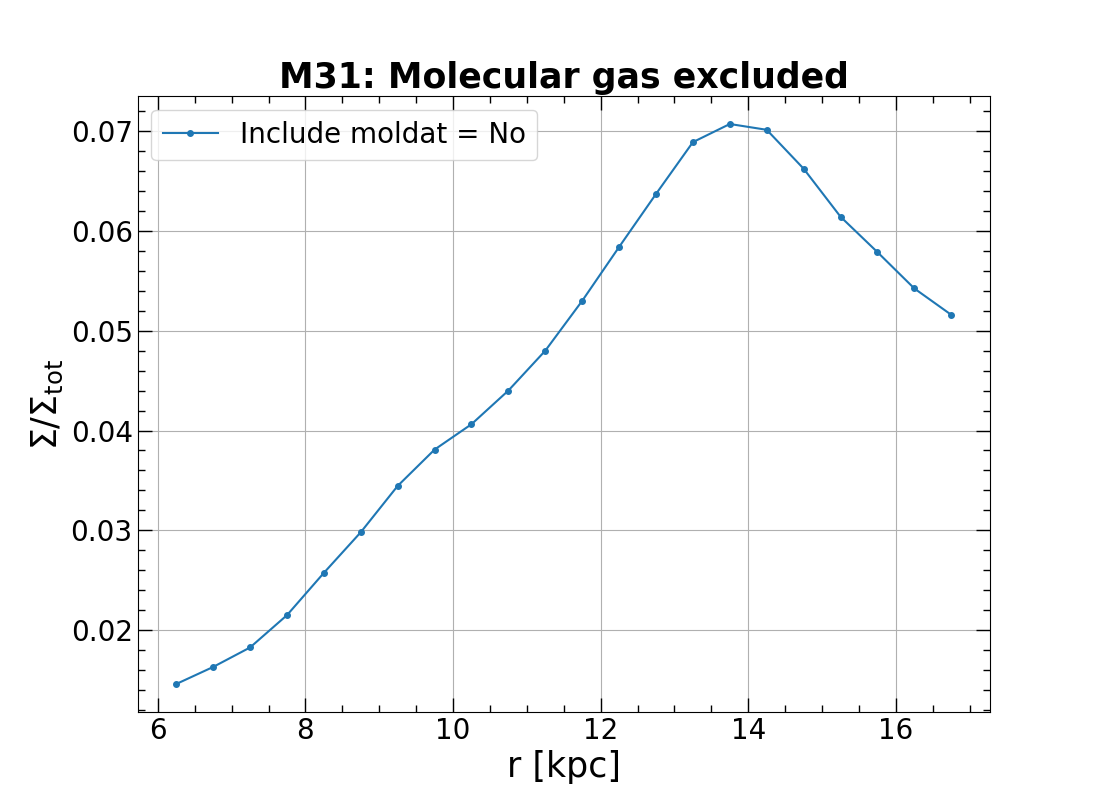}
    \includegraphics[width=8cm,keepaspectratio]{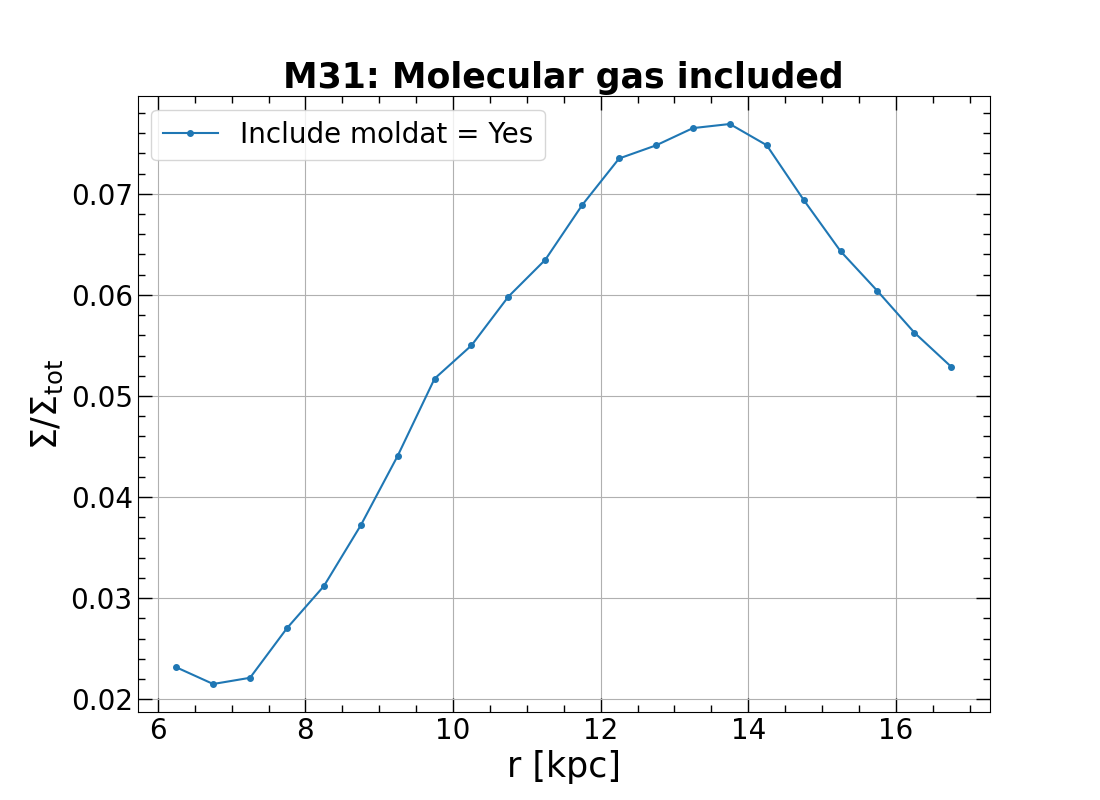}\\ 
    \includegraphics[width=8cm,keepaspectratio]{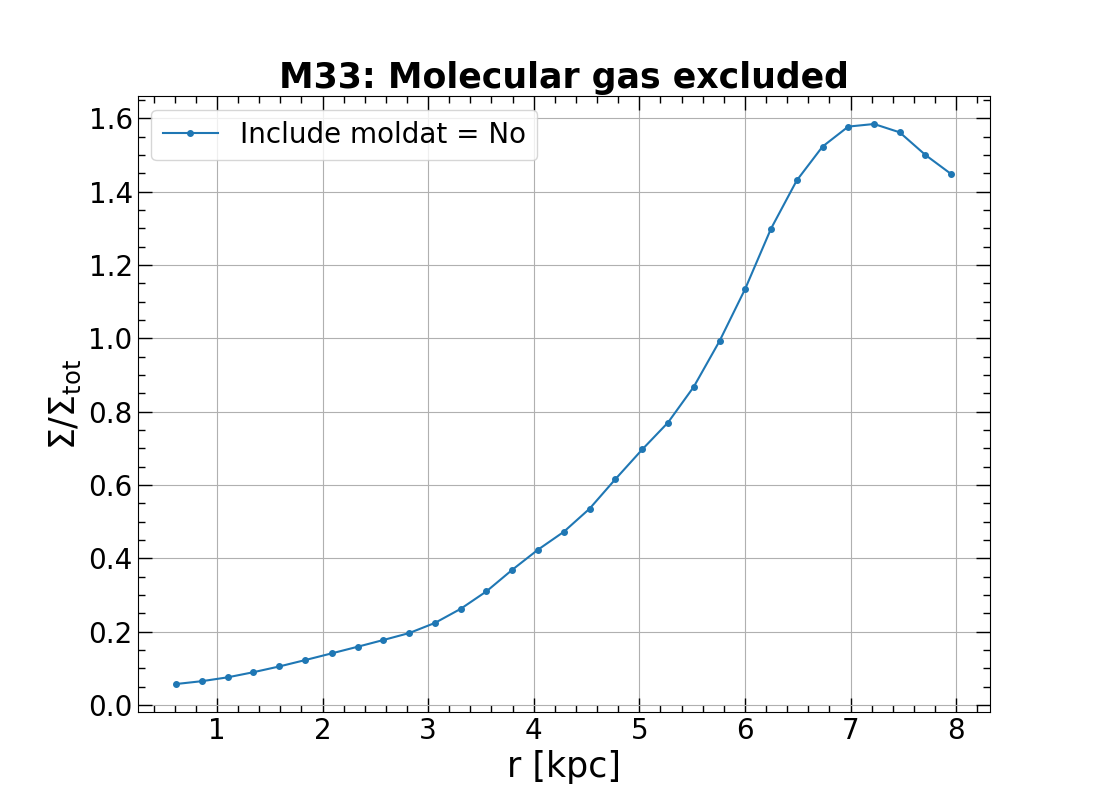}
    \includegraphics[width=8cm,keepaspectratio]{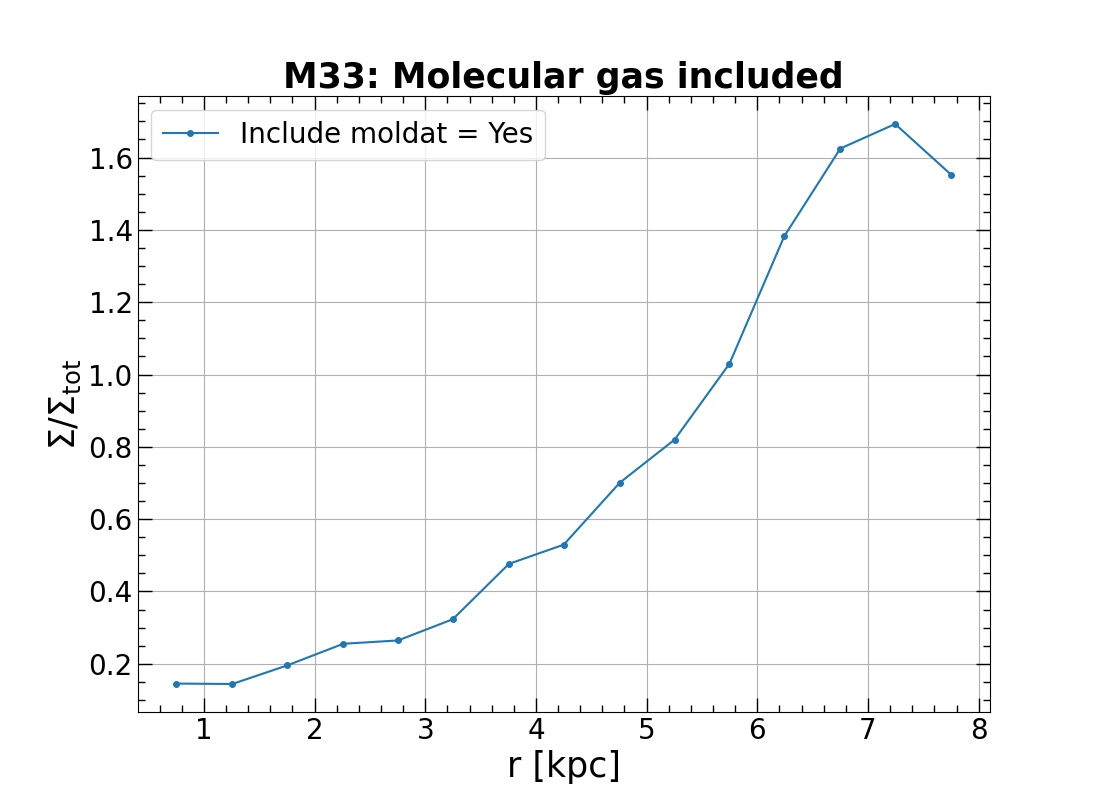}\\ 
    \includegraphics[width=8cm,keepaspectratio]{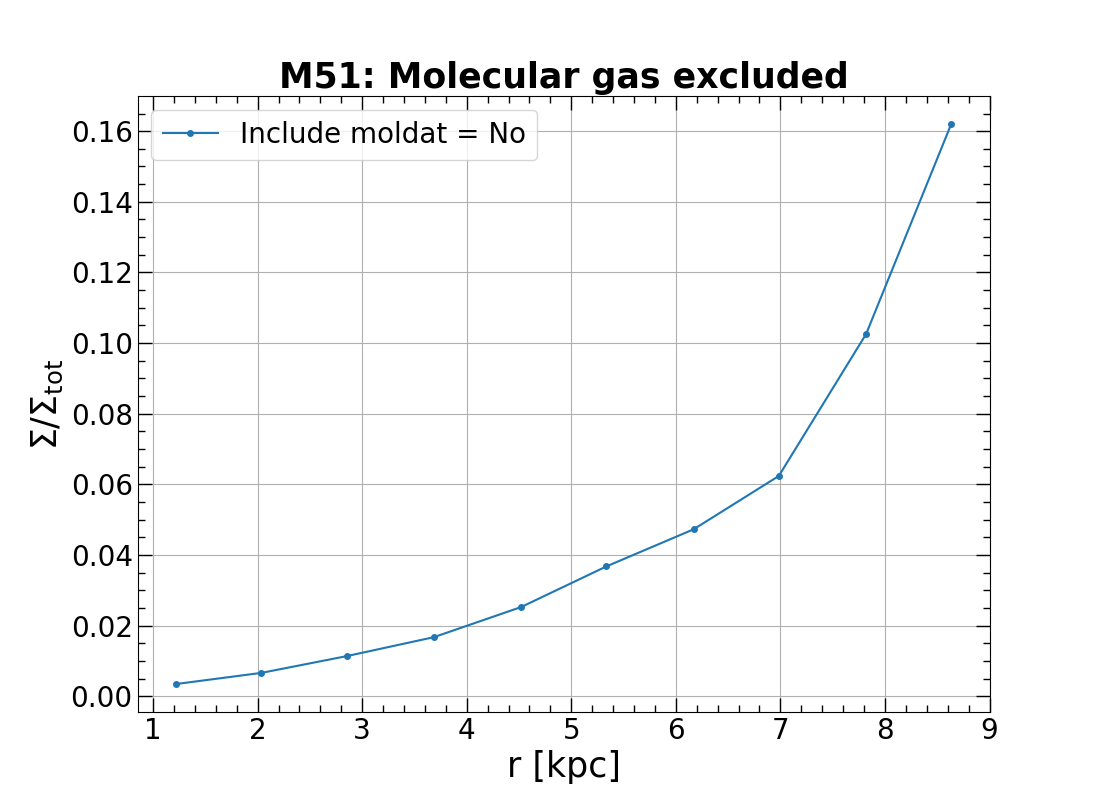}
    \includegraphics[width=8cm,keepaspectratio]{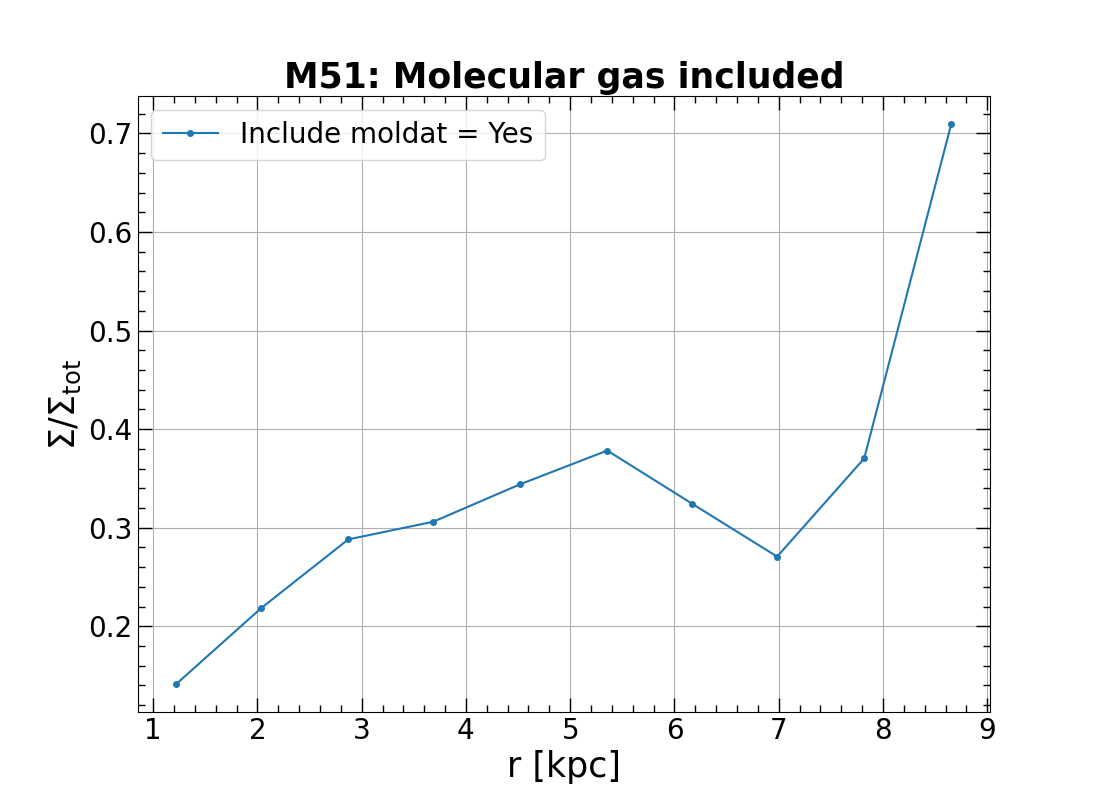}\\ 
    \includegraphics[width=8cm,keepaspectratio]{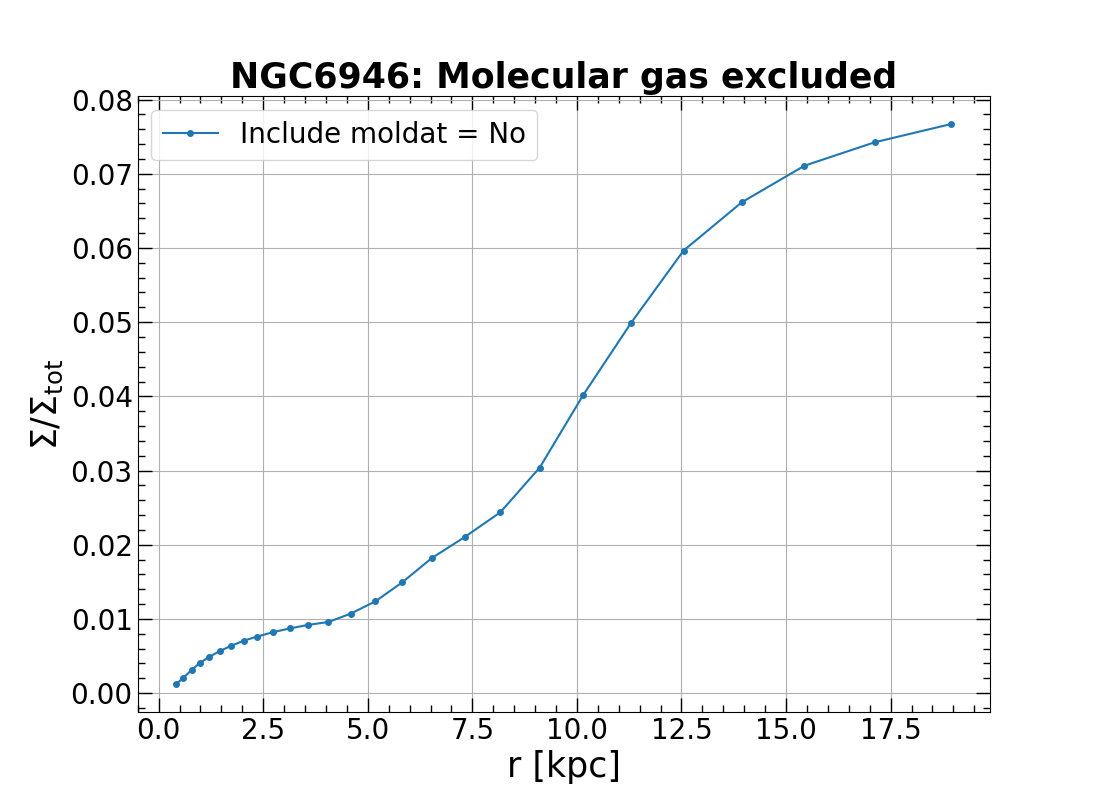}
    \includegraphics[width=8cm,keepaspectratio]{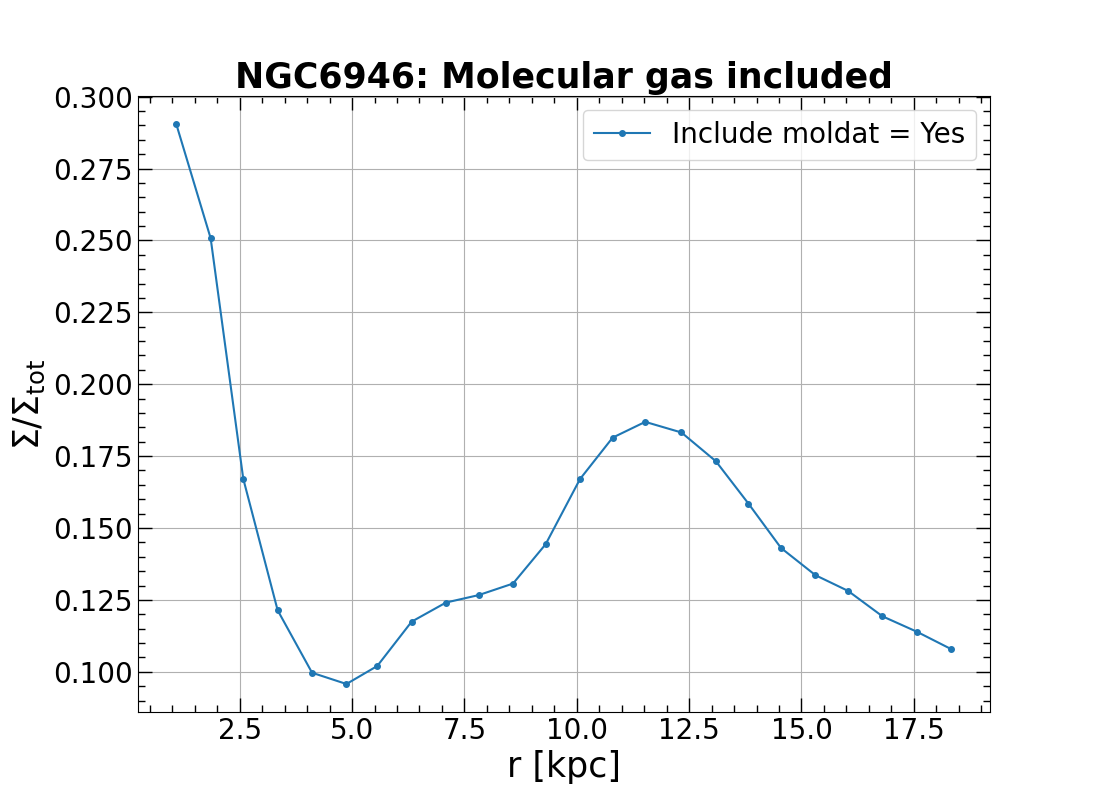}
    
    \caption{Ratio of gas to stellar/total surface density.
    }
    \label{fig:gas_to_star_ratio}
\end{figure*}
\end{center}

Figure \ref{fig:gas_to_star_ratio} plots the gas to star surface density ratios found using the interpolated data against the radius for each galaxy. The left panels consider $\Sigma
  = \frac{3 \mu}{4-\mu} \Sigma_{\mathrm{HI}}$ (molecular gas excluded), and the right panels take $\Sigma = \frac{3 \mu}{4-\mu} \Sigma_{\mathrm{HI}} + \frac{\mu^{\prime}}{4-\mu^{\prime}} \Sigma_{\mathrm{H_2}}$. 

\end{document}